\documentclass[12pt]{article}
\usepackage{epsf}
\usepackage[usenames]{color}
\usepackage{pstricks,fancybox}
\usepackage{amsmath,amssymb,euscript,graphicx,cite}
\fontfamily{pnc}\selectfont
\def\N{{\cal N}}
\def\ms{{\mathfrak M}}

\newcommand{\EQ}[1]{\begin{equation} #1 \end{equation}}

\newcommand{\SP}[1]{\begin{equation}\begin{split} #1 \end{split}\end{equation}}

\begin{document}
\setlength{\parskip}{2ex} \setlength{\parindent}{0em}
\setlength{\baselineskip}{3ex}
\newcommand{\onefigure}[2]{\begin{figure}[htbp]
         \caption{\small #2\label{#1}(#1)}
         \end{figure}}
\newcommand{\onefigurenocap}[1]{\begin{figure}[h]
         \begin{center}\leavevmode\epsfbox{#1.eps}\end{center}
         \end{figure}}
\renewcommand{\onefigure}[2]{\begin{figure}[htbp]
         \begin{center}\leavevmode\epsfbox{#1.eps}\end{center}
         \caption{\small #2\label{#1}}
         \end{figure}}
\newcommand{\comment}[1]{}
\newcommand{\myref}[1]{(\ref{#1})}
\newcommand{\secref}[1]{sec.~\protect\ref{#1}}
\newcommand{\figref}[1]{Fig.~\protect\ref{#1}}
\def\sl2z{SL(2,\Z)}
\newcommand{\mathbold}[1]{\mbox{\boldmath $\bf#1$}}
\newcommand{\mJ}{\mathbold{J}}
\newcommand{\momega}{\mathbold{\omega}}
\newcommand{\bz}{{\bf z}}
\newcommand{\be}{\begin{equation}}
\newcommand{\ee}{\end{equation}}
\newcommand{\bea}{\begin{eqnarray}}
\newcommand{\eea}{\end{eqnarray}}
\newcommand{\nn}{\nonumber}
\newcommand{\unit}{1\!\!1}
\newcommand{\R}{\bf R}
\newcommand{\X}{{\bf X}}
\newcommand{\T}{{\bf T}}
\newcommand{\PP}{\bf P}
\newcommand{\CC}{\bf C}
\newdimen\tableauside\tableauside=1.0ex
\newdimen\tableaurule\tableaurule=0.4pt
\newdimen\tableaustep
\def\phantomhrule#1{\hbox{\vbox to0pt{\hrule height\tableaurule width#1\vss}}}
\def\phantomvrule#1{\vbox{\hbox to0pt{\vrule width\tableaurule height#1\hss}}}
\def\sqr{\vbox{%
\phantomhrule\tableaustep
\hbox{\phantomvrule\tableaustep\kern\tableaustep\phantomvrule\tableaustep}%
\hbox{\vbox{\phantomhrule\tableauside}\kern-\tableaurule}}}
\def\squares#1{\hbox{\count0=#1\noindent\loop\sqr
\advance\count0 by-1 \ifnum\count0>0\repeat}}
\def\tableau#1{\vcenter{\offinterlineskip
\tableaustep=\tableauside\advance\tableaustep by-\tableaurule
\kern\normallineskip\hbox
    {\kern\normallineskip\vbox
      {\gettableau#1 0 }%
     \kern\normallineskip\kern\tableaurule}%
  \kern\normallineskip\kern\tableaurule}}
\def\gettableau#1 {\ifnum#1=0\let\next=\null\else
  \squares{#1}\let\next=\gettableau\fi\next}

\tableauside=1.0ex \tableaurule=0.4pt

\def\N{{\cal N}}
\def\ms{{\mathfrak M}}
\def\T{{\bf T}}
\def\Z{{\bf Z}}
\def\CC{{\bf C}}
\def\R{{\bf R}}
\def\X{{\bf X}}
\def\S{{\bf S}}
\def\PP{{\bf P}}

\bibliographystyle{utphys}
\setcounter{page}{1} \pagestyle{plain}
 \begin{titlepage}
\begin{center}
\hfill hep-th/0310272\\ \vskip 1cm {\Large{\sc Matrix Models,
Geometric Engineering and  Elliptic Genera}} \vskip 0.5cm \large{\sc
Timothy Hollowood$^{1}$\,,\,\,Amer\,\, Iqbal$^{2}$\,,\,\,Cumrun
Vafa$^{2}$}\vskip 0.5cm {\small $^{1}$Department of Physics,\\
University of Wales Swansea,\\
Swansea, SA2 8PP, UK.\\
\vskip 0.5cm
$^{2}$Jefferson Physical Laboratory,\\ Department of Physics,\\
Harvard University,\\ Cambridge, MA 02138, U.S.A.\\}\end{center}

\begin{abstract}{ We compute the
prepotential of ${\cal  N}=2$ supersymmetric gauge theories in four
dimensions obtained by toroidal compactifications of gauge theories
from 6 dimensions, as a function of K\"ahler
 and complex moduli of $\T^2$.  We use three different methods
to obtain this: matrix models, geometric engineering and instanton calculus.
Matrix model approach involves summing up planar diagrams of an associated
gauge theory on $\T^2$.  Geometric engineering involves considering
F-theory on elliptic threefolds, and using topological vertex to sum up
worldsheet instantons.  Instanton calculus involves computation of elliptic
genera of instanton moduli spaces on $\R^4$.
We study the compactifications of ${\cal N}=2^*$ theory
in detail and establish equivalence of all these three approaches in this case.
As a byproduct we geometrically engineer theories with massive
adjoint fields.
 As one application, we show that the moduli space of mass deformed
M5-branes wrapped on ${\bf T^2}$ combines the K\"ahler and complex moduli of
$\T^2$ and the mass parameter into the period matrix of a genus 2 curve.}
\end{abstract}
\end{titlepage}

\baselineskip 0.69cm

\section{Introduction} String theory has been rather successful in
providing insights into the dynamics of supersymmetric gauge theories in 4
dimensions.  In particular essentially all questions involving vacuum
geometry can be settled exactly for a large class of gauge theories;
all the F-terms are exactly computable.  Topological strings on Calabi-Yau
geometries have played a key role in this regard.  In particular
consideration of type IIA (and topological A-model) strings on local
Calabi-Yau threefolds leads to exact results, via geometric engineering,
to questions involving a large class of ${\cal N}=2$ supersymmetric
gauge theories in 4 dimensions \cite{KKV,Katz:1997eq}.  Also consideration
of type IIB (and topological B-model) geometries with wrapped and spacetime
filling branes leads to exact results for ${\cal N}=1$ supersymmetric gauge
theories \cite{DV1, Cachazo:2001jy}, which is also equivalent to the matrix model
realization of a perturbative window into non-perturbative dynamics of these
theories.  This approach can also be used to address questions involving
${\cal N}=2$ supersymmetric theories, as this is a special case of
${\cal N}=1$ supersymmetric gauge theories. There has been another approach
developed recently \cite{Nekrasov:2002qd} for answering F-term questions
involving ${\cal N}=2$ supersymmetric gauge theories. This involves the
development of an  instanton calculus, and can be viewed as an efficient
method to do the relevant integration over the instanton moduli space.

 One  can also ask questions about the dynamics of higher dimensional supersymmetric
gauge theories, which will be the main focus of this paper.  Moreover we
will focus mainly on the overlap of these approaches that relate to theories
with $8$ supercharges.   All these three approaches can be extended to higher
dimensions and more specifically to dimensions 5 and 6. In the geometric
engineering approach to go from $4\rightarrow 5 \rightarrow 6$ one has to
consider the chain of duality  between type IIA on Calabi-Yau $\X $ with
M-theory on $\X\times {\bf S}^1$ and F-theory on $\X\times \T^2$.  The latter
duality requires ellipticity of $\X$ \cite{Vafa:1996xn,Morrison:1996na,Morrison:1996pp}
and this gets related to the fact that only special 6D gauge theories with 8
 supercharges are anomaly free. In the Matrix model approach to go from
 $4\rightarrow 5 \rightarrow 6$ one considers associated gauge theories
 in $0,1,2$ dimensions respectively, corresponding to geometry of point,
 ${\bf S}^1$, $\T^2$ \cite{DV4}. In the instanton calculus approach one
 replaces (for the case of ${\cal N}=2^*$) the measure from $1$
 to arithmetic genus $\chi$ and then to
 elliptic genus, in going from 4 to $5$ and then $6$ dimensions
 (the last point will be explained in this paper).
\onefigure{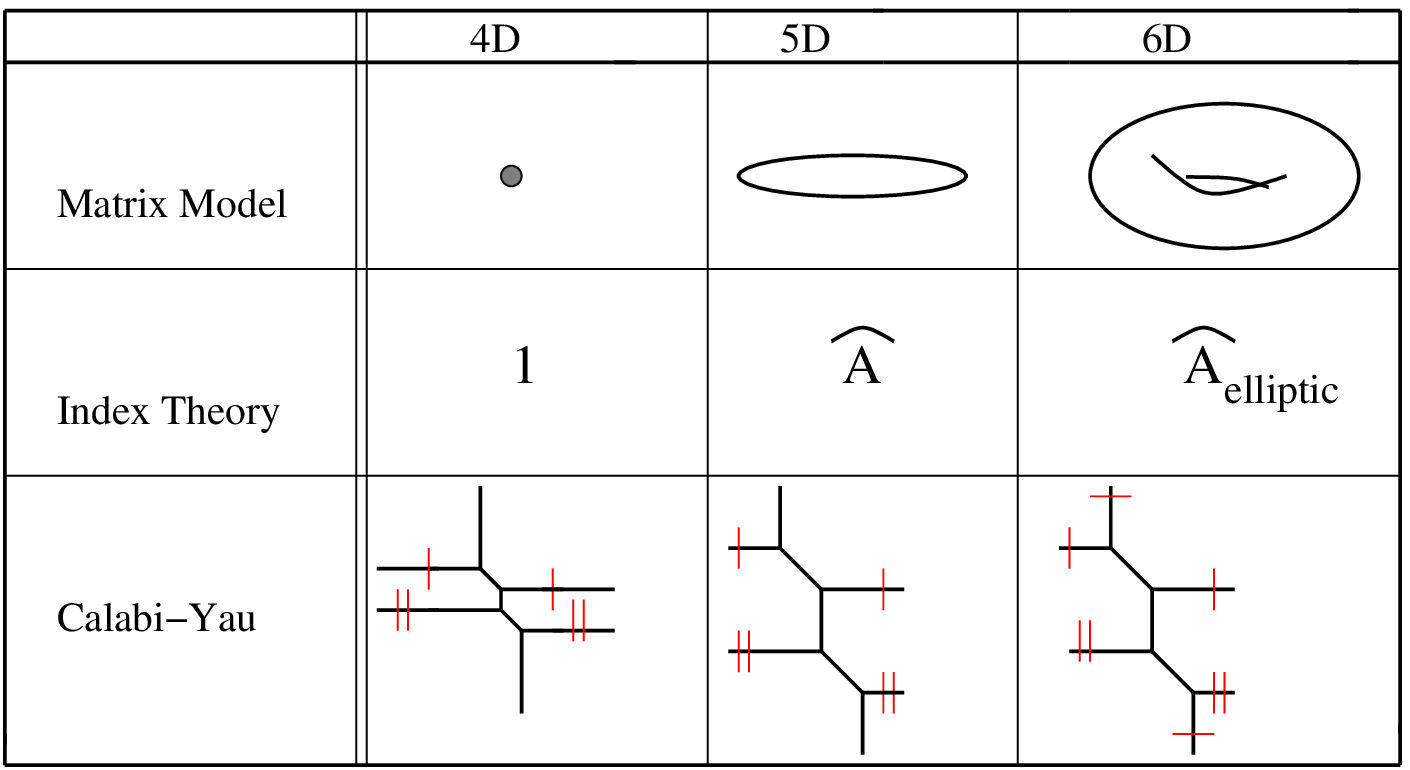}{F-term computations for
supersymmetric gauge theories from the view point
of matrix models, instanton calculus and geometric engineering,
in 4, 5 and 6 dimensions. The vertical and horizontal line segments on the
external line of the web shown in the figure indicate gluing of those lines.}

We will restrict to a special class of gauge theories, namely those which do
exist as anomaly free theories in 6 dimensions. In particular we
will focus mainly on $U(N)$ coupled with an adjoint matter,
known as ${\cal N}=2^*$; we also discuss as a further example how these
generalize
to the theory with $2N$
fundamental hypermultiplets
in the terminology of ${\cal N}=2$ supersymmetric theories in 4 dimensions.

In the course of implementing these ideas we solve a number of related problems:
we find a nice way to summarize the integrality predictions
\cite{Gopakumar:1998ii,Gopakumar:1998jq} of topological string free energy $F$
in terms of the integrality of the partition function $Z=\exp F$.   We also use the
more refined information of instanton calculus \cite{Nekrasov:2002qd}\ to shed
light on the meaning of it in terms of curve counting for toric Calabi-Yau.  We
apply the topological vertex
to double elliptically fibered Calabi-Yau (the possibility of doing this
was noted in \cite{AKMV})
and in doing so we end up geometrically engineering theories with adjoint matter
(${\cal N}=2^*$) on the one hand, and lifting theories from 5-dimensional M-theory,
to 6 dimensional F-theory (with elliptic 3-folds) on the other. Moreover we show
that, for the
simplest gauge theory with gauge group $U(1)$,
the relevant local model involves
combining the K\"ahler class of the two elliptic fibrations as the elliptic moduli
of the ``two tori'' of a genus 2 curve. In relating these to the instanton calculus
approach we end up studying the (equivariant) elliptic genus on the moduli space of
instantons on $\R^4$.  For the case of $U(1)$ gauge theory this gets related to
the elliptic genus for symmetric products of $\R^4$.  Elliptic genera
of symmetric
products have been studied \cite{Dijkgraaf:1996xw}\ and it turns out
that there, the
double ellipticity (coming from the elliptic genus on the one hand, and the
parameter counting the number of copies of the symmetric product on the other)
and the appearance of genus $2$ curve was already apparent.  The study of elliptic
genera of
symmetric products of instanton moduli spaces in \cite{Dijkgraaf:1996xw}\
was motivated by the question of 5D black hole entropy \cite{Strominger:1996sh}\
(see also \cite{Katz:1999xq}). As for the matrix model approach,
going from $4\rightarrow 5\rightarrow 6$ involves
changing the spectral plane from $\CC$ to $\CC^*$ and
then to $\T^2$. In the case of the ${\cal N}=2^*$
theory with gauge group $U(1)$, a genus 2 curve arises naturally as well.

The organization of this paper is as follows.
In Section 2 we apply the matrix model techniques to
study aspects of gauge theories in 5 and 6 dimensions.
In Section 3 we review basic aspects of geometric engineering
in 4, 5 and 6 dimensions, including theories with adjoint matter.
In Section 4 we review
topological A-model strings,
and the integrality structure of its partition function. We also
discuss how to use topological vertex to compute these amplitudes.
  In Section 5 we apply topological vertex techniques to
calculate prepotentials for gauge theories in 4, 5 and 6 dimensions.
As examples we take ${\cal N}=2^*$
gauge theories as well as $U(N)$ theories with $2N$
fundamentals in these dimensions (for explicit example we take the cases of
$N=1,2$). In Section 6 we review aspects of instanton calculus and apply
it to the theories under consideration.  We explain how elliptic genus of
moduli space of instantons arises in studying gauge theory questions in 6
dimensions. In Section 7 we relate the ${\cal N}=2^*$ theory lifted to 6 dimensions
to the deformed theory of the M5-brane, or NS5-brane wrapped on $\T^2$,
but deformed
with a mass parameter.  We discuss the implication of the appearance of the genus 2
curve from this perspective.

\section{The matrix model approach}

In this section we discuss how we can obtain results for prepotential
of ${\cal N}=2$ supersymmetric theories in 4 dimensions, obtained
from compactification of gauge theories in 6 dimensions on $\T^2$
using matrix model techniques \cite{DV3} adapted to higher dimensional gauge
theories \cite{DV4}. The idea is to consider deformations of ${\cal N}=2$
theory by an ${\cal N}=1$ preserving superpotential. This superpotential
is just a convenience which allows one to probe a particular point on the
Coulomb branch and at the end its strength may be taken to zero
\cite{Cachazo:2002pr}.  Thus we start with a gauge theory on $\T^2$
which encodes the superpotential of the corresponding ${\cal N}=1$ theory,
as in \cite{DV4}\ and compute the glueball superpotential by studying the
planar diagrams of that theory. We then extremize it to find the superpotential
and the $U(1)$ gauge theory coupling constants which are encoded by the
geometry (period matrix) of the resolvent curve.  Since in this paper we
would be mostly interested in the ${\cal N}=2$ aspects of the theory,
we will mainly keep track of the geometry of the curve because it is a
feature that survives the limit when the superpotential is turned off
and so pertains to the $\N=2$ theory \cite{Cachazo:2002pr}.
  We will consider one main example with gauge group $U(N)$,
 to illustrate these ideas: $\N=2^*$ ({\it i.e.\/}~the ${\cal N}=2$
 theory with a massive adjoint hypermultiplet).  Note that
 the choice of the gauge theory should be
  such that it is anomaly free for the 6 dimensional
  chiral theory and these two classes are consistent with that.   These
 techniques can be easily generalized to many other examples, which we leave
 to the interested reader.

\subsection{Engineering the Curve from the Matrix Model}

In this section, we show how the curve for the six-dimensional theory
can be engineered from a matrix model applying the techniques
developed in \cite{DV4,Cachazo:2002pr}. More precisely, we will consider the
six-dimensional $U(N)$ gauge theory with $\N=(1,1)$ supersymmetry
compactified on a torus $\T^2$ defined by
\bea
\T^2=\big\{y\ \big|\
y\thicksim y+\frac\beta{2{\rm Im}\rho}(p+q\rho)\ ,
\quad p,q\in{\bf Z}\big\}\ ,
\eea
where $\beta$ is a length scale and $\rho$ is the complex
structure of the torus.
The effective theory in four dimensions will be the $\N=4$ gauge
theory. However, we can also incorporate a mass for an adjoint
hypermultiplet in the compactification. If $\T^2$ has finite volume
then the effective theory in four dimensions with be generalization of
the $\N=2^*$ theory involving all the Kaluza-Klein modes of the fields
on the torus.

However, if we break this effective four-dimensional
to $\N=1$ by adding an arbitrary superpotential for the one massless
adjoint chiral multiplet then we can use the higher-dimensional
generalization of the holomorphic matrix
integral approach, described in \cite{DV4},
to find the effective superpotential. In other words, we need to
generalize \cite{Hollowood:2003gr}, which considered the
five-dimensional lift of the $\N=1^*$ theory, to the six-dimensional
lift. Correspondingly, we have to lift the matrix quantum mechanics to
a two-dimensional matrix field theory, {\it i.e.\/}~a two-dimensional
gauge theory.

{}From the point-of-view of the
effective four-dimensional theory there are
 3 adjoint chiral fields $\Phi_i$,
$i=1,2,3$. One of the fields, say
$\Phi_3$, is now interpreted as the holomorphic
component of the six-dimensional gauge field along
the compactification torus. According to the general procedure of
\cite{DV4}, after breaking to $\N=1$,
the superpotential of the effective
four-dimensional theory is determined by a two-dimensional gauge theory
involving the fields $\Phi_i(y,\bar y)$ and defined by the
partition function
\EQ{
Z=\int\prod_{i=1}^3[d\Phi_i]\,\exp\big(-g_s^{-1}\int
d^2y\,W(\Phi_i)\big)\ ,
\label{part}
}
where $g_s$ is a coupling constant.
The action of the matrix model is a generalization of the one
 that describes
the $\N=1^*$ deformation of the four dimensional theory \cite{mm1,mm2}:
\EQ{
W(\Phi_i)={\rm
  Tr}\big(\Phi_1D_{\bar y}\Phi_2+m\Phi_1\Phi_2+V(\Phi_3)\big)
\label{act}
}
where the covariant derivative is
$D_{\bar y}\Phi_2=\partial_{\bar
  y}\Phi_2+[\Phi_3,\Phi_2]$. If we want to engineer the Seiberg-Witten
curve of the six-dimensional theory on a torus, then
the potential $V(\Phi_3)$ has to be chosen to be suitably generic in
order that its critical points allow one to track across the Coulomb
branch of the $\N=2^*$. At the end, the strength of $V(\Phi_3)$ can
then be taken to zero and results regarding the $\N=2^*$ theory are
obtained. We will make a suitable choice for $V(\Phi_3)$ later.

In order to complete the description of the theory we need to
specify the measure for the integrals in \eqref{part}.
Part of the matrix model approach involves
interpreting the integrals in a holomorphic way. To be concrete, we
can subject the matrices to particular reality
conditions.
In the present case, we take $\Phi_1^\dagger=\Phi_2$, or equivalently
$\Phi_1+\Phi_2$ and $i(\Phi_1-\Phi_2)$ are Hermitian.
In particular, the measure for the latter combination of fields is the
appropriate measure for Hermitian fields.
The gauge field component $\Phi_3(y,\bar y)$
is treated in a somewhat different
manner since it is the anti-holomorphic component of a gauge field on
$\T^2$. First of all,
local gauge transformations on the torus can be used to transform
$\Phi_3$ into a constant diagonal matrix:
\EQ{
UD_{\bar y}U^{-1}={\rm
  diag}\big(\phi_1,\ldots,\phi_N\big)\ .
}
This fixes all of the gauge group apart from permutations of the
diagonal elements and
large gauge transformations on the torus $\T^2$ in the abelian $U(1)^N$
subgroup. These latter group elements are
\EQ{
U_i=
\exp\Big(\frac{2i\pi}\beta\big((p_i+\rho q_i)\bar
y-(p_i+q_i\bar{\rho})y\big)\Big)\ ,\qquad
p_i,q_i\in{\bf Z}
}
for $i=1\,\ldots,N$. These transformations have the effect of shifting
\EQ{
\phi_i\to\phi_i+\frac{2\pi i}{\beta}(p_i+q_i\rho)\ .
\label{doubp}
}
In other words, the $\phi_i$ are naturally defined on the
dual to the compactification torus which
we denote $\tilde \T^2$:
\EQ{
\tilde \T^2=\big\{x\ \big|\
x\thicksim x+\frac{2\pi i}\beta(p+q\rho)\ ,\quad p,q\in{\bf Z}\big\}\ .
\label{deftt}
}
This torus also has a complex structure $\rho$.

Following the logic of \cite{DV2,me1,me2}, we integrate out the
fields $\Phi_1(y,\bar y)$ and
$\Phi_2(y,\bar y)$ since they appear Gaussian in \eqref{part} and
gauge fix $\Phi_3$ in the way described above.
We end up with a (zero-dimensional) matrix integral involving the
quantities $\phi_i$:
\EQ{
Z=\int \prod_{i=1}^Nd\phi_i\,
\frac{{\rm Det}'(D_{\bar y})}{
{\rm Det}(D_{\bar y}+m)}
\exp\big(-g_s^{-1}v\sum_{i=1}^NV(\phi_i)\big)\ ,
\label{mtin}
}
where $v$ is the volume of $\T^2$.
The determinant in the numerator is the gauge-fixing Jacobian while
the one in the denominator arises from integrating out $\Phi_{1,2}$.
For consistency, we now see that the probe potential
$V(\phi)$ must respect the double-periodicity
of the torus $\tilde \T^2$ \eqref{doubp}:
\EQ{
V(x)=V(x+\frac{2\pi i}\beta(p+q\rho))\ ,\qquad p,q\in{\bf Z}\ .
}

It is straightforward to evaluate the ratio of determinants in \eqref{mtin}.
To start with, consider the simplified quantity
\EQ{
{\rm Det}(\partial_{\bar y}+C)\ ,
}
where $C$ is a constant. Take the eigenvalue equation
\EQ{
\big(\partial_{\bar y}+C\big)\psi(y,\bar y)=\lambda\psi(y,\bar y)\ .
}
The eigenvectors and eigenvalues can be found explicitly:
\EQ{
\psi(y,\bar y)=
\exp\Big(\frac{2i\pi}\beta\big((p+q\rho)\bar y-(p+q\bar
{\rho})y\big)
\Big)\qquad
p,q\in{\bf Z}\ ,
}
and
\EQ{
\lambda=C+\frac{2\pi i}\beta(p+q\rho)\ .
}
Therefore the determinant, up to an infinite
factor which will cancel between the denominator and numerator in
\eqref{mtin}, is
\EQ{
{\rm Det}(\partial_{\bar y}+C)\thicksim\prod_{p,q}\Big(C+
\frac{2\pi i}\beta(p+q\rho)\Big)\ .
}
Using the identities
\EQ{
\sin x=x\prod_{n=1}^\infty\big(1-\tfrac{x^2}{\pi^2n^2}\big)
\
,\qquad\theta_1(z|\tau)=q^{1/4}e^{iz}\prod_{n=1}^\infty(1-q^{2n})
(1-q^{2n-2}e^{-2iz})(1-q^{2n}z^{2iz})
}
we can write the ratio of the determinants in \eqref{mtin} in terms of
elliptic theta functions:
\EQ{
\frac{{\rm Det}'(D_{\bar y})}{
{\rm Det}(D_{\bar y}+m)}\thicksim
\frac{\prod_{i\neq
    j}\theta_1\big(\tfrac{\beta}{2i}(\phi_i-\phi_j)
\big|\rho\big)}{\prod_{i
    j}\theta_1\big(\tfrac{\beta}{2i}(\phi_i-\phi_j+m)
\big|\rho\big)}\ ,
}
up to a $\phi_i$ independent multiplicative factor.

Now we are ready to perform a large-$N$ saddle-point evaluation of the
remaining matrix model around a critical point. In order to engineer
the Seiberg-Witten curve for this theory, $V(x)$ must have at
least $N$ critical points. Given this, one expands around a critical point
where there is one eigenvalue $\phi_i$ in a subset of $N$ of the
critical points. We will make a convenient choice for $V(x)$ later.
As usual in the matrix model we replace $N\to\hat N$ and
introduce a degeneracy $\hat N_i$ at each of the $N$ critical points
inhabited by a field theory eigenvalue. We then take the limit
$N_i\to\infty$, $g_s\to 0$ with $S_i=g_s\hat N_i$ fixed.
In the large-$\hat N$ limit, the eigenvalues $\phi_i$ form a continuum and
condense onto $N$ open contours on the dual torus $\tilde \T^2$.
We define these contour by specifying the end-points:
\EQ{
{\cal C}_i=[a_i,b_i]\ .
}
We also define the union
\EQ{
{\cal C}=\bigcup_{i=1}^N{\cal C}_i\ .
}
The configuration is described by the
density of eigenvalues $\varrho(x)$, a function which has support only along
the $N$ contours, and which we normalize according to
\EQ{
\int_{{\cal C}}\varrho(x)\,dx=1\ .
}

\begin{figure}
\begin{center}
\includegraphics[scale=0.6]{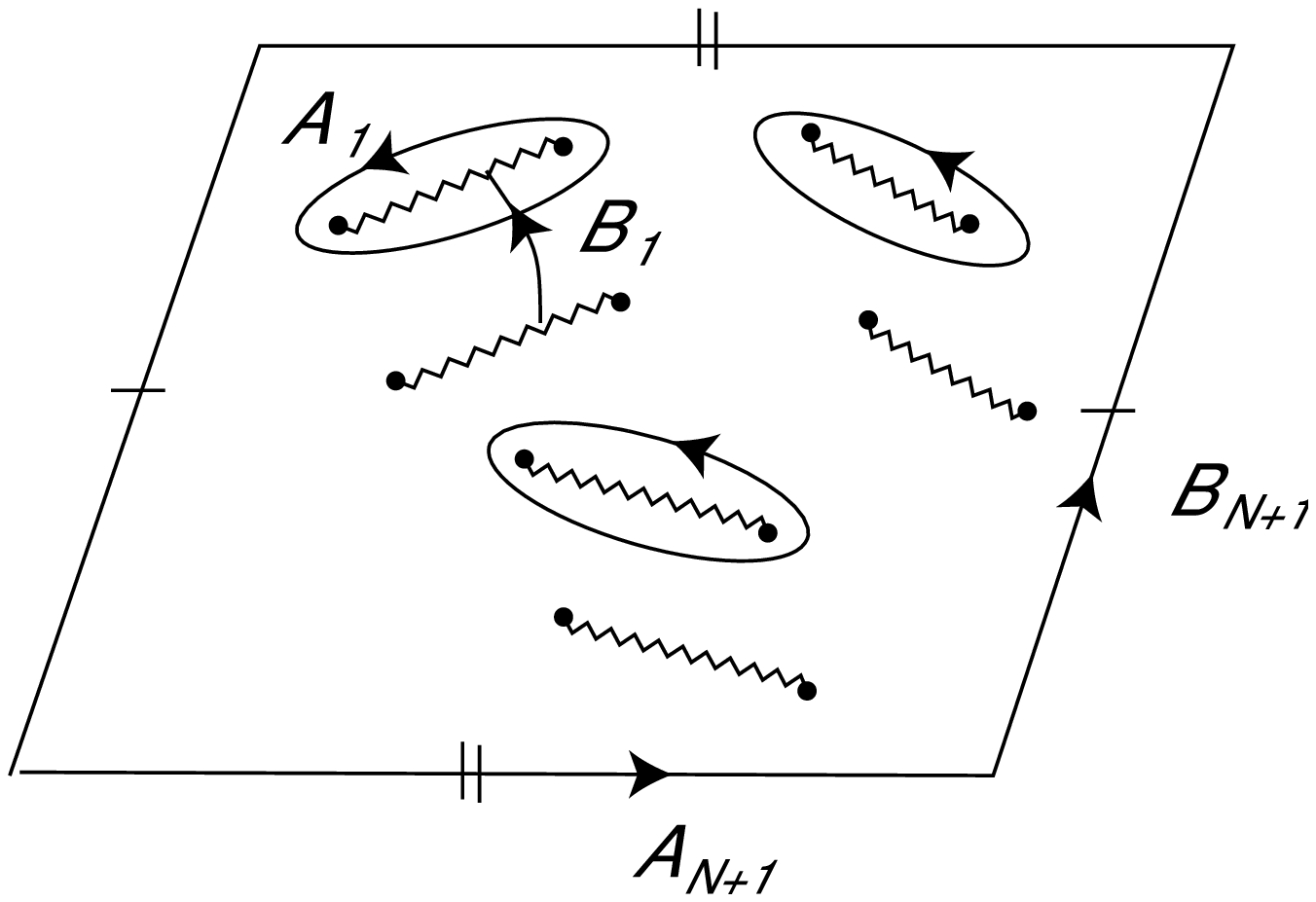}
\end{center} \caption{\small The cut torus on which $G(x)$ is
defined for the case $N=3$. Each pairs of cuts is identified. The
cycles $A_i$ and $B_i$,
 $i=1,\ldots,N$ are associated to each pair of cuts and $A_{N+1}$ and
$B_{N+1}$ are the cycles of the torus $\tilde \T^2$.}
\label{mm1}\end{figure}

\begin{figure}
\begin{center}
\includegraphics[scale=0.5]{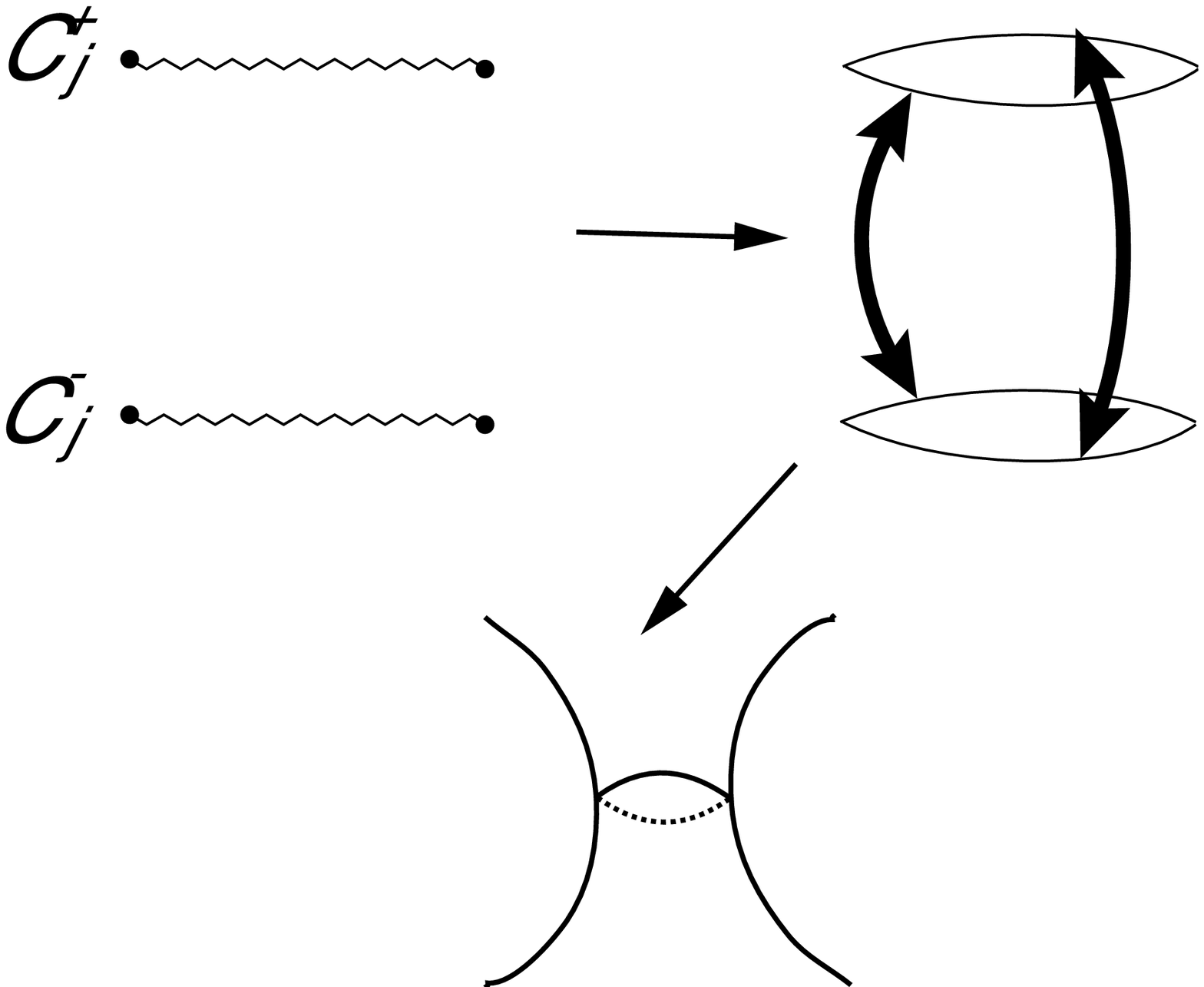}
\end{center}
\caption{\small The top (bottom) of ${\cal C}_j^+$ is identified with
the bottom (top) of ${\cal C}_j^-$. The figure shows how this
generates a handle in the surface on which $G(x)$ is defined.}
\label{mm2}\end{figure}

The saddle-point equation is most conveniently formulated after defining the
resolvent function
\EQ{
\omega(x)=\int_{{\cal C}}dy\,
\varrho(y)\partial_x\log\theta_1\big(\tfrac{\beta}{2i}(x-y)\big|
\rho\big)\ .
\label{defres}
}
This function is a multi-valued function on the torus $\tilde \T^2$,
\EQ{
\omega(x+2\pi i/\beta)=\omega(x)\
,\qquad\omega_2(x+2\pi i\rho/\beta)=\omega(x)-\beta^{-1}\ ,
}
except cuts along the $N$ contours ${\cal C}_i$.
The matrix model spectral density
$\rho(x)$ is then equal to the discontinuity across the cut
\EQ{
\omega(x+\epsilon)-\omega(x-\epsilon)=2\pi
i\varrho(x)\ ,\qquad x\in{\cal C}\ .
}
In this, and following equations, $\epsilon$ is a suitable
infinitesimal such that $x\pm \epsilon$ lies infinitesimally
above and below the cut at $x$.
The saddle-point equation expresses the condition of
zero force on a test eigenvalue in the presence of the large-$N$
distribution of eigenvalues along the cut:
\EQ{
\frac{vV'(x)}{S}=\omega(x+\epsilon)+\omega(x-\epsilon)
-\omega(x+m)-\omega(x-m)\ ,\qquad x\in{\cal C}\ .
\label{spe}
}
This equation can be re-written in terms of the function
\EQ{
G(x)=U(x)+iS\big(
\omega(x+\tfrac{m}2)-\omega(x-\tfrac{m}2)\big)\
,\label{defg}
}
where $U(x)$ is determined by the finite-difference equation
\EQ{
U(x+\tfrac{m}2)-U(x-\tfrac{m}2)=ivV'(x)\ .
}
{}From its definition, one can see that $G(x)$ is now single-valued on
$\tilde \T^2$ with $N$ pairs of cuts
\EQ{
{\cal C}_i^\pm=[a_i\pm\tfrac{m}2,b_i\pm\tfrac{m}2]\ .
}
This is illustrated in Fig.~\ref{mm1}.
In terms of $G(x)$, the matrix model saddle-point
equation \eqref{spe} is \EQ{ G(x+ \tfrac{m}2\pm
\epsilon)=G(x-\tfrac{m}2\mp \epsilon)\qquad x\in{\cal C}\ .
\label{glue} } These equations can be viewed as conditions which
glue the top (bottom) of ${\cal C}_i^+$ to the bottom (top) of its
partner ${\cal C}_i^-$. This generates a handle as illustrated in
Fig.~\ref{mm2}.
In other words $G(x)$ naturally defines a genus
$N+1$ Riemann surface $\Sigma_{\rm mm}$ defined as the torus
$\tilde \T^2$ with $N$ pairs of cuts ${\cal C}_i^\pm$ which
are glued together in pairs to create $N$ additional handles.

It appears that the resulting Riemann surface has $2N$ moduli provided
by the positions $\{a_i,b_i\}$ of the
ends of the cuts $C_i$. In fact, let us call ${\EuScript M}$ the
moduli space of surfaces defined in this way with (complex) dimension ${\rm
  dim}\,{\EuScript M}=2N$.\footnote{We count all dimensions as complex.}
However, the requirement that a
meromorphic function $G(x)$ exists on the surface with a suitable
polar divisor actually means that the actual moduli space of the
matrix model curve is only an
$N$ (complex) dimensional subspace ${\EuScript M}_{\rm mm}\subset{\EuScript M}$. To see
this notice that $V'(x)$ has by hypothesis at least $N$ zeros and
hence a polar divisor of order at least $N$ on the torus $\tilde
\T^2$. This can be arranged, for example, by
taking $U(x)$ to have a pole of order $N+2$ at a single point
$x_0$ on the torus $\tilde \T^2$. It follows that $V'(x)$ will have a
polar divisor of order $2N+4$ and hence have $2N+4$ zeros.
The reason for this choice is purely
based on convenience as will emerge shortly. We can write our choice
for $U(x)$ explicitly as
\EQ{
U(x)=\frac{\prod_{l=1}^{N+2}\theta_1(\tfrac\beta{2i}(x-c_l)|\rho)}
{\theta_1(\tfrac\beta{2i}(x-x_0)|\rho)^{N+2}}\
,\quad\sum_{l=1}^{N+2}c_l=(N+2)x_0\ .
}
Note that this function is single-valued on $\tilde \T^2$.
With the above choice, $G(x)$ must also have a pole of order $N+2$ at
$x_0$ on $\Sigma$.
For generic $x_0$, the Riemann-Roch Theorem, guarantees that $G(x)$
will be unique up to an overall scaling. Hence, matching the
singular part of $G(x)$ with $U(x)$ at $x_0$ leads to $N$ conditions
on the moduli of the surface. Consequently, the dimension of the moduli space
of matrix model curves ${\EuScript M}_{\rm mm}$ is $N$ as claimed.
Of course, the same counting of moduli will also work for other choices of
$U(x)$ for which $V'(x)$ has at least $N$ zeros, but our choice was
a convenient one.

The $N$ moduli of the surface are encoded in the quantities
$S_j=g_s\hat N_j$ which can be expressed as the
following contour integrals:
\EQ{
S_j=S\int_{{\cal C}_i}dx\,\varrho(x)=-\frac S{2\pi
  i}\oint_{A_j}dx\,\omega(x)
= -\frac1{2\pi}\oint _{A_j} G(x)dx\ ,\qquad j=1,\ldots,N\ ,
\label{wowa}
}
where $A_j$ encircles the cut ${\cal C}_j^+$ as in Fig.~\ref{mm1}.

The other ingredient required to determine the
glueball superpotential of the six-dimensional QFT compactified on the torus
is the variation of the genus zero free energy ${\cal F}_0$ of the
matrix model in transporting a test eigenvalue from infinity to one of
the original $N$ cuts ${\cal C}_j$.
This is obtained by integrating the force on a
test eigenvalue, which
can be expressed in terms of the function $G(x)$ as
\EQ{
-i\big(G(x+\tfrac{m}2)-G(x-\tfrac{m}2)\big)\ ,
}
from infinity to a point on the cut ${\cal C}_j$. This can be written as an
integral of $G(x)$ itself along a contour starting at a point on the
lower cut ${\cal C}_j^-$
going off to infinity and then back to a point on the upper cut
${\cal C}_j^+$. This
can be deformed into the contour
running from a point on ${\cal C}_j^-$ to the image point on
$C_j^+$ related by a shift in $x$ by $m$.
Since the 1-form $G(x)dx$ is single-valued on $\Sigma_{\rm mm}$
this integral is in fact around the closed cycle $B_j$ on $\Sigma_{\rm
  mm}$ conjugate to the cycle $A_j$ defined above: see Fig.~\ref{mm1}. Hence,
\EQ{
{\partial {\cal F}_0\over \partial S_j}= -i\oint_{B_j}
G(x)dx\ ,\qquad j=1,\ldots,N\ .
\label{dfds}
}

According
to the matrix model approach \cite{DV1,DV2,DV3,DV4}, the effective glueball
superpotential in this vacuum where the degeneracies are $N_i=1$
is given by
\EQ{
W_{\rm eff}(S_i)=\sum_{j=1}^N
\Big({\partial {\cal F}_0\over \partial S_j}-2\pi i
\tau S_j\Big)\ ,
}
where $\tau$ is the usual complexified coupling of the
supersymmetric gauge theory in four dimensions.

A critical point of $W_{\rm eff}(S_j)$ corresponds to
\bea
\sum_{j=1}^N\frac{\partial^2{\cal F}_0}{\partial S_k\partial
  S_j}=2\pi i\tau\qquad k=1,\ldots,N\ .
\label{CP} \eea This equation can be written in a more suggestive
way by noticing that
$\omega_j=-\frac1{2\pi}\frac{\partial}{\partial S_j}G(x)dx\,$
$j=1,\ldots,N$ are a subset of the holomorphic 1-forms on
$\Sigma_{\rm mm}$. The reason is that the singular part of
$G(x)dx$ at $x_0$ depends only on $U(x)$ and so is manifestly
independent of the moduli $\{S_j\}$. Furthermore, the $\omega_j$
are normalized so that \EQ{ \oint_{A_j}\omega_k=\delta_{jk}\ . }
Hence \EQ{ \frac{\partial^2{\cal F}_0}{\partial S_k\partial
  S_j}=2\pi i\oint_{B_j}\omega_k=2\pi i\Pi_{jk}\ ,
}
where $\Pi_{jk}$ are
elements of the period matrix of $\Sigma_{\rm
  mm}$ excluding the last row and column.
Consequently the critical point equations are
\EQ{
\sum_{j=1}^N\Pi_{jk}=\tau\qquad k=1,\ldots,N\ .
\label{conc}
}
Given that ${\EuScript M}_{\rm mm}$ is $N$-dimensional, these $N$ conditions
completely fix the geometry of the Riemann surface $\Sigma_{\rm mm}$
in terms of the parameters of the probe potential $V(x)$.

The remaining elements of the period matrix $\Pi$ are fixed in the following
way. Notice that the remaining holomorphic 1-form $\omega_{N+1}$ is
identified with $\beta dx/(2\pi i)$ since
\EQ{
\oint_{A_j}dx=0\ ,\quad j=1,\ldots,N\
,\quad\oint_{A_{N+1}}dx=\frac{2\pi i}{\beta}\ .
}
Hence
\EQ{
\Pi_{N+1,j}=\Pi_{j,N+1}=\oint_{B_j}\omega_{N+1}=\frac\beta{2\pi i}
\int_{x-\tfrac{m}2}^{x+
\tfrac{m}2}dx=\frac{\beta m}{2\pi i}\ ,\quad j=1,\ldots,N\
,
}
while
\EQ{
\Pi_{N+1,N+1}=\oint_{B_{N+1}}\omega_{N+1}=\frac\beta{2\pi i}\int_{0}^{2\pi
  i\rho/\beta}dx=\rho\ .
}
Hence, the period matrix of $\Sigma_{\rm mm}$ at a critical point is
\EQ{
\Pi=\begin{pmatrix} \Pi_{11}&\cdots&\Pi_{1N}& \tfrac{\beta m}{2\pi i}\\
\vdots&\ddots&\vdots & \vdots\\ \Pi_{N1}&\cdots&\Pi_{NN}&
\tfrac{\beta m}{2\pi i}\\
\tfrac{\beta m}{2\pi i}&\cdots&\tfrac{\beta m}{2\pi i}&\rho
\end{pmatrix}\ ,\quad\sum_{j=1}^N\Pi_{jk}=\tau\ .
\label{psw}
}

\subsection{Extracting the Seiberg-Witten curve}

We now show how to extract the Seiberg-Witten curve for the compactified
six-dimensional theory $\Sigma$. The idea is that
this curve at some point in its moduli space
is simply identified with the matrix model curve
$\Sigma_\text{mm}$. By
changing the potential we can move around in the moduli space of the
curve $\Sigma$. In
other words, the Seiberg-Witten curves $\Sigma$ are the
curves in ${\EuScript M}$ subject to the $N$ conditions \eqref{conc}.

The crucial observation is that the
curve $\Sigma$ admits the two multi-valued
functions. Firstly, the
critical point equations \eqref{CP} imply that $z$ defined by
\EQ{
z(P)=\int_{P_0}^P\sum_{j=1}^N\omega_N\ ,
\label{mzd}
}
for an arbitrary point $P_0$, is a multi-valued function on
$\Sigma_\text{mm}$ with
\SP{
&A_j:\quad z\to z+1\ ,\quad B_j:\quad z\to z+\tau\ ,\quad j=1,\ldots,N\\
&A_{N+1}:\quad z\to z\ ,\quad B_{N+1}:\quad z\to
  z+\tfrac{N\beta m}{2\pi i}\ .
\label{multiz}
}
In addition to this we also have the multi-valued function
$x$
\EQ{
x(P)=\frac{\beta}{2\pi i}\int_{P'_0}^P\omega_{N+1}\ ,
\label{mxd}
}
defined with respect to some other, possibly different,
base point $P'_0$, with
\SP{
&A_j:\quad x\to x\ ,\quad B_j:\quad x\to x+m\
,\quad j=1,\ldots,N\\
&A_{N+1}:\quad x\to x+\frac{2\pi i}\beta\ ,\quad B_{N+1}:\quad x\to
  x+\frac{2\pi i}\beta\rho\ .
\label{multix}
}

{}From these monodromy properties
it follows that $\Sigma$ is holomorphically embedded in a
slanted 4-torus $\T^4$. Introducing complex coordinates for $\CC^2$
\EQ{
z_1=z\ ,\qquad z_2=\frac{\beta N}{2\pi i}x\ ,
}
then we can write
\EQ{
\T^4=\Big\{z_i\in\CC^2\ \Big|\
z_i\thicksim z_i+\sum_{\alpha=1}^4\Omega_{i\alpha}p_\alpha\ ,
\quad p_\alpha\in{\bf Z}\Big\}\ ,
\label{defas}
}
where the $2\times4$-dimensional period matrix is
\EQ{
\Omega=\begin{pmatrix} 1 & 0 & \tau & \tfrac{N\beta m}{2\pi i}\\
0 & N & \tfrac{N\beta m}{2\pi i} & N\rho \end{pmatrix}\ .
}
In fact, the form of the period matrix implies that $\T^4$
is an abelian surface, or 2-dimensional abelian variety \cite{GH,LB}.

\begin{figure}
\begin{center}
\includegraphics[scale=0.5]{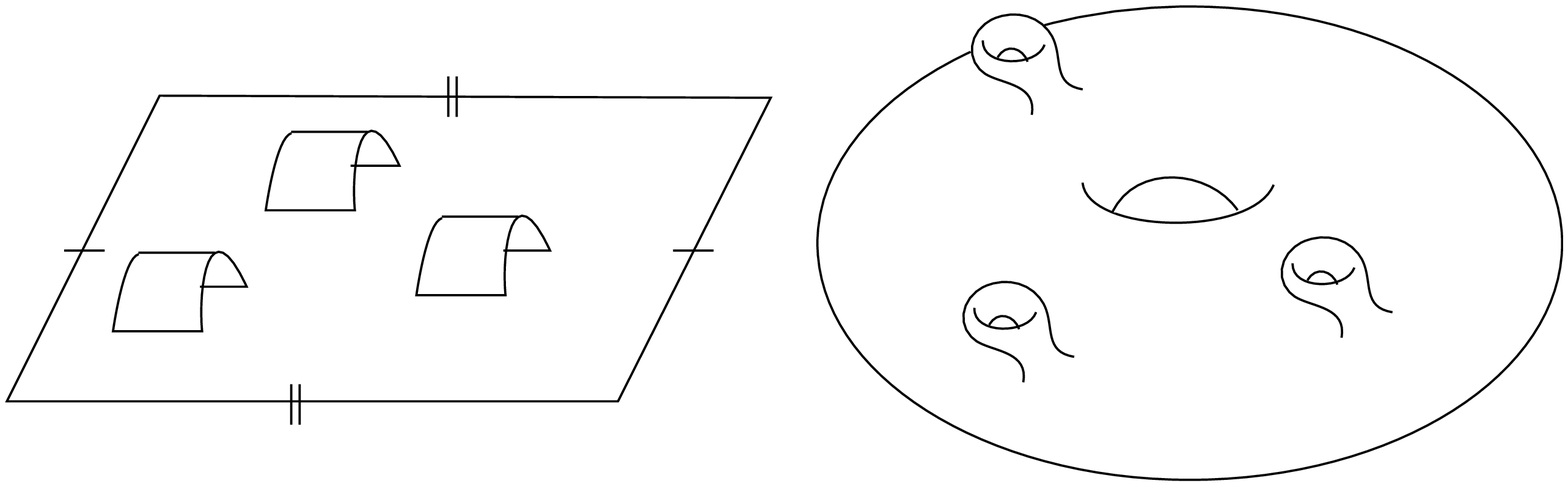}
\end{center}
\caption{\small On the left, the
surface $\Sigma$ realized as the cut $x$-torus $\tilde
\T^2$. The cuts in each of the $N$ pairs are separated by $m$ and are
glued together as in \figref{mm2}. On the right, an impression of the
surface realized as $N$ handles on the $x$-torus.}
\label{mm3}\end{figure}

\begin{figure}
\begin{center}
\includegraphics[scale=0.5]{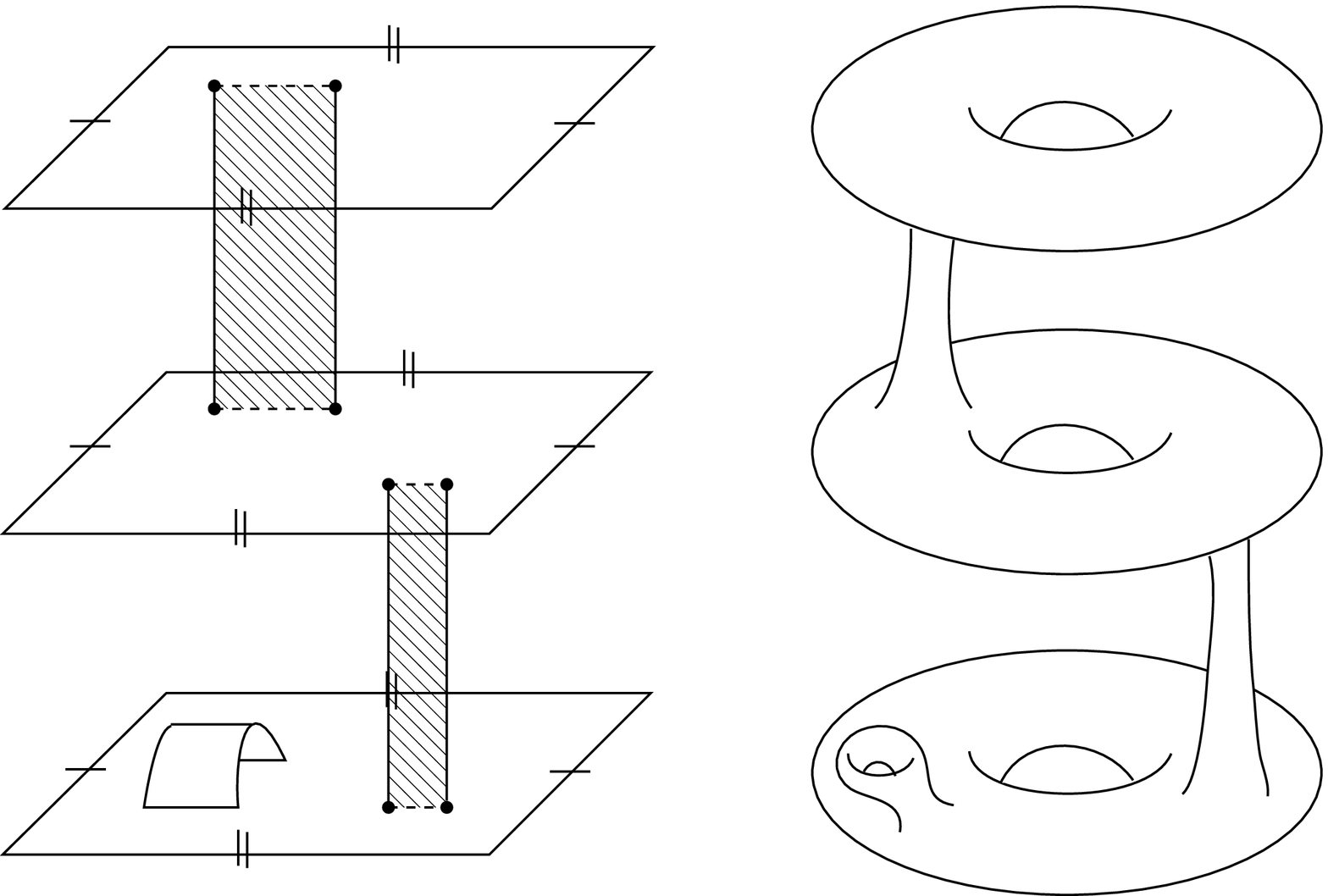}
\end{center}
\caption{\small On the left, the
surface $\Sigma$ realized as $N$ copies of the
$z$-torus connected by $N-1$ branch cuts. On one of the
sheets there is an additional pairs of cuts separated by $N\beta
m/(2\pi i)$ which are glued together as in
\figref{mm2}.
On the right, is
an impression of the surface illustrating the $N$ copies of the $z$-torus
plumbed together along with the additional handle on one of the sheets.}
\label{mm4}\end{figure}

We can picture the curve in two ways. Firstly, as already presented in
the matrix model, as
a torus in the $x$-plane with periods $(2\pi i/\beta,2\pi
i\rho/\beta)$ and with $N$ pairs
of cuts across which $x$ jumps by $\pm m$
whose edges are identified to create a handle as in Fig.~\ref{mm2}.
This is illustrated in \figref{mm3}.
The second representation consists of $N$ copies of
a torus in the $z$-plane with periods $(1,\tau)$ joined by $N-1$
branch cuts. On the face of it, such a surface would have genus $N$ but
on one of the sheets there is a pair of cuts across which $z$ jumps by
$\pm N\beta m/(2\pi i)$ whose edges are identified to create an extra
handle. This is illustrated in Fig.~\ref{mm4}.

Since $\T^4$ is an abelian surface, it turns out that there is an
explicit realization of the curve in terms of generalized
theta-functions associated to $\T^4$ \cite{GH}.
In our conventions, these are defined as
\EQ{
\Theta\left[{\delta\atop\epsilon}\right](Z|\Pi)
=\sum_{m\in{\bf Z}^g}\exp\big(\pi
  i(m+\delta)\cdot\Pi\cdot(m+\delta)
+2\pi i(Z+\epsilon)\cdot(m+\delta)\big)\ .
\label{defgt}
}
In this definition, $Z$, $\delta$, $\epsilon$ and $m$ are $g$-vectors
and $\Pi$ is a $g\times g$ matrix. In the present case, $g=2$ and the
curve can then be written as\footnote{For more details, see \cite{bh}.}
\EQ{
\sum_{j=0}^{N-1}A_j\Theta\left[\begin{matrix}0&\tfrac jN\\
0&0\end{matrix}\right]\left(z\quad\frac
{N\beta x}{2\pi i}\Big|\begin{matrix}  \tau& \tfrac{N\beta m}{2\pi i}\\
\tfrac{N\beta m}{2\pi i}&N\rho\end{matrix}\right)=0\ .
\label{curvb}
}
The coefficients $A_j$ are moduli of the curve. Since the overall
scale of the $A_j$ is unimportant, the moduli are actually valued in ${\bf
  P}^{N-1}$. There are two other moduli corresponding to
moving the curve as a whole in $\T^4$. In all
there are $N+1$ moduli which matches the number of moduli of the
matrix model curve when we include $P_0$ the arbitrary fixed point in
the definition of $z$ in \eqref{mzd}. It can be shown that $\Sigma$ is
in the homology class dual to
\EQ{
Ndy_1\wedge dy_3+dy_2\wedge dy_4\ ,
}
where $y_\alpha$ are real coordinates, $0\leq y_\alpha<1$, with
$z_i=\sum_{\alpha=1}^4\Omega_{i\alpha}y_\alpha$. This is
interpreted as meaning that the curve is
wrapped $N$ times around the $z$ torus and once around the $x$ torus,
as in clear from Figs.~\ref{mm3} and \ref{mm4}.
Similar curves which wrap $k$ times
around the $x$ torus describe the $U(N)^k$ quiver theories in six
dimensions. It is interesting to note that the construction of our
curve is identical to the curve that appears in \cite{Ganor:2000un}
describing instantons in non-commutative gauge theory on $\T^4$. The
relation between the two problems can be made by compactifying our
effective four-dimensional theory down to 3 dimensions, in other words
the six-dimensional theory is on a 3-torus \cite{Cheung:1998wj}.
This is precisely the
philosophy of \cite{Dorey:2001qj} which formulates the problem of
finding the vacuum states of the theory when broken to $\N=1$ in terms
of equilibrium configurations of an integrable system
\cite{Hollowood:2003ds}. This line of thinking leads
to the question of what integrable system lies behind the compactified
six-dimensional theory which generalizes the $N$-body elliptic
Calogero-Moser system, for the four-dimensional theory, and the $N$-body
elliptic Ruijsenaars-Schneider system, for the compactified
five-dimensional theory? It turns out that the resulting system is not
the ``doubly elliptic system'' of
\cite{Fock:1999ae,Braden:2001yc}, rather it is an $N$-body system
where the momenta and positions $(q_i,p_i)$ as complex 2-vectors lie
in the 4-torus $\T^4$ \cite{bh}.

The form of the curve \eqref{curvb} can be re-caste in
the following way which makes the reduction to five and four
dimensions more immediate \cite{bh}:
\EQ{
\sum_{n=0}^\infty\frac1{n!}\Big(\frac m{2\pi i}\Big)^n
\partial_z^n\theta_1\big(\pi z\big|\tau\big)
\partial_x^n H(x)=0\ ,
\label{curva}
}
where
\EQ{
H(x)=\prod_{j=1}^N\theta_1\big(\tfrac\beta
{2i}(x-\zeta_i)\big|\rho\big)\ .
}
Here, $\zeta_i$ are $N$ of the $N+1$ moduli and the remaining one corresponds to
shifting $z$ by a constant.
To go from the six to the five-dimensional curve one takes
$\rho\to i\infty$ in which case
\EQ{
H(x)\to\prod_{j=1}^N\sinh \tfrac{\beta}{2}(x-\zeta_i)\ ,
}
and from the five to the four-dimensional curve one takes
$\beta\to 0$ giving rise to
\EQ{
H(x)\to\prod_{j=1}^N(x-\zeta_i)\ .
}
The curve of the four-dimension theory is identical to the curve
described by Donagi and Witten \cite{Donagi:1995cf}. It is well-known
that this is the spectral curve of the $N$-body elliptic
Calogero-Moser integrable system
\cite{Martinec:1995qn,D'Hoker:1997ha}.
The curve of the
five-dimensional theory can be shown to be the spectral curve of the
Ruijsenaars-Schneider integrable system as predicted by Nekrasov
\cite{nekrasov4}. The relation between this integrable system and the matrix
quantum mechanical system has already been established in
\cite{Hollowood:2003gr}.

The form of the curve \eqref{curva} is
very natural from Type IIA/M Theory elliptic
brane construction \cite{Witten:1997sc}. Using the representation
\EQ{
\theta_1(z|\tau)=\sum_{n\in{\bf
    Z}}(-1)^{n-1/2}e^{i\pi\tau(n+1/2)^2}e^{i(2n+1)z}
}
\eqref{curva} can be written
\EQ{
\sum_{n\in{\bf
    Z}}(-1)^{n-1/2}e^{i\pi\tau(n+1/2)^2}e^{i(2n+1)\pi
  z}H(x+m(n+1/2))=0\ .
}
For the four-dimensional case where $H(x)=
\prod_{i=1}^N(x-\zeta_i)$ we
recognize $z$ and $x$ with the spacetime coordinates as follows
\EQ{
z=(x_{10}+ix_6)/R_{10}\ ,\qquad x=x_4+ix_5\ ,
\label{coords}
}
where $R_{10}$ is the size of the M-theory circle.
The parameters $\zeta_i$ are nothing but the positions of the $N$
D4-branes and the curve takes account of the periodicity in the $x_6$
direction by including an infinite set images each shifted by an
integer multiple of $m$ which identifies $m$ as the hypermultiplet mass.
The five and six-dimensional curves result from compactifying $x=x_4+ix_5$ on
a circle and torus $\tilde \T^2$, respectively.
The replacement $H_{4D}(x)\to H_{5D}(x)\to H_{6D}(x)$
takes account of the compactification by including all the images of
the D4-branes.

To summarize, the Seiberg-Witten curve of the six-dimensional
$\N=(1,1)$ theory compactified on a torus is a Riemann surface
embedded holomorphic in the abelian surface.
In the M-theory formulation, the M5-brane is wrapped on this
Riemann surface.

\section{Geometric engineering of gauge theories}

Calabi-Yau manifolds have played an important role in the study of
supersymmetric gauge theories in various dimensions. The geometry
of CY3-folds has been the source of important insights for gauge
theories. The geometries we will consider in this paper give rise
to gauge theories with $U(N)$ gauge group and fundamental or
adjoint hypermultiplets via geometric engineering as we will
explain later. In this section we will review the geometric
engineering of four \cite{KKV,Katz:1997eq}, five
\cite{Douglas:1996xp,Morrison:1996xf,
Intriligator:1997pq, Leung:1998tw} and six
\cite{Morrison:1996na,Morrison:1996pp}\ dimensional SYM gauge
theories with eight supercharges from CY3-folds.   The basic idea
is to use F-theory compactification on elliptic threefolds times
$\T^2$, and its equivalence to M-theory on the 3-fold times an
$\S^1$ and type IIA on the 3-fold.  Moreover one has to choose
special threefolds which admit appropriate loci of $A_{N-1}$
singularities, to engineer $U(N)$ gauge theories with some matter
content encoded by the geometry. We will also solve a puzzle in
the geometric engineering approach by showing how theories with
massive adjoint matter can be engineered.  In order to motivate
this it is convenient to also review the $(p,q)$ 5-brane web
construction of some of these theories \cite{Kol:1997fv}\ and how
to realize adjoint matter in the brane constructions
\cite{Witten:1997sc}\ and reading off the equivalent CY geometry
from the resulting webs \cite{Leung:1998tw}.

\subsection{ ${\cal N}=4$ $D=4$} Let us begin by considering the
well known case of pure $U(N)$ gauge theory with ${\cal N}=4$
supersymmetry. Type IIA superstrings in the background of $A_{N-1}$ singularity
inside a $K3$ realizes $U(N)$ gauge theory with $16$ supercharges in 6 dimensions.
The $D2$ branes wrapped over the 2-cycles of the blown up geometry realize
the charged fields of the vector multiplet.  Type IIB on the $A_{N-1}$
singularity leads to tensionless strings and is equivalent to $N$ copies of NS5-branes of IIA
\cite{Ooguri:1995wj}.
Now, consider compactifying type IIA strings in the background
of $A_{N-1}$ to 4 dimensions. Depending on how the $A_{N-1}$ geometry
is fibered over the extra 2 dimensions we get various
 kinds of gauge theories.

If we consider a trivial fibration on $\T^2$ we get ${\cal N}=4$
supersymmetric theory in 4 dimensions.  The gauge coupling constant
in 4 dimensions is given by the volume of $\T^2$.
Note that this is also equivalent
to type IIB on the same geometry by doing a T-duality on $\T^2$ exchanging
K\"ahler and complex structures on $\T^2$. The Montonen-Olive duality
is realized in this context by the modular group $SL(2,\Z)$ acting on the
complex
structure of $\T^2$.
Perhaps the most well known way to realize this
theory is on a set of $N$ coincident D3-branes in flat space. By a
chain of dualities this configuration of D3-branes is related to
the set of type IIA NS5-branes wrapped on a  $\T^{2}$.

The prepotential of this 4D theory gets only classical contributions
which, in terms of geometry of $A_{N-1}$, is proportional to
the triple intersection
numbers of the 4-cycles, which include 2-cycles of $A_{N-1}$ times $\T^2$.
To see this note that
$H_{2}(A_{N-1},\Z)$ is isomorphic to the root lattice of
$A_{N-1}$ Lie algebra. The holomorphic curves in $A_{N-1}$ 2-fold
are in one to one correspondence with positive roots of $A_{N-1}$
algebra. Let us denote by $a_{i}$ the moduli of the Coulomb branch such that
$\sum_{i=1}^{N}a_{i}=0$ and by $\phi_{i}=a_{i}-a_{i+1}$ the area of the
curve $F_{i}$ corresponding to the $i^{\rm th}$ simple root, $1\leq i\leq N-1$.
The intersection
number of $F_{i}$ is given by the Cartan matrix $A_{ij}$ {\it i.e.\/}~,
\bea
F_{i}\cdot F_{i}=-2\,,\,\,
F_{i}\cdot F_{i+1}=1\,,\,\,i=1,\ldots, N-2\,.
\eea
Then the prepotential is given by
\bea
{\cal F}=\tfrac{\tau}{2}\,F\cdot F\,,\,\,\,F=\sum_{i,j}\phi_{i}(A^{-1})_{ij}F_{j}\,.
\eea
Thus the geometry of the 2-fold encodes the prepotential is a simple way.
This is also holds for ${\cal N}=2$ 4D theories: the classical contribution
to the prepotential is given by classical intersection numbers of the
CY geometry.   In the case of ${\cal N}=4$ the classical result is exact.

\subsection{${\cal N}=2$, $D=4$ pure $SU(N)$ theory:}
After this brief review of ${\cal N}=4$ theory let us consider
${\cal N}=2$ 4D pure $SU(N)$ theory. The engineering of an
${\cal N}=2$ $SU(N)$ gauge theory requires a singularity of
$A_{N-1}$ type to produce the appropriate gauge symmetry and
another two dimensional space over which $A_{N-1}$ is fibered
to get four non-compact dimensions. However, the 2D space cannot
be arbitrary since the total space has to be CY3-fold. In
the case of ${\cal N}=4$ this was $\T^{2}$ and the CY3-fold was a
product $A_{N-1}\times \T^{2}$ space. To break supersymmetry down
to ${\cal N}=2$ (eight supercharges) the surface should have no
holomorphic one forms and therefore has to be a $\PP^{1}$.
However, the total geometry cannot be a product of $A_{N-1}$ and
$\PP^{1}$ anymore since it is not Calabi-Yau threefold. To obtain
a CY3-fold the $A_{N-1}$ is fibered non-trivially over the $\PP^{1}$. The
details of the ${\cal N}=2$ theory obtained by type IIA
compactification on such a CY3-fold depends on the way $A_{N-1}$
is fibered over the $\PP^{1}$. In the 4D field theory limit, which we
will describe later, all such 3-folds give the same theory after appropriate
identification of parameters.

This theory can also be realized using NS5-branes similar to the case of
${\cal N}=4$. In this case the NS5-branes are wrapped on $\PP^{1}$ inside
$T^{*}\PP^{1}$ (${\cal O}(-2)$ bundle over $\PP^{1}$, (\figref{nsgeometry2})).
\onefigure{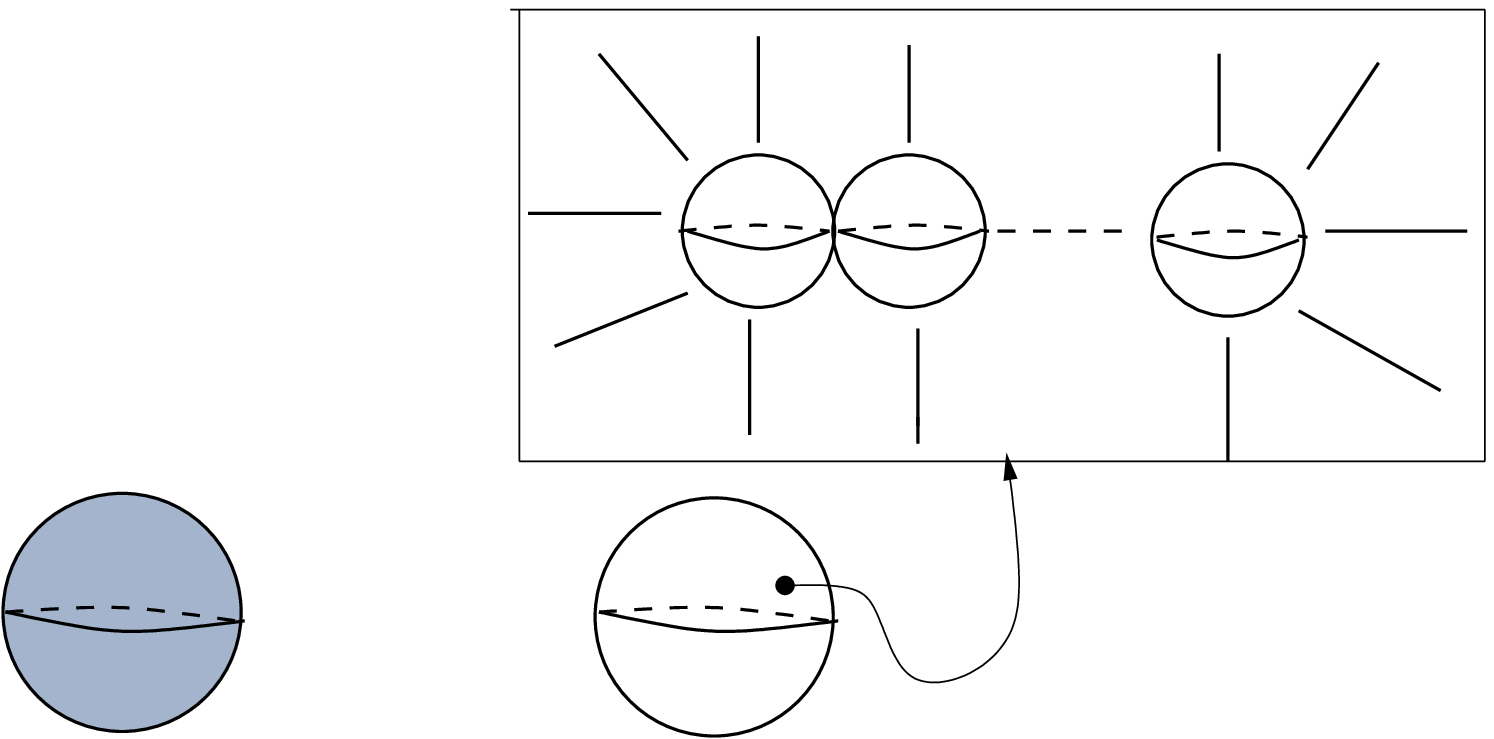}{Realization of pure ${\cal N}=2$ D=4 theory on the
worldvolume of $N$ NS5-branes wrapped on a $\PP^{1}$ (inside $T^{*}\PP^{1}$)
(a) and its dual description in terms of  $A_{N-1}$ fibered CY3-fold (b).}

The four dimensional field theory limit is obtained by taking the string
scale to infinity. By the relations of the base and fiber K\"ahler
parameters to the gauge coupling and W-boson masses, these
parameters must be scaled as \cite{KKV}
\begin{eqnarray} \label{fieldtheorylimit}
Q_{b}:=e^{-T_{b}}=\Big(\tfrac{\beta
\Lambda}{2}\Big)^{2N}\,,\,\,\,Q_{F_{i}}:=e^{-T_{F_{i}}}=e^{-\beta
(a_{i}-a_{i+1})}\,\,i=1,\ldots,N-1\,,
\end{eqnarray}
where $T_b$ denotes the volume of the base $\PP^1$ and $T_{F_i}$
denote the volumes of the fiber $\PP^1$'s.
$\Lambda$ in the above denotes the quantum scale in four
dimensions, $a_{i}$ are the moduli of the Coulomb branch and the
parameter $\beta$ is introduced such that the
four dimensional field theory limit corresponds to $\beta\rightarrow 0$.

The ${\cal N}=2$ prepotential has both 1-loop perturbative and
non-perturbative (instanton) contributions, \bea {\cal F}={\cal
F}_\text{classical}+{\cal F}_\text{1-loop}+\sum_{k=1}^{\infty}
c_{k}(a_{i})\Lambda^{2Nk}\,. \eea In the geometric engineering
picture this is obtained from the genus zero topological string
amplitude of the CY3-fold on which type IIA string theory is
compactified. By considering (\ref{fieldtheorylimit}), it then
becomes clear that the $k$-th gauge instanton contributions stem
from worldsheet instantons that wrap the base $\PP^1$ of our
geometries $k$-times.   As we will discuss later the instanton
contribution are encoded in the classical geometry of type IIB on the mirror
Calabi-Yau, which in turn is equivalent to NS5-branes compactified on a
Riemann surface $\Sigma$  \cite{Klemm:1996bj}.  This can also
be viewed, from the M-theory perspective as M5-brane with worldvolume
of $\Sigma$. In the next section
we will review the brane construction that directly leads to the M5-brane.

\underline{\bf Brane description:}
The D-brane construction of this theory is well known \cite{Witten:1997sc}\ and
involves D4-branes and NS5-branes. Consider type IIA string theory
with two NS5-branes and $N$ D4-branes. The NS5-branes are extended in the
$x^{0,1,2,3,4,5}$ directions, being located at equal values in $x^{7,8,9}$ and
separated in the $x^{6}$ direction by a distance $L$. The D4-branes span
the $x^{0,1,2,3}$ and $x^{6}$ directions, being finite in the $x^{6}$ direction
in which they are suspended between the NS5-branes as shown in
\figref{branes4d}.
\onefigure{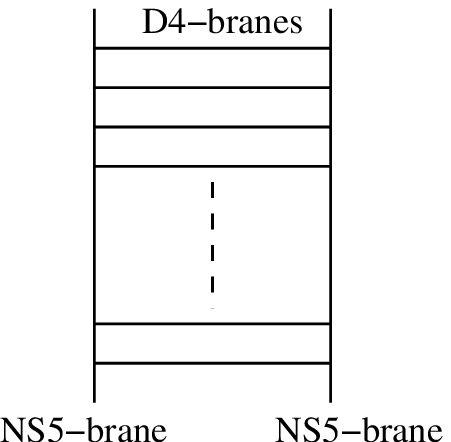}{The brane configuration giving rise to four dimensional
pure ${\cal N}=2$ $U(N)$ theory.}
When the D4-branes are coincident the effective
worldvolume theory is $D=4$ ${\cal N}=2$ $SU(N)$ gauge theory with
coupling constant given by $1/L$. The picture of the geometry shown above is
only approximate and is a good description for small $g_{s}$ since in this
limit we can ignore the effect of D4-brane ending on the NS5-brane. In
general the  NS5-brane will get curved due to the D4-brane. In this limit it is
more useful to lift the above configuration to M-theory and so that
the above configuration of D4-brane and NS5-branes becomes a single M5-brane
wrapped on a Riemann surface (the Seiberg-Witten curve) of genus $N-1$
given by
\bea
y^{2}=\prod_{i=1}^{N}(x-\phi_{i})^{2}-4\,(\tfrac{\Lambda}{2})^{2N}\,.
\eea

\underline{\bf Five dimensional:} As mentioned before in the field theory limit
($\beta \rightarrow 0)$ all CY3-folds which are given by $A_{N-1}$ fibration
over $\PP^{1}$
give the same four dimensional field theory. However, it is possible to
distinguish between these different CY3-folds if we instead consider the
five dimensional $SU(N)$ gauge theory obtained via M-theory compactification
on the same Calabi-Yau times an ${\bf S}^1$.  The parameter $\beta$ gets identified
with the perimeter of ${\bf S}^1$.
Recall that in this case these different CY3-folds are distinguished by
the Chern-Simons coefficient, $k$, of the five dimensional theory {\it i.e.\/}~,
the coefficient of the term
\bea
\int_{\R^{5}} \mbox{Tr} A\wedge F\wedge F\,,
\eea
where $A$ is the gauge field and $F$ is the corresponding field
strength. The cubic part of the prepotential of the five dimensional
$SU(N)$ gauge theory with Chern-Simons coefficient $k$,
in the limit $\beta \rightarrow \infty$,
is given
by the triple intersection numbers of the corresponding CY3-fold
\cite{Intriligator:1997pq},
\bea
{\cal F}^\text{cubic}_{5D}=
\tfrac{1}{6}\Big[\sum_{i,j}\phi_{i}(A^{-1})_{ij}S_{j}\Big]^{3}=
\tfrac{k}{6}\sum_{i}a_{i}^{3}+
\tfrac{1}{6}\sum_{i>j}|a_{i}-a_{j}|^{3}\,.
\eea
$S_{i}$ are the various non-trivial 4-cycles in the CY3-fold.
The Chern-Simons  coefficient takes values from $-N$ to $+N$.
The geometry of the corresponding CY3-fold can be seen easily from
the toric diagram or the corresponding dual web of $(p,q)$ 5-branes
of type IIB \cite{Kol:1997fv,Leung:1998tw} (\figref{web}).
\onefigure{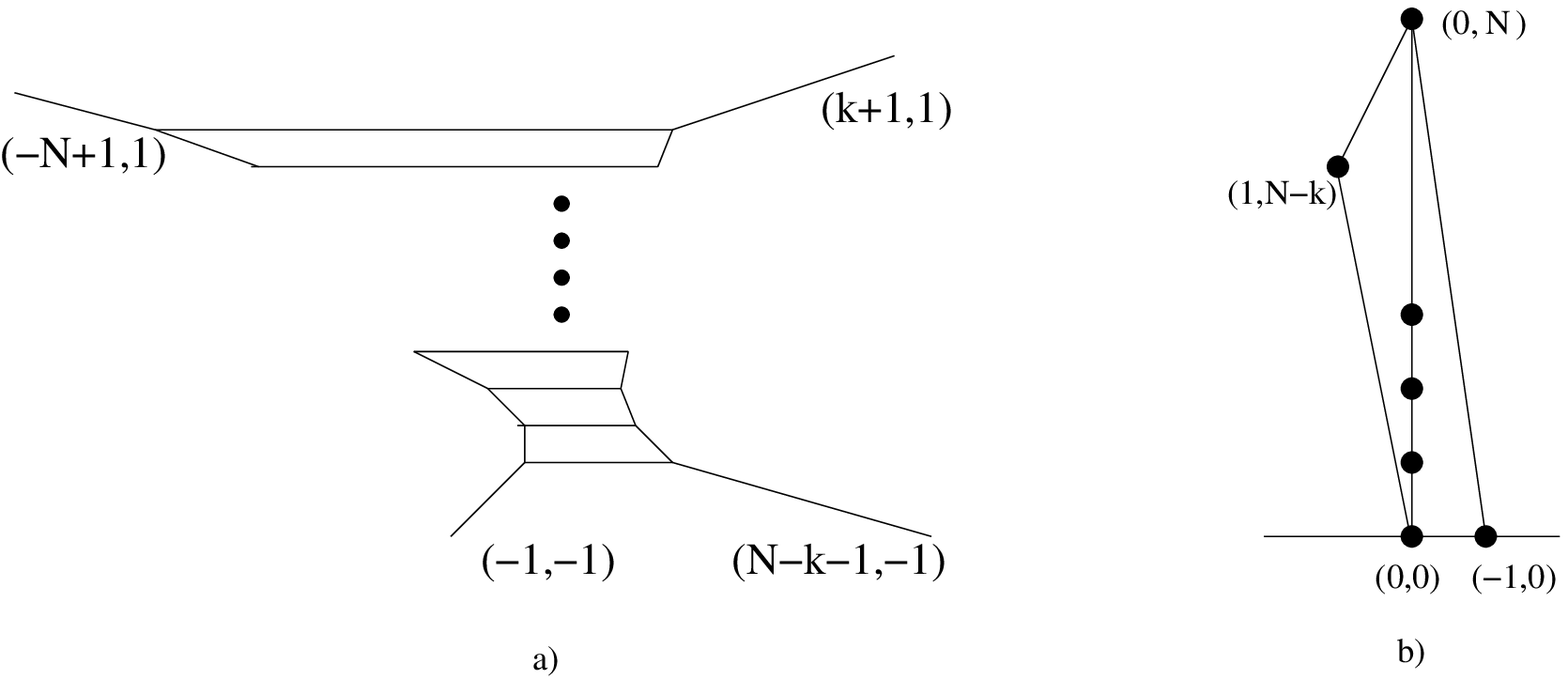}{$(p,q)$ 5-brane web (a), and the corresponding toric diagram (b),
which realizes five dimensional pure ${\cal N}=1$ theory with Chern-Simons coefficient $k$. }
Recall that for non-compact toric threefolds the five dimensional
theory obtained via M-theory compactification is dual to the five
dimensional theory living on a $(p,q)$ 5-brane web in type IIB.
This is a consequence of the duality between M-theory on $\T^{2}$
and type IIB on a circle. Since the non-compact toric CY3-folds
have a $\T^{2}$ fibration which degenerates on a planar tri-valent
graph therefore using the M-theory/IIB duality adiabatically one
can replace the degeneration locus with 5-branes. The $(p,q)$
charge of the 5-brane is determined by the degenerate cycle of the
$\T^{2}$. Holomorphicity of the CY3-fold implies that the
orientation of the 5-brane is correlated with its charge {\it i.e.\/}~,
$(p,q)$ 5-brane is oriented in the direction $(p,q)$ (for type IIB
coupling constant $\tau=i$). This web diagram can be obtained
directly from the toric diagram as its dual.

As an example consider the case of ${\cal O}(-1)\oplus {\cal
O}(-1)$ over $\PP^{1}$. This is a  non-compact toric 3-fold with
one K\"ahler parameter, the size of the $\PP^{1}$, which we will
denote as $r$. The linear sigma model description \cite{Witten:1993yc}
of this geometry is given by
\bea
&&|\Phi_{1}|^{2}+|\Phi_{2}|^{2}-|\Phi_{3}|^{2}-|\Phi_{4}|^{2}=r\,,\\
\nn &&(\Phi_{1},\Phi_{2},\Phi_{3},\Phi_{4})\sim
(\Phi_{1}e^{i\alpha},\Phi_{2}e^{i\alpha},\Phi_{3}e^{-i\alpha}
\Phi_{4}e^{-i\alpha})\,. \eea The base of this geometry
parameterized by $(|\Phi_{1}|^{2},|\Phi_{3}|^{2},|\Phi_{4}|^{2})$
is shown in \figref{toricconifold}(a).
\onefigure{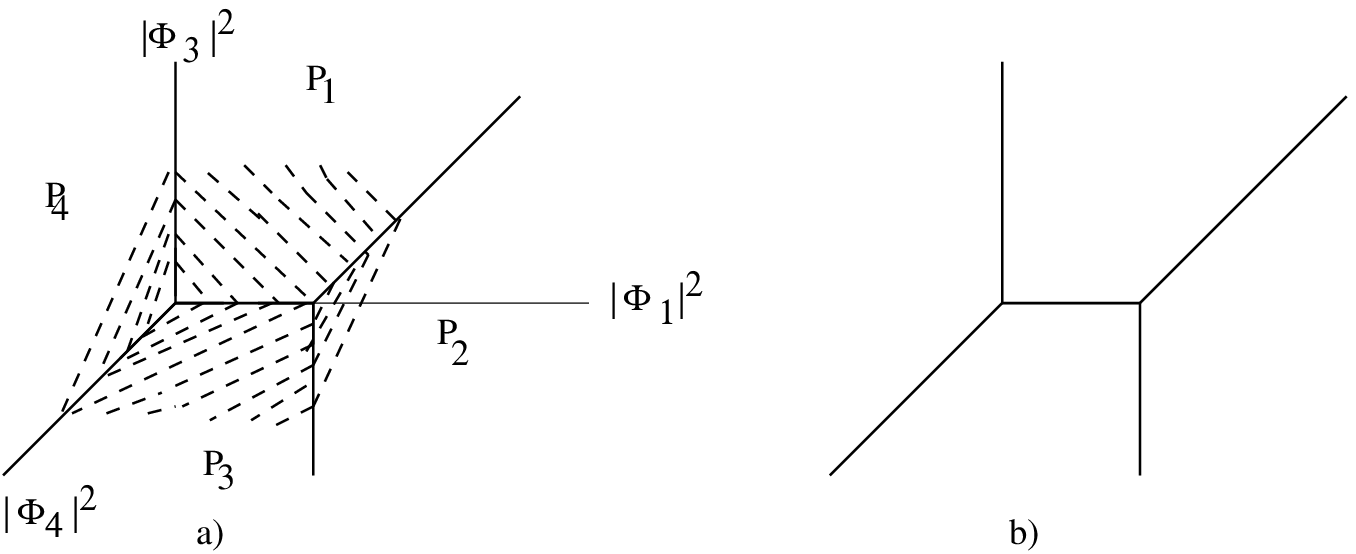}{The linear sigma model description of the
base of the deformed conifold in $\R^{3}_{+}$ (a) and the corresponding
web in in $\R^{2}$ (b). }
The base is the three dimensional convex region bounded by the planes $P_{1,2,3,4}$.These
2 dimensional planes $P_{1},P_{2},P_{3},P_{4}$ are given by $\Phi_{4}=0,\Phi_{2}=0,\Phi_{3}=0$
 and $\Phi_{1}=0$ respectively. We can project this geometry onto a two dimensional plane
and since the locus where the various planes intersect each other
has a degenerate $\T^{2}$ we see that in the two dimensional plane
the $(p,q)$ cycle of the $\T^{2}$ fibration degenerate over line
which is projection of the intersection of two planes and is
oriented in the $(p,q)$ direction. This is the corresponding web
diagram and is shown in \figref{toricconifold}(b).

The various geometries which give rise to $SU(3)$ five dimensional
gauge theory are shown in \figref{su3}(a).
\onefigure{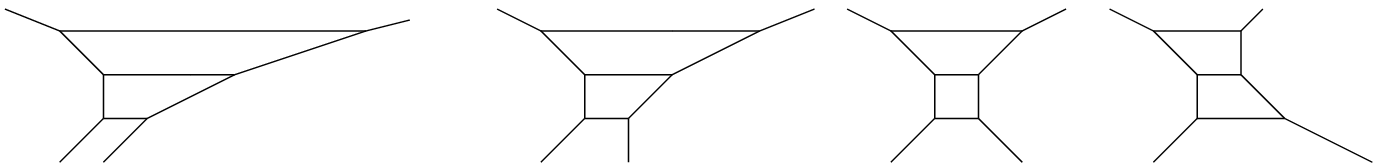}{The web diagram of various CY3-fold geometries which
realize the five dimensional pure $U(3)$ theory.}

For a more concrete connection with the four dimensional gauge
theory consider compactifying the five dimensional theory on a
circle of radius $\beta$. Then for small $\beta$ it is more useful to lift
the web of 5-branes to M-theory on $\T^{2}$ (with the area of
$\T^{2}$ equal to $1/\beta$) such that the web becomes a single
M5-brane wrapped on a Riemann surface $\Sigma$. The Riemann
surface $\Sigma$ is embedded in $\R^{2}\times \T^{2}$ where $\R^{2}$
is the plane in which the original 5-brane web lived and $\T^{2}$
is dual to the circle of type IIB. The Riemann surface $\Sigma$ is
given by just thickening the original graph of the web and its
equation can be read easily from the toric diagram, \bea
Y+\alpha\,\tfrac{X^{N-k}}{Y}+P_{N}(X)=0\,,\,\,\,X,Y\in \CC^{*}\,,
\eea where $P_{N}(X)$ is a polynomial of degree $N$. This Riemann
surface actually is the non-trivial part of the CY3-fold which is
mirror to the CY3-fold which geometrically engineers the five
dimensional theory via M-theory compactification. To see this note
that the mirror of the CY3-fold \cite{Klemm:1996bj,Hori:2000kt,Hori:2000jk}
with toric diagram given by
\figref{web}(b) is \bea
F(X,Y):=Y+e^{-t_{B}}\tfrac{X^{N-k}}{Y}+P_{N}(X)=uv\,,\,\,\,X,Y\in
\CC^{*}\,,\,\,u,v\in \CC\,. \eea The Calabi-Yau can be viewed
as $\CC^{*}$ fibration over the $X,Y$ plane where the circle
fibration degenerates over the Riemann Surface $F(X,Y)=0$.   The
periods of this CY3-fold reduce to integrals of a 1-form over
1-cycles of $\Sigma$. The complex structure parameters of the
mirror CY, the complex coefficients in the above equation, are
related to the K\"ahler parameters $t_{B}$ (size of the base
$\PP^{1}$) and $t_{F_{i}}$ (size of the i-th fiber $\PP^{1}$). The
geometry of the  degeneration of the circle fibration maps this to the
geometry of type IIA NS5-brane wrapped on the same Riemann
surface \cite{Klemm:1996bj}, which can then be lifted up to
an M-theory M5-brane. Thus we see that the M5-brane wrapped on the
mirror Riemann surface gives the compactified five dimensional
theory. This Riemann surface becomes the Seiberg-Witten curve if
we redefine the variables $X,Y$ suitably in the field theory
limit. For this note that we can write the equation of the mirror
Riemann surface as
\bea
y^{2}=\prod_{i=1}^{N}(X-e^{\phi_{i}})^{2}-4e^{-t_{B}}\,X^{N-k}\,.
\eea Now define \bea X=e^{\beta x}\,,\,\,\, e^{\phi_{i}}=e^{\beta
(a_{i}-a_{i+1})}\,,\,\,e^{-t_{B}}=(\tfrac{\beta
\Lambda}{2})^{2N}\,. \label{limit} \eea then in the limit $\beta
\rightarrow 0$ (4D limit) the equation of the mirror Riemann surface
becomes the equation for the SW curve, \bea
y^{2}=\prod_{i=1}^{N}(x-a_{i,i+1})^{2}-4(\tfrac{\Lambda}{2})^{2N}\,.
\eea Note that since $X$ was a $\CC^{*}$ variable, $x$ takes
value on a cylinder of radius $1/\beta$. Thus in the limit
$\beta\rightarrow 0$, $x$ becomes a $\CC$ variable. Also the integer
$k$ has disappeared from the equation
 reflecting the fact that in this limit the geometries becomes
equivalent. From Eq.~\eqref{limit} it is clear that in the limit of four dimensional
theory the base $\PP^{1}$ grows and the fiber $\PP^{1}$ shrink with a scaling
given by \eqref{fieldtheorylimit}.

\subsection{${\cal N}=2$, $SU(N)$ with $N_{f}=2N$}

Once pure $SU(N)$ gauge theory has been engineered it is
relatively simple to modify the CY3-fold geometry to include
hypermultiplets in the fundamental representation. A fundamental
hypermultiplet of mass $m$ appears in the gauge theory if we
blowup the CY3-fold of the pure $SU(N)$ gauge theory such that the
mass of the hypermultiplet $m$ is proportional to the area of the
blown up curve.

Let us begin by considering the case of 5-dimensional $U(1)$ theory with
 $N_{f}=2$. The
geometry giving rise to pure $U(1)$ theory is that of resolved conifold {\it i.e.\/}~, the
total space of ${\cal O}(-1)\oplus {\cal O}(-1)$ bundle over $\PP^{1}$.
The size of the $\PP^{1}$, $t$, is proportional to the length of the
internal line as shown in \figref{conifold}(a) and is inversely proportional
to the gauge theory coupling constant. To introduce fundamental hypermultiplets
we need to blow up this geometry at two points as shown in
\figref{conifold}(b).
\onefigure{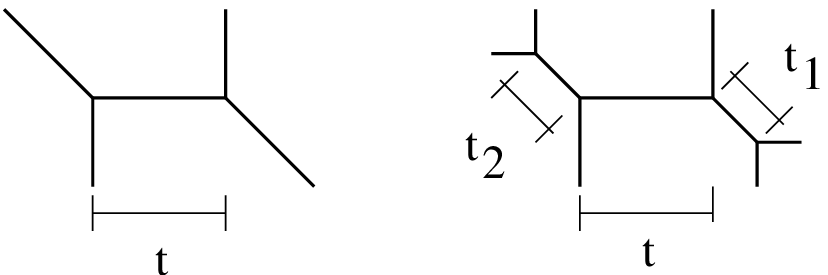}{a) ${\cal O}(-1)\oplus {\cal O}(-1)$ bundle over $\PP^{1}$, b)
blow up of ${\cal O}(-1)\oplus {\cal O}(-1)$ bundle over $\PP^{1}$ at two points.}
In this case the 4 dimensional field theory limit is given by \bea
e^{-t_{a}}=e^{-\beta m_{a}}\,,\,\,q=e^{-\beta}\,,\,\,\beta \rightarrow 0\,, \eea where
$m_{1,2}$ is the mass of the two hypermultiplets in the 4 dimensional field theory
and $t_{1,2}$ is the area of the blown up rational curves. If we do
not take this limit we
obtain five dimensional theory compactified on a circle of size $\beta$.

Generalization to 5-dimensional $U(N)$ with matter is
straightforward. All we have to do is blow up the geometry giving
rise to pure $U(N)$ theory at $N_{f}$ points as shown in
\figref{sunmatter}(b) for the case $N_{f}=2N$. The corresponding
$(p,q)$ 5-brane web in type IIB is shown in \figref{sunmatter}(a)
and consists of intersecting D5-branes and NS5-branes. This is the
case when the hypermultiplets have zero mass. To introduce non-zero
mass one has to resolve the intersection locus of D5-brane and the
NS5-brane. In terms of CY3-fold geometry each exceptional curve
generated by the blow up gives rise to a fundamental hypermultiplet
such that the mass of the hypermultiplet is proportional to the area
of the corresponding curve. \onefigure{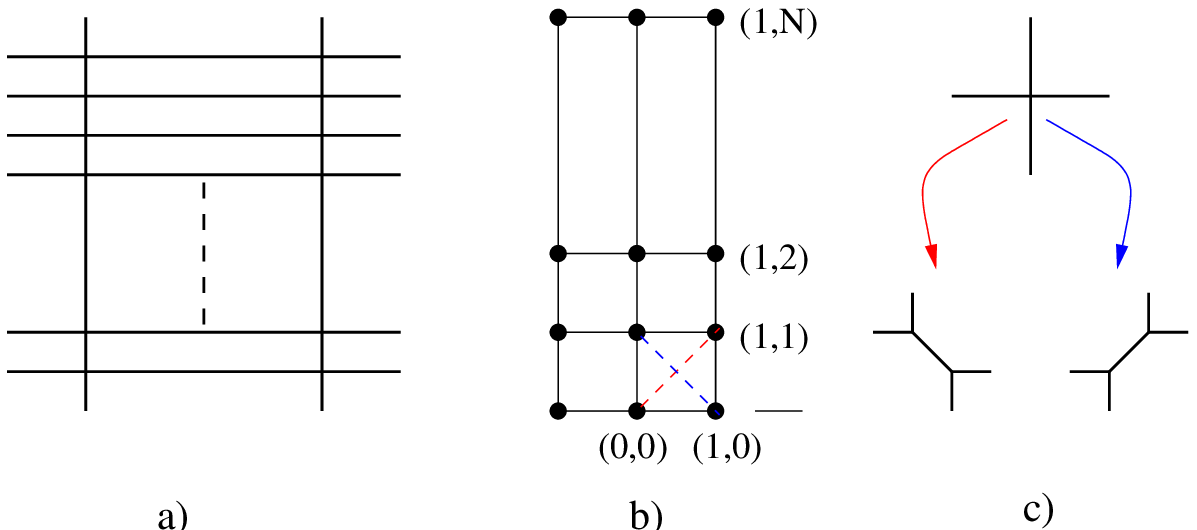}{a) The web diagram
of the geometry relevant for five dimensional $U(N)$ theory with
$N_{f}=2N$, b) the corresponding toric diagram and c) the flop
transition reflecting the choice of the triangulation of the toric
diagram.} The limit of 4 dimensional field theory is given by
\bea\nn  e^{-T_{b}}=(\tfrac{\beta
\Lambda}{2})^{2N-N_{f}}\,,\,\,\,e^{-T_{F_{i}}}= e^{-\beta
(a_{i}-a_{i+1})}\,,\,\,e^{-t_{a}}=e^{-\beta m_{a}}\,,\,\,
a=1,\ldots, N_{f}\,,\,\,\,\beta \rightarrow 0\, \eea where $t_{a}$
is the area of the $a^{\rm th}$ exceptional curve generated by the
blow up.

{}From the toric diagram, \figref{sunmatter}(b), it is clear that
there are $4N$ distinct triangulations of the diagram. Different
triangulations of the toric diagram give geometries related to
each other by flop transitions. In the web description this
corresponds to two possible resolutions into tri-valent graph at
each of the $2N$ points where the $(1,0)$ line  meets the $(0,1)$
line. In the gauge theory language this is given by the choice of
the sign of the mass term. Thus for zero size of the blown up
$\PP^{1}$ the geometry is unique. The case of $SU(2)$ is
illustrated in \figref{su2matter}.
\onefigure{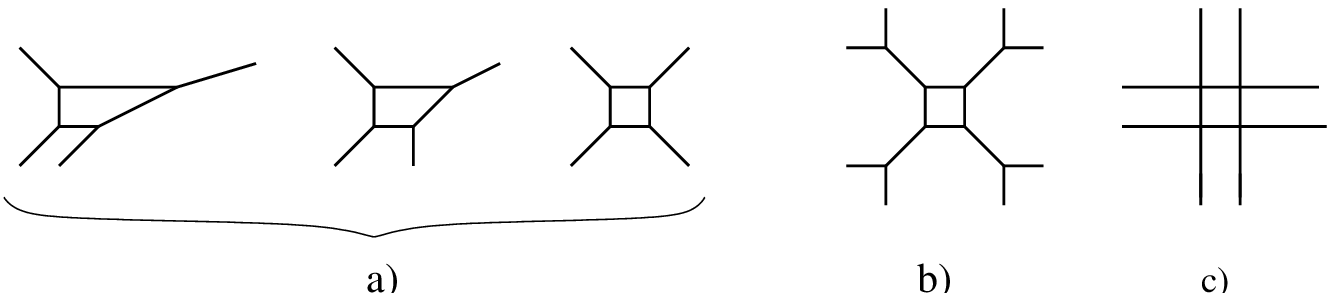}{a) The web diagrams of the CY3-fold
which realize pure 5D $U(2)$ theory, b) its blow up at four points
for a particular choice of the triangulation and c) the unique web
diagram obtained by blowing up four points with zero area.}
The various geometries which give rise to $SU(2)$ theory
(\figref{su2matter}(a)) are related to the same geometry
(\figref{su2matter}(b)) by flop transitions once four points have
been blown up.
In all these case the classical part of the prepotential
is given by the triple intersection numbers of the corresponding CY3-fold.

\underline{\bf 6D Theory:} In the previous section we saw that 4D theory can
be obtained as a limit of the 5D theory. A natural question that arises here
is whether the 5D theory can be obtained as some limit of a compactified 6D
theory. One way of answering this is to consider F-theory on $\X\times \T^{2}$.
But for this to work the CY3-fold $\X$ must be an elliptic fibration with a
section. This is related to the fact that the 6D theory must be anomaly
free and therefore must have a very specific matter content. For the case of
$U(N)$ theory with $N_{f}$ fundamental hypermultiplets it requires that
$N_{f}=2N$. The elliptic 3-folds which give rise to these theories
where constructed in \cite{Bershadsky:1997sb}.

To construct the non-compact 3-folds with elliptic fibration
relevant for $U(N)$ theory with $N_{f}=2N$ let us first consider
the simple geometry of the deformed conifold given by \bea
x_{1}x_{2}-x_{3}x_{4}=\varepsilon\,. \eea We can write the above as \bea
x_{1}x_{2}=z\,,\,\,\,x_{3}x_{4}=z-\varepsilon\,. \eea Then we see that the
deformed conifold is given by two $\CC^{*}$ fibrations over the
complex $z$-plane as shown in \figref{twocstar}(a). The $\S^3$ in
the geometry, formed by the two circles in the $\CC^{*}$
fibration and the line segment joining the points $z=0$ and $z=\varepsilon$,
has size given by $\varepsilon$. In terms of the $(p,q)$ 5-brane web the
$\varepsilon\neq 0$ deformation is given by separating D5-brane and the
NS5-brane from each other by a length $\varepsilon$ as shown in
\figref{twocstar}(b).
\onefigure{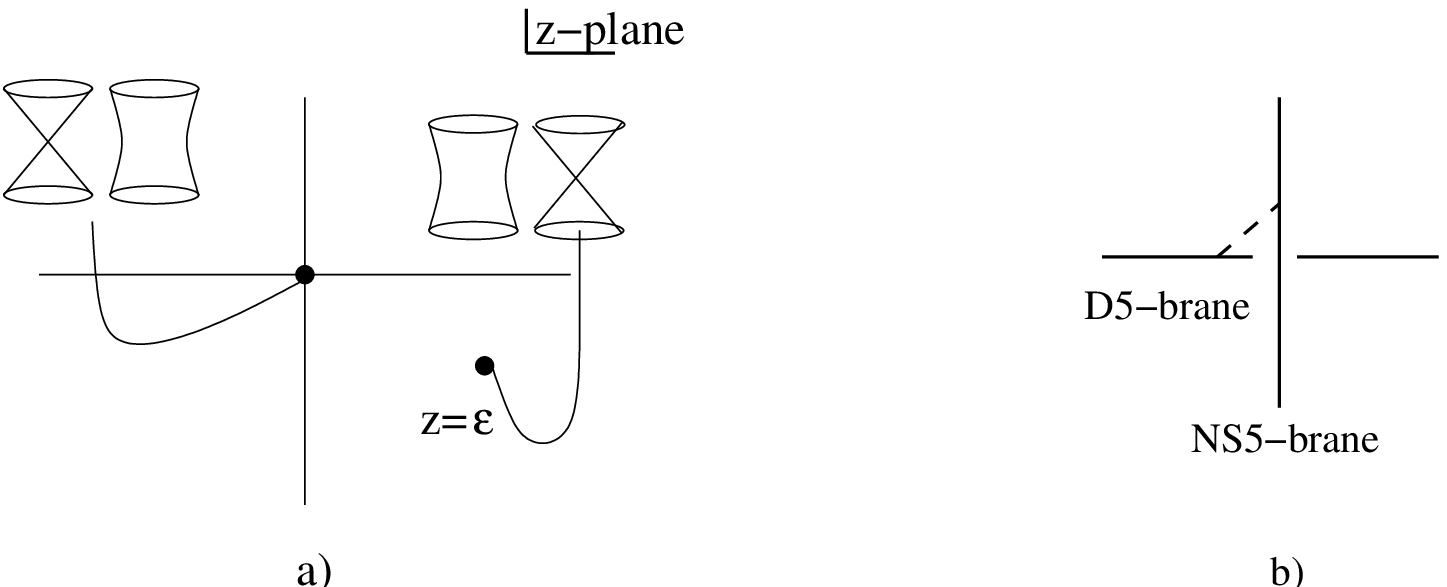}{a) The deformed conifold as the
double $\CC^{*}$ fibration over the z-plane and b)
the corresponding 5-brane web configuration }

Now given this picture of the geometry in terms of two $\CC^{*}$
fibration it is easy to see that one can get an elliptically
fibered CY3-fold by compactifying one of the $\CC^{*}$ fibers to
a $\T^{2}$, as shown in \figref{5dadjointu1}(a), such that \bea
x_{1}x_{2}&=&z\,,\\ \nn y^{2}&=&x^{3}+f(z,\varepsilon)x+
g(z,\varepsilon)\,. \eea The
second equation above defines the elliptic fibration over the
$z$-plane which degenerates at $z=\varepsilon$. The corresponding type IIB
configuration of 5-branes is now such that the NS5-brane is
wrapped on a circle and for $\varepsilon=0$ the D5-brane intersects the
circle as shown in \figref{5dadjointu1}(b). The circle on which
the NS5-brane is wrapped is exactly the circle created by the
compactification of the $\CC^{*}$ fiber to $\T^{2}$ and its size
is related to the K\"ahler class of the compactified $\T^{2}$.
\onefigure{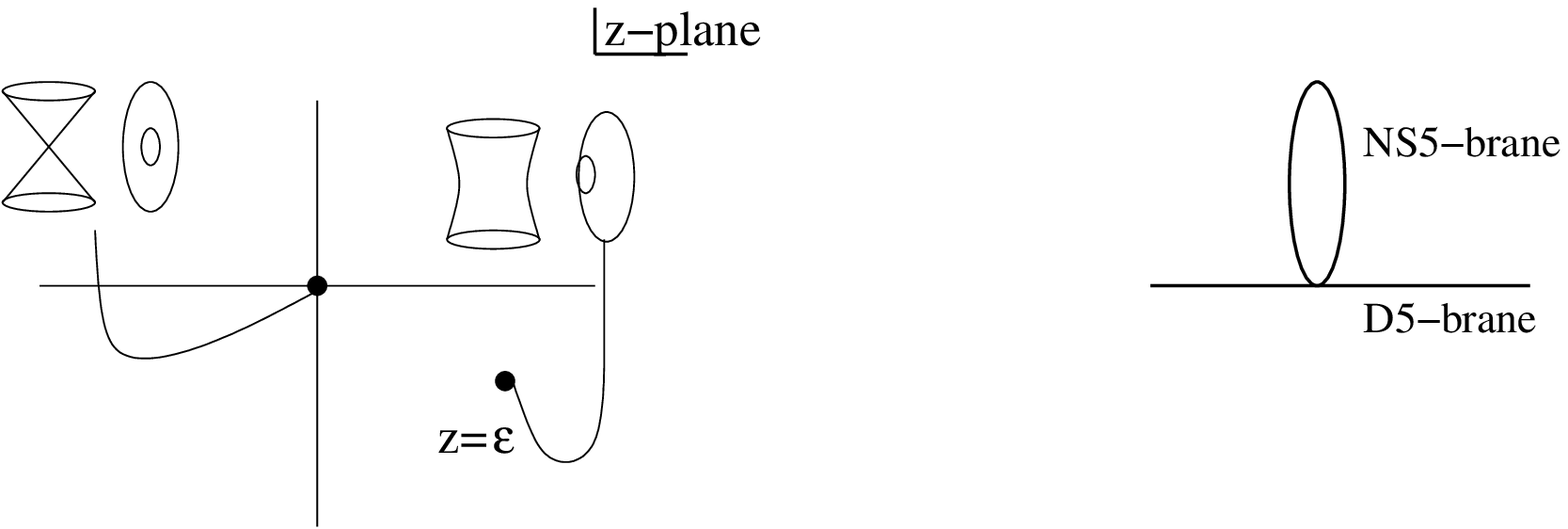}{ a) The partial compactification of the
deformed conifold geometry by replacing one of the $\CC^{*}$ with
a $\T^{2}$ and b) the corresponding 5-brane web description.}
Now given this web description we can consider two NS5-branes
wrapped on the circle but separated in along the D5-brane. In this
case it is easy to see that when the radius of the circle goes to
infinity we get 5-dimensional $U(1)$ theory with $N_{f}=2$ living
on the D5-brane and intersecting the two NS5-branes. Therefore the
six dimensional theory compactified on a circle is given by a
$(p,q)$ 5-brane web which consists of a D5-brane and two
NS5-branes wrapped on a circle such that the D5-brane intersects
the circle at a point and the NS5-brane are separated along the
D5-brane as shown in \figref{6du1hyper}. The distance between the
two NS5-branes, along the D5-brane, is inversely proportional to
the coupling constant of the 6 dimensional gauge theory.
\onefigure{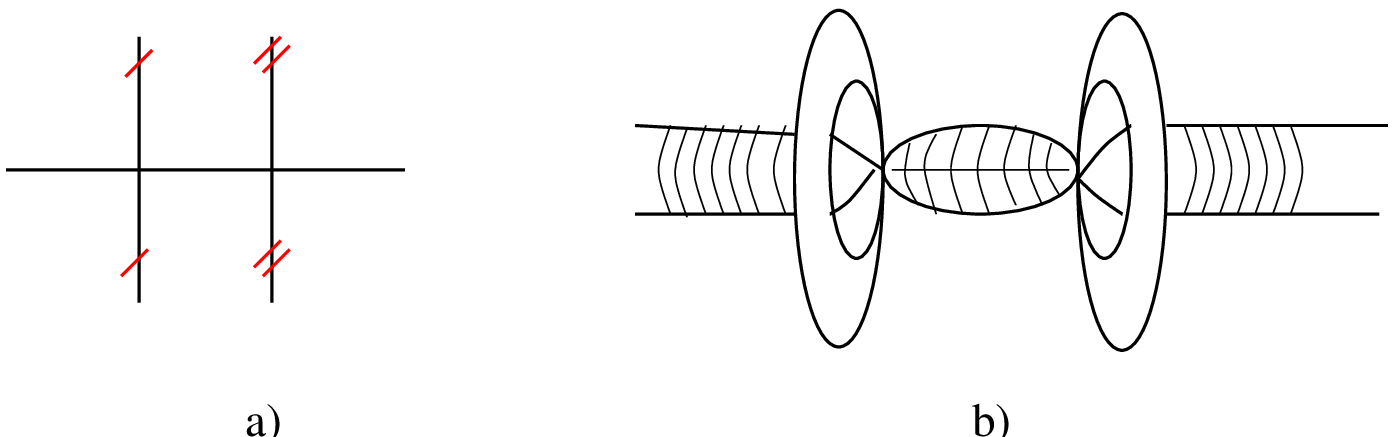}{a) The 5-brane web description of the
partial compactification of the resolved conifold blown up at
two points and b) the geometry as an elliptic fibration which
degenerates on the base $\PP^{1}$.}

Generalization to geometries giving rise to $U(N)$ theory with
$N_{f}=2N$ is straightforward. Instead of single D5-brane we
consider $N$ D5-branes intersecting the circle on which two
NS5-branes are wrapped as shown in \figref{6dunhyper}.\onefigure{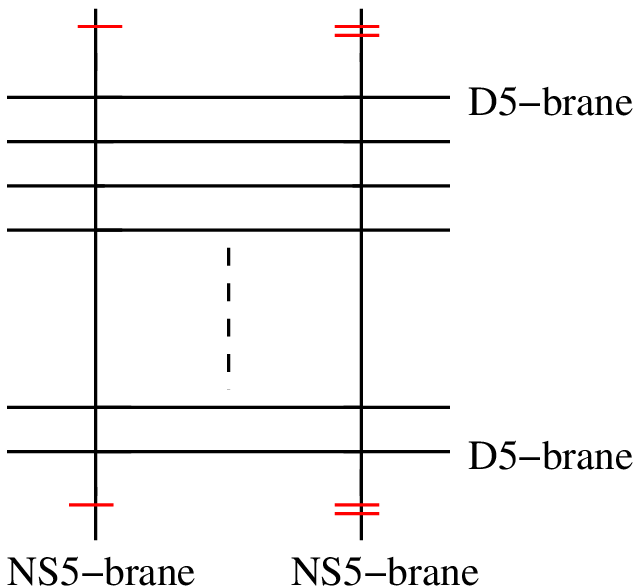}{The 5-brane web description of the
compactified 6D $U(N)$ theory with $N_{f}=2N$.}

In the limit that the size of the circle goes to
infinity we get back the $(p,q)$ web configuration giving rise to 5
dimensional $U(N)$ theory with $N_{f}=2N$.

\begin{figure}
\begin{center}
\includegraphics[scale=0.6]{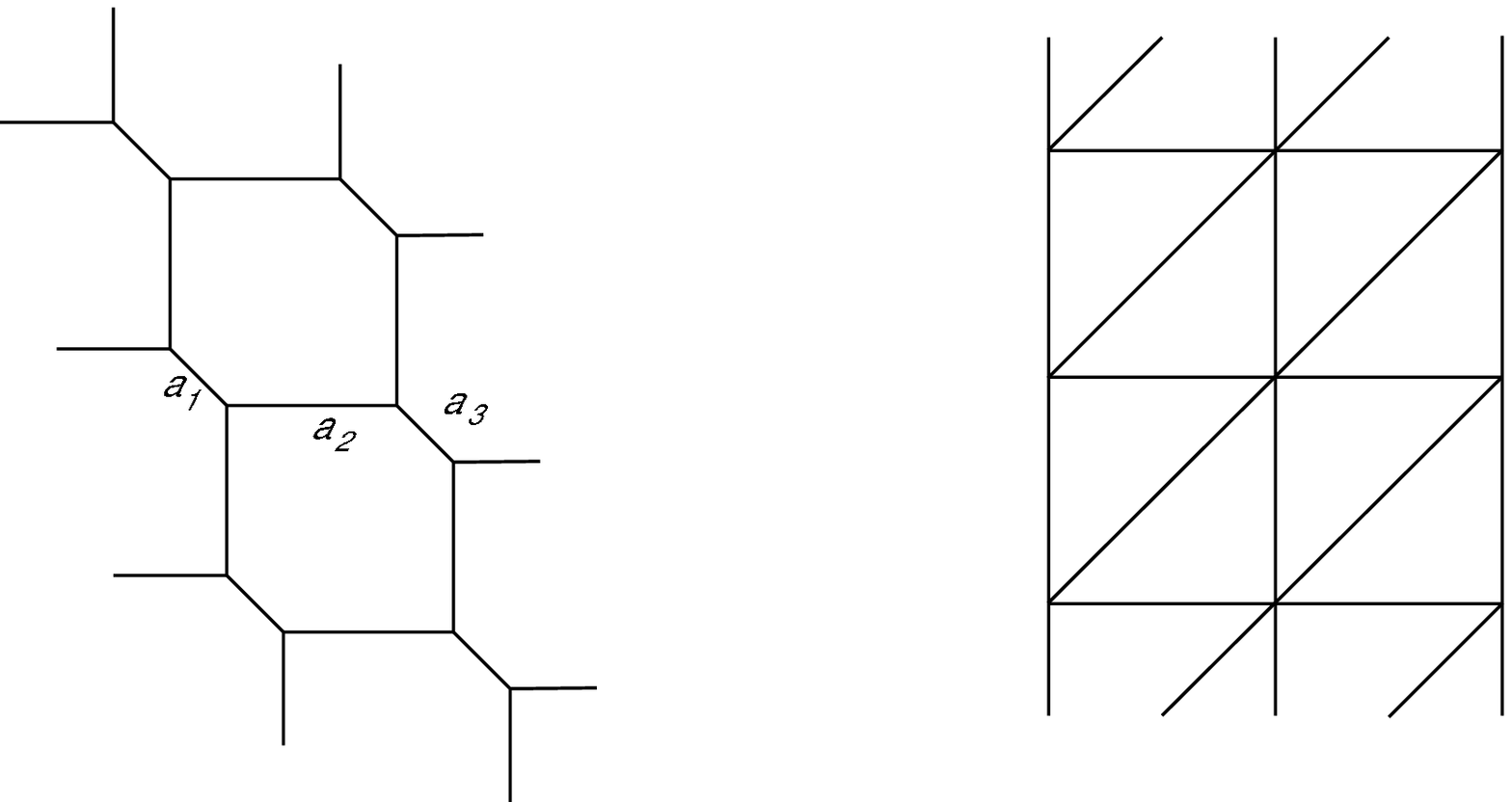}
\end{center} \caption{\small The web diagram for the 6D
  $N=1$ and $N_f=2$ theory where periodicity is implemented by
  including an infinite set of images. On the right is the associated
  grid diagram.}
\label{webgridf}\end{figure}

The exact form of the curve can be extracted from the web diagrams
using the rules of  \cite{Kol:1997fv} or that of local
mirror symmetry applied to toric threefolds
\cite{Katz:1997eq,Hori:2000kt,Hori:2000jk}. To apply this method to the 6D
theories we have to take account of the periodicity of the web
by including all its images under the periodic shift.
For the $N=1$ the web diagram with
all its images is shown in Fig.~\ref{webgridf}. On the right is the
associated grid diagram.
The associated curve can be written $F(X,Y)=0$, where
$F(X,Y)$ is the sum of monomials, one for each vertex of the grid
diagram. If a vertex is at $(k,l)$ then the monomial is simply
$A_{kl}X^kY^l$. The $A_{kl}$ is a potential modulus of the curve;
however, there are conditions that must be satisfied that restrict
them. These conditions
arise as follows. Consider a link on the grid that joins $(k,l)$ with
$(u,v)$. Each such link is uniquely associated to a
link on the web to which it is orthogonal. Suppose the link of
the web is described by the equation $py=qx+\alpha$, then the
orthogonality condition is
\EQ{
(k,l)-(u,v)=(-q,p)\ .
}
For each such link, we then have the constraint
\EQ{
py=qx+\alpha:\qquad A_{kl}=e^{\beta\alpha}A_{uv}\ .
\label{rule}
}
In the present case, we have monomials $A_{kl}X^kY^l$, $l\in\Z$ and $k=0,1,2$.
Applying the rules,  we find that all the coefficients are fixed in
terms of the parameters $a_{1,2,3}$ in Fig.~\ref{webgridf}, up to
an unimportant overall scaling and a choice of origin:
\EQ{
A_{0,l}=e^{\beta L l(l-1)}\ ,\quad A_{1,l}=A_{0,l}\,e^{a_1(1-l)}
\ ,\quad A_{2,l}=A_{0,l}\,e^{a_1(2-l)+ a_2+ a_3(1-l)}\ ,
}
where $L$ is the period of the web in the vertical
direction. This parameter is related to $\rho$ via
\EQ{
L=\frac{2\pi i\rho}\beta\ .
}
We will identify the parameters $a_1$ and $a_3$ in terms of the masses,
\EQ{
a_1=-\beta m_1\ ,\qquad a_3=\beta m_2\ .
}
This leaves one remaining degree-of-freedom in $a_2$
which corresponds to the bare coupling. So after a rescaling of $Y$ the curve becomes
\EQ{
F(X,Y)=\sum_{l=-\infty}^{\infty}
e^{2\pi il(l-1)\rho}Y^l
\big(e^{-\beta m_1(l-1/2)}+X+e^{ a_2-\beta(m_1+m_2)/2}
e^{-\beta m_2(l-1/2)}X^2\big)=0\ .
}
Identifying $Y=e^{\beta x}$ and $X=e^{2\pi iz}$, and with some
re-scalings of $X$ and $Y$, the curve becomes
\EQ{
\frac c4\,\theta_1(\tfrac\beta{2i}(x-m_1)|\rho)+X
\theta_1(\tfrac{\beta}{2i}x|\rho)+
X^2\theta_1(\tfrac\beta{2i}(x-m_2)|\rho)=0\ ,\quad \frac
c4=e^{a_2-\beta(m_1+m_2)/2}\ .
}
Defining $y=2X+\theta_1(\tfrac{\beta}{2i}x)/
\theta_1(\tfrac\beta{2i}(x-m_2)$, the curve becomes
\EQ{
y^2=\theta_1(\tfrac{\beta}{2i}x|\rho)^2-
c\,\theta_1(\tfrac\beta{2i}(x-m_1)|\rho)
\theta_1(\tfrac\beta{2i}(x-m_2)|\rho)\ .
}
The constant $c=c(a_2)$ represents the freedom to change the bare
coupling of the theory.

For $N>1$, using an identical approach we find a curve that can be
written
\EQ{
y^2=\prod_{i=1}^N\theta_1\big(\tfrac{\beta}{2i}
(x-\zeta_i)\big|\rho\big)^2-c
\prod_{f=1}^{2N}\theta_1\big(\tfrac{\beta}{2i}(x-m_f)
\big|\rho\big)\ .
\label{geom6}
}
where the $m_f$ are the masses and the $\zeta_i$ are the moduli of the
Coulomb branch. In the four-dimensional limit, this reduces to the
well-known hyper-elliptic geometry. We note that our curve is very simply
related to the spectral curve of an XYZ spin chain suggested in
\cite{Gorsky:2000px}.

\subsection{${\cal N}=2$, $SU(N)$ with adjoint hypermultiplet}

In this section we will review the brane configurations and the CY3-fold
geometry that realizes the ${\cal N}=2$ $U(N)$
gauge theory with an adjoint hypermultiplet in 4, 5 and 6 dimensions.

\underline{\bf Brane construction:} The basic type
IIB setup which realizes the 5-dimensional theory is similar to the elliptic
models which realize 4 dimensional theory with adjoint hypermultiplet
\cite{Witten:1997sc}. The only difference is that instead
of D4-branes we have D5-branes in this case. There is a single NS5-brane
and the D5-branes are wrapped on a circle: the NS5-brane is extended in
the $x^{0,1,2,3,4,5}$ direction. The D5-branes span the
$x^{0,1,2,3,4}$ and $x^{6}$ direction. The direction $x^{6}$ is
taken to be compact so that the D5-branes wrap the circle and intersect the
NS5-brane which is a point on the circle (\figref{elliptic}(a)).
\onefigure{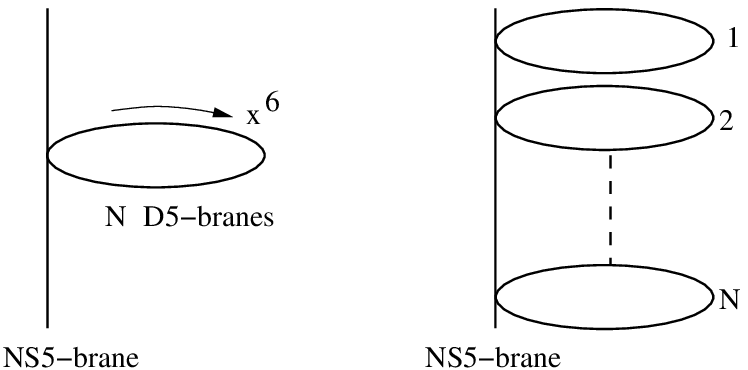}{Elliptic models involving 5-branes which
realize 5D $U(N)$ theory with a massless adjoint.}
On the 5-dimensional non-compact part of the D5-brane
worldvolume there is an ${\cal N}=1$ (8 supercharge)
$U(N)$ gauge theory with a
massless adjoint {\it i.e.\/}~, it has ${\cal N}=2$ (16 supercharge)
supersymmetry reflected
by the fact that since there is a single NS5-5brane it can be moved
away from the circle leaving behind $N$ D5-branes wrapped on a
circle. The theory on such a set of $N$ D5-branes is ${\cal N}=2$
$U(N)$ gauge theory. Turning on the mass of the adjoint breaks the
supersymmetry down to ${\cal N}=1$ and therefore the corresponding
branes configuration must be such that NS5-brane cannot be moved
away from the circle. This is achieved as in \cite{Witten:1997sc} by
changing the geometry so that as one goes around the $x^6$ circle
$x^5$ shifts by $m$.\footnote{In the five-dimensional theory $m$ is real.}
Obviously this corresponds to resolving
the intersection of D5-branes and the NS5-branes into a tri-valent web of
$(p,q)$ 5-branes in the $x^{5,6}$ plane
as shown in \figref{elliptic2}(b,c). In this case the
separation distance is equal to the mass of the adjoint.
\onefigure{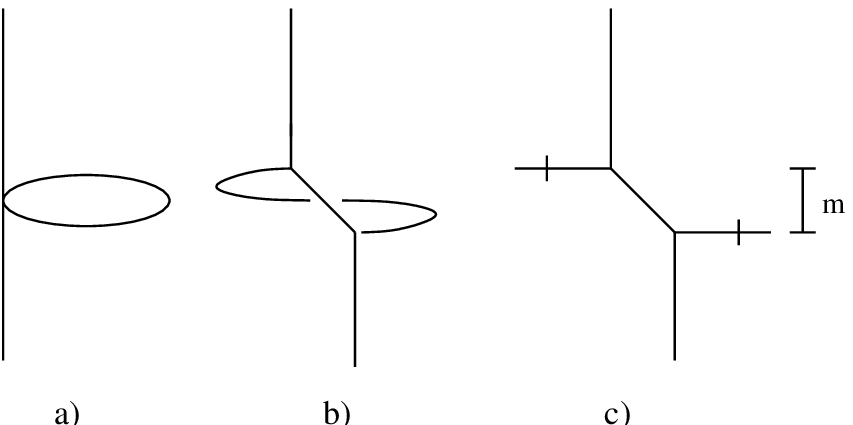}{a) Elliptic model realizing $U(1)$ theory
with massless adjoint, b) adjoint can be made massive by deforming the
brane configuration and c) the same web diagram drawn
with identifications.}
Once the adjoint acquires non-zero mass we can Higgs the gauge group by
separating the D5-branes from each other (\figref{elliptic}).
\onefigure{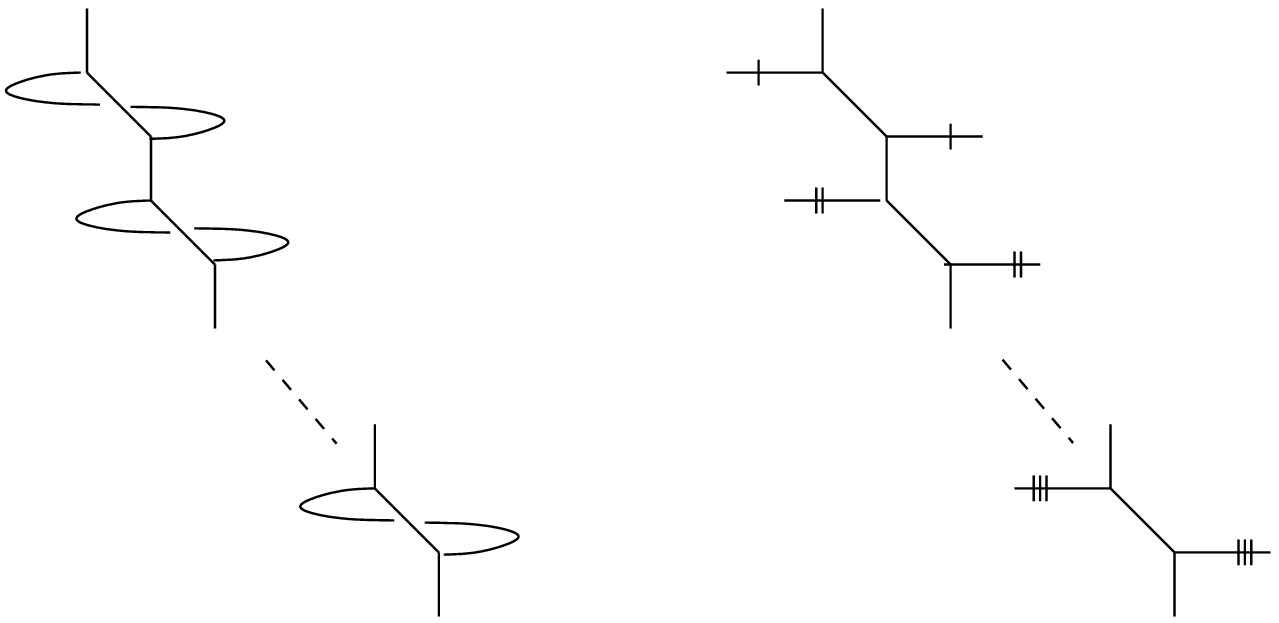}{Brane configuration which realizes $U(N)$ theory with adjoint mass.}
The $(p,q)$ 5-brane web given above also defines the CY3-fold geometry
which in this case has a elliptic fibration since the web is
compactified on a circle. By compactifying one of the direction
perpendicular to the web, say $x^4$, on a circle of size $\beta$ we can take a
limit in which we get the 4 dimensional field theory:
\bea
e^{-A(\T^{2})}=e^{2\pi i\tau}\,,\,\,q:=e^{-g_{s}}=e^{-\beta}\,,\,\,
e^{-t}=e^{-\beta m}\,,\,\beta \rightarrow 0\,,
\eea
where $A(\T^{2})$ is the area of the compactified $\T^{2}$ in the CY3-fold
geometry, $t$ is the area of the exceptional curves in the geometry
and $\tau$ is the coupling of the four-dimensional theory.

To describe the 6 dimensional geometry we start with the $U(1)$
case. The corresponding 5-dimensional geometry was discussed in
the last section and is given by compactifying one of the two
$\CC^{*}$ fibers so that we have a $\CC^{*}\times \T^{2}$
fibration over the $z$-plane. The $\CC^{*}$ fiber degenerates at
$z=0$ and the elliptic fibration degenerates at $z=\varepsilon$. The two
fibrations together define an $\S^3$ in the geometry. Shrinking
this $\S^3$ produces a singularity which when resolved gives the
picture dual to the picture of D5-brane and NS5-brane intersecting
and then deformed into a tri-valent web of $(p,q)$ 5-branes. The
$(1,1)$ 5-brane introduced by this resolution is dual to the
exceptional curve produced by the resolution of singularity. The 6
dimensional compactified theory can be obtained by compactifying
both the NS5-brane and the D5-brane on two different circles so that
the plane of the web is a 2-torus. This
corresponds to compactifying both $\CC^{*}$ fibrations
as shown in \figref{6dadjointu1}(a).
\onefigure{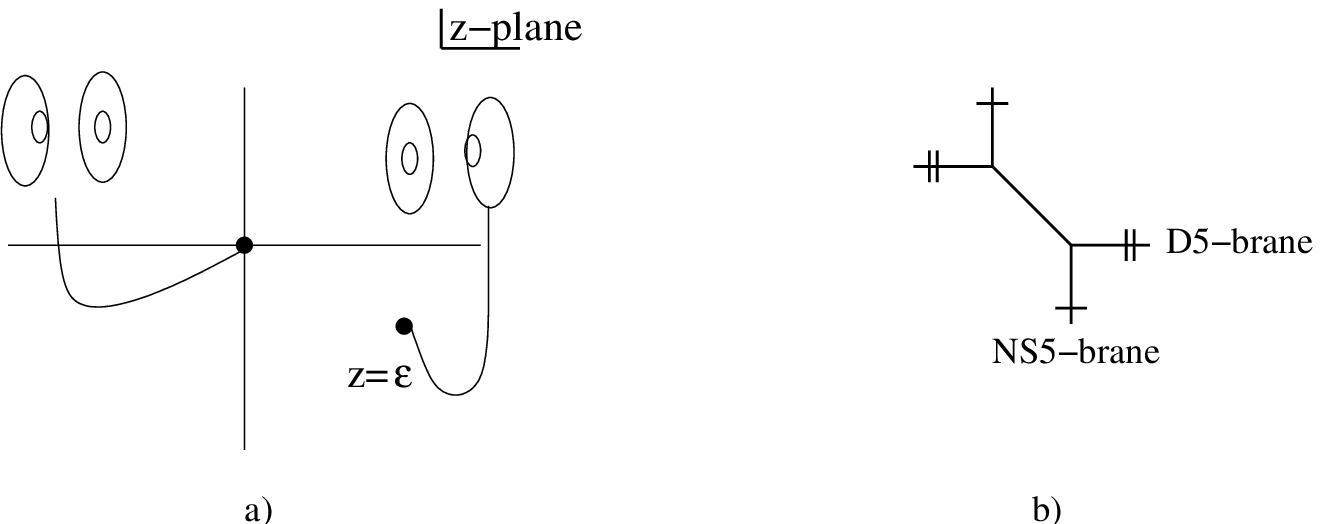}{a) The elliptically fibered CY which gives
$U(1)$ theory with massive adjoint and b) the corresponding
5-brane web diagram.}
Recall that the compactified 5-dimensional theories on the $(p,q)$
5-brane webs can be lifted to M-theory where the theory lives on
an M5-brane with worldvolume $\R^{4}\times \Sigma$. The Riemann
surface $\Sigma$ is obtained from the $(p,q)$ 5-brane web by
``thickening'' the lines and is embedded in the 4-dimensional space
$x^{4,5,6,10}$, where $x^{10}$ is the M-theory dimension. This is
locally of the form $\S^1\times\R\times\T_\tau^2$, where $\T^2_\tau$
is the torus with complex structure $\tau$ in the $x^{6,10}$
directions and $\S^1$ is the T-dual of the original $x^4$ circle,
{\it i.e.\/}~now having size $\beta^{-1}$. Globally one must take
account of the mass so that as one goes around the $x^6$-cycle of
$\T^2_\tau$, $x^5$ shifts by $m$.
In the six dimensional case, where both directions along the web are
now periodic, we can lift the $(p,q)$ 5-brane web to obtain
an M5-brane wrapped on a Riemann surface $\Sigma$. In
this case $\Sigma$ can also be
obtained from the web by thickening it but it now is embedded in
a slanted 4-torus in the $x^{4,5,6,10}$ space. This 4-torus is identified with
the abelian surface $\T^4$ in Eq.~\eqref{defas}.\footnote{In this context $m$
  can be taken to be complex.} It is easy to see that $\Sigma$ is a
genus two curve which degenerates into two elliptic curves when
the mass of the adjoint goes to zero (\figref{genus2}). This is precisely as one would
predict from the analysis in Section 2.
\onefigure{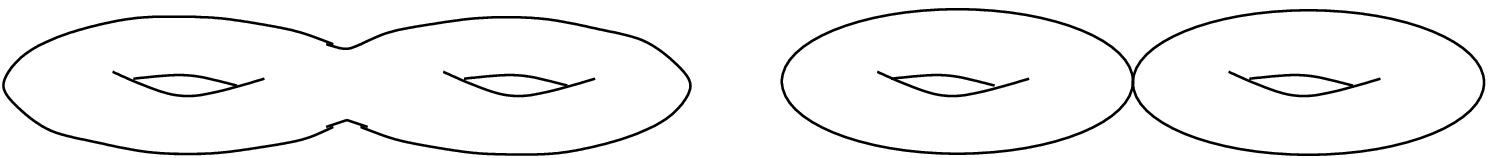}{a) The genus 2 curve which describes the geometry of
  the 6D theory with an adjoint for $N=1$, b) when the mass $m\to 0$,
the genus 2 curve  degenerates into 2 elliptic curves with complex structures $\tau$ and
  $\rho$.}

\begin{figure}
\begin{center}
\includegraphics[scale=0.6]{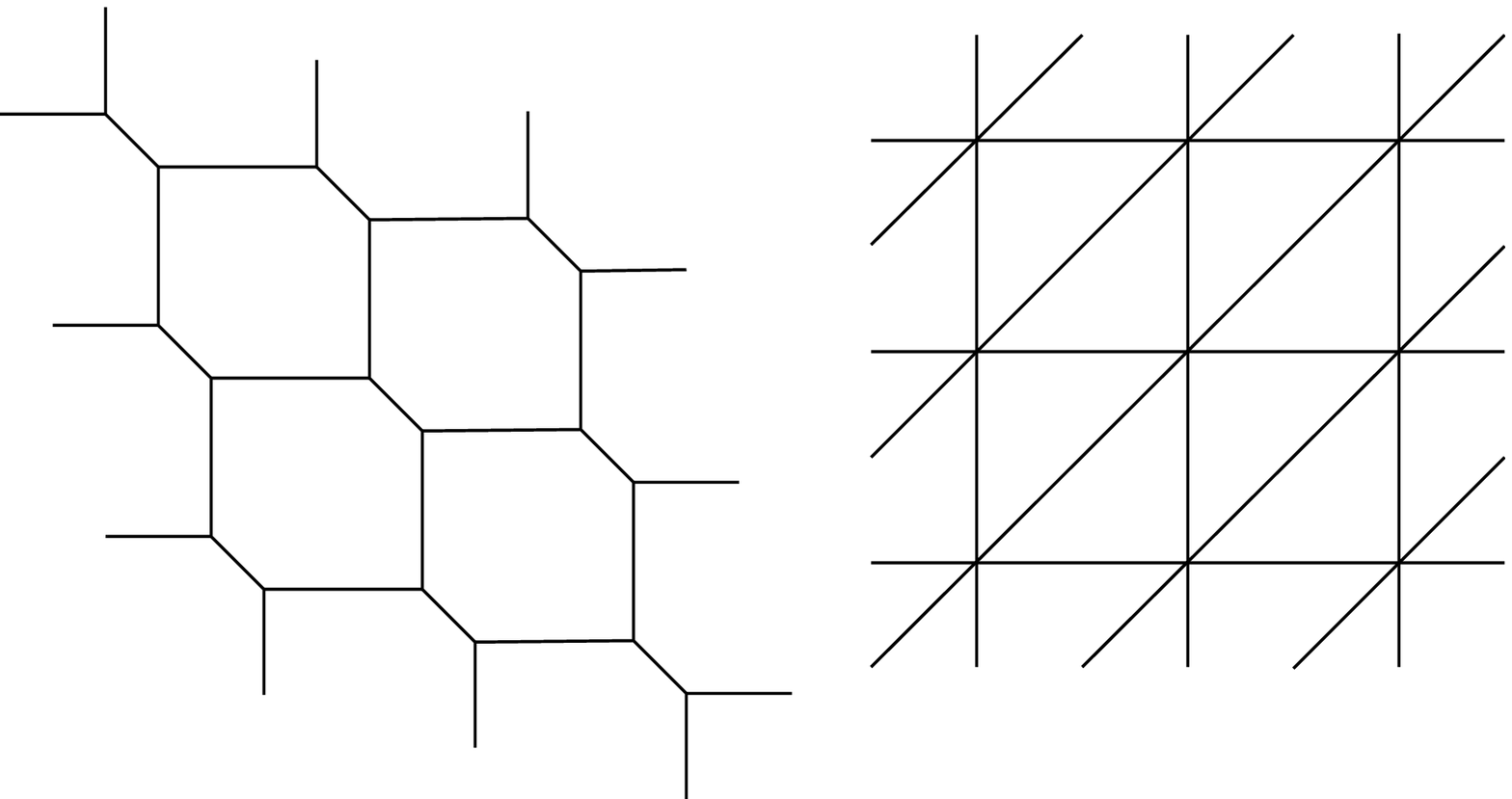}
\end{center} \caption{\small The web diagram for the 6D $N=1$ theory
  with adjoint
including all the images under the two periodic directions.
On the right is the associated grid diagram.}
\label{webgrid}
\end{figure}

The exact form for the curve can be obtained from the web diagram
using the rules established in \cite{Kol:1997fv} or local
mirror symmetry applied to toric Calabi-Yau
 \cite{Katz:1997eq,Hori:2000kt,Hori:2000jk} that we summarized at
the end of Section 3.3. As in Section 3.3,
in order to apply them to webs with periodicities
we employ the method of
images, although now we have double periodicity.
We will only consider the $N=1$ case in any detail but it
should be clear to reader how to extend the method to $N>1$.
When we include all the images, the web tessellates the plane as
illustrated in Fig.~\ref{webgrid}. This figure also shows the
associated grid diagram.
The associated curve can be written $F(X,Y)=0$, where
$F(X,Y)$ is the sum of monomials $A_{kl}X^kY^l$, $k,l\in\Z$,
one for each vertex of the grid diagram. The $A_{kl}$
are potential moduli which
can be fixed using the connection with the web diagram
as in Section 3.3,
that fixes them, up to an overall scaling. Using the rule \eqref{rule},
we find the following recursion relation for the elements:
\EQ{
A_{k+1,l}=e^{\beta(L_hk+ml)}A_{kl}\ ,\quad
A_{k,l+1}=e^{\beta(L_vl+mk)}A_{kl}\ ,
\label{recur}
}
where $L_h$ and $L_v$ are the periods of the web in the vertical and
horizontal direction. These parameters are related to $\rho$ and
$\tau$ via
\EQ{
L_v=\frac{2\pi i\rho}\beta\ ,\qquad L_h=\frac{2\pi i\tau}\beta\
,
}
at least for $\rho$ and $\tau$ purely imaginary. However, the
results extend holomorphically to arbitrary values of $\rho$
and $\tau$. Solving \eqref{recur}, we find
\EQ{
A_{kl}=e^{2\pi ik(k-1)\tau+2\pi il(l-1)\rho+kl\beta m}
}
up to an unimportant overall factor. Notice that once the rule \eqref{rule}
has been imposed there are no remaining moduli apart
from the overall position of the web: just as one expects for the
$N=1$ curve.  So the curve becomes
\EQ{
F(X,Y)=\sum_{kl=-\infty}^\infty e^{2\pi ik(k-1)\tau+2\pi
il(l-1)\rho+kl\beta m}X^kY^l=0\ ,
}
which, after identifying $X=e^{2\pi iz}$ and $Y=e^{\beta x}$, and
after suitable re-scalings, is simply
\EQ{
\Theta\left[\begin{matrix}0&0\\
0&0\end{matrix}\right]\left(z\quad\frac
{\beta x}{2\pi i}\Big|\begin{matrix}  \tau& \tfrac{\beta m}{2\pi i}\\
\tfrac{\beta m}{2\pi i}&\rho\end{matrix}\right)=0\ .
}
This is precisely the $N=1$ version of \eqref{curvb}. Notice that the
form of the curve ensures that
under the identification
$$X\rightarrow X e^{2i\pi \tau}\quad Y\rightarrow Y e^{m}$$
$$Y\rightarrow Y e^{2i\pi \rho}\quad X\rightarrow X e^m$$
$F(X,Y)=0$ is invariant, since $F(X,Y)\rightarrow X^aY^b
F(X,Y)$ for some $a,b$ (note that $X,Y\not=0$).

Note that from the above construction of the curve from the toric diagram
we have a cyclic symmetry between the parameters. To see this consider the
transformation
\bea
X\mapsto X\,,\,\,Y\mapsto XY\,,
\eea
which has the effect of changing the basic grid in the toric diagram.
The curve $F(X,Y)$ after this transformation becomes
\bea
F(X,XY)=\Theta\left[\begin{matrix}0&0\\
0&0\end{matrix}\right]\left(z+\tau\quad\frac
{\beta x}{2\pi i}\Big|\begin{matrix}  \tau& \tfrac{\beta m}{2\pi i}-\tau\\
\tfrac{\beta m}{2\pi i}-\tau&\tau+\rho-\tfrac{\beta m}{2\pi
  i}\end{matrix}\right)=0\ .
\eea
The period matrix is given by
\bea
\Pi =\begin{pmatrix}  \tfrac{\beta m}{2\pi i}+\widehat{\tau}& -\widehat\tau\\
-\widehat{\tau}&\widehat{\rho}+\widehat{\tau}\end{pmatrix}\ ,
\eea
where $\tau=\widehat{\tau}+\tfrac{\beta m}{2\pi i}$ and $\rho=\widehat{\rho}+\tfrac{\beta m}{2\pi i}$.
After an $Sp(4,\Z)$ transformation we can write this as
\bea
S^{-1}\Pi S=\begin{pmatrix}\widehat{\rho}+\widehat{\tau}& \widehat{\tau}\\
\widehat{\tau}&\tfrac{\beta m}{2\pi i}+\widehat{\tau}\end{pmatrix}=
\begin{pmatrix}
\tau+\rho+2\tfrac{\beta m}{2\pi i}& \tau-\tfrac{\beta m}{2\pi i}\\
\tau-\tfrac{\beta m}{2\pi i} &\tau\end{pmatrix}\ .
\eea
Now consider the transformation
\bea
X\mapsto XY\,,\,\,Y\mapsto Y\,\ ,
\eea
which again changes the basic grid. Also note that the choices of the basic grid
in the toric diagram that we used are the only possibilities.
In this case the curve is given by
\bea
F(XY,Y)=\Theta\left[\begin{matrix}0&0\\
0&0\end{matrix}\right]\left(z-\tau+\tfrac{\beta m}{2\pi i}\quad\frac
{\beta x}{2\pi i}\Big|\begin{matrix}  \tau+\rho-2\tfrac{\beta m}{2\pi i}& \tfrac{\beta m}{2\pi i}-\rho\\
\tfrac{\beta m}{2\pi i}-\rho&\rho\end{matrix}\right)=0\ .
\eea
The period matrix is given by
\bea
\Pi=\begin{pmatrix}  \widehat{\tau}+\widehat{\rho}& -\widehat{\rho}\\
-\widehat{\rho}&\widehat{\rho}+\tfrac{\beta m}{2\pi i}\end{pmatrix}\ .
\eea
After an $Sp(4,\Z)$ transformation we get
\bea
S^{-1}\Pi S=\begin{pmatrix}  \tfrac{\beta m}{2\pi i}+\widehat{\rho}& \widehat{\rho}\\
\widehat{\rho}&\widehat{\tau}+\widehat{\rho}\end{pmatrix}\ .
\eea
Thus the period matrix has a cyclic symmetry ($\widehat{\tau}
\mapsto \widehat{\rho}\mapsto \tfrac{\beta m}{2\pi i}$)
with three period matrices given by
\bea
\Pi:=\begin{pmatrix}  \widehat{\tau}+\tfrac{\beta m}{2\pi i}& \tfrac{\beta m}{2\pi i}\\
\tfrac{\beta m}{2\pi i}&\widehat{\rho}+\tfrac{\beta m}{2\pi i}\end{pmatrix}\,,\,\,
\begin{pmatrix}\widehat{\rho}+\widehat{\tau}& \widehat{\tau}\\
\widehat{\tau}&\tfrac{\beta m}{2\pi i}+\widehat{\tau}\end{pmatrix}\,,\,\,
\begin{pmatrix}  \tfrac{\beta m}{2\pi i}+\widehat{\rho}& \widehat{\rho}\\
\widehat{\rho}&\widehat{\tau}+\widehat{\rho}\end{pmatrix}\ .
\eea
This symmetry is quite clear from the web diagram and is also present in the
corresponding partition function as we will see in
a later section.

This analysis can be extended to $N>1$ by applying the method of
images as above. In this case the curve should be invariant (up to
$F(X,Y)\mapsto X^{a}Y^{b}F(X,Y)$) under the transformations
\EQ{
X\mapsto Xe^{2\sigma}\,,\,\,Y\mapsto Ye^{N\lambda}\,\,\,\,\, \mbox{and}\,,
X\mapsto X e^{N\lambda}\,,\,\, Y\mapsto Y e^{2N\sigma}\,.
\label{tra}
}
The curve is given by summing over all the monomials associated with
the vertices of the toric diagram $(k,l)\mapsto X^{k/N}Y^l$,
\EQ{
F(X,Y):=\sum_{k,l=-\infty}^{+\infty}A_{kl}X^{\tfrac{k}{N}}Y^{l}
= \sum_{j=0}^{N-1}\,\sum_{m,l}A^{j}_{ml}X^{m+\frac{j}{N}}Y^{l}\,,
}
where in the final expression above we have defined $j=k\, (\text{mod } N)$.
Using Eq.~(\ref{tra}) we see that
\bea
A^{j}_{m l}=A_{j}\,e^{m(m-1)\mu+l(l-1)\sigma+mlN\lambda+\frac{2\mu jm}{N}+j\lambda n}\,.
\eea
So the curve becomes
\bea
F(X,Y)&=&
\sum_{j=0}^{N}A_{j}
\sum_{m,l}e^{m(m-1)\mu+l(l-1)\sigma+mlN\lambda+\frac{2\mu jm}{N}+
j\lambda n}X^{m+j/N}Y^{l}\,,\\ \nn
&=&\sum_{j=0}^{N-1}A_{j}\Theta\left[\begin{matrix}0&\tfrac{j}{N}\\
0&0\end{matrix}\right]\left(z \quad\frac{N\beta x}{2\pi i}\Big|\begin{matrix}  \tau& \tfrac{N\beta m}{2\pi i}\\
\tfrac{N\beta m}{2\pi i}&N\rho\end{matrix}\right)\,,
\eea
where $X=e^{N\beta x}$ and $Y=e^{2\pi i z}$. This agrees with the matrix model
calculation of the last section.

\onefigure{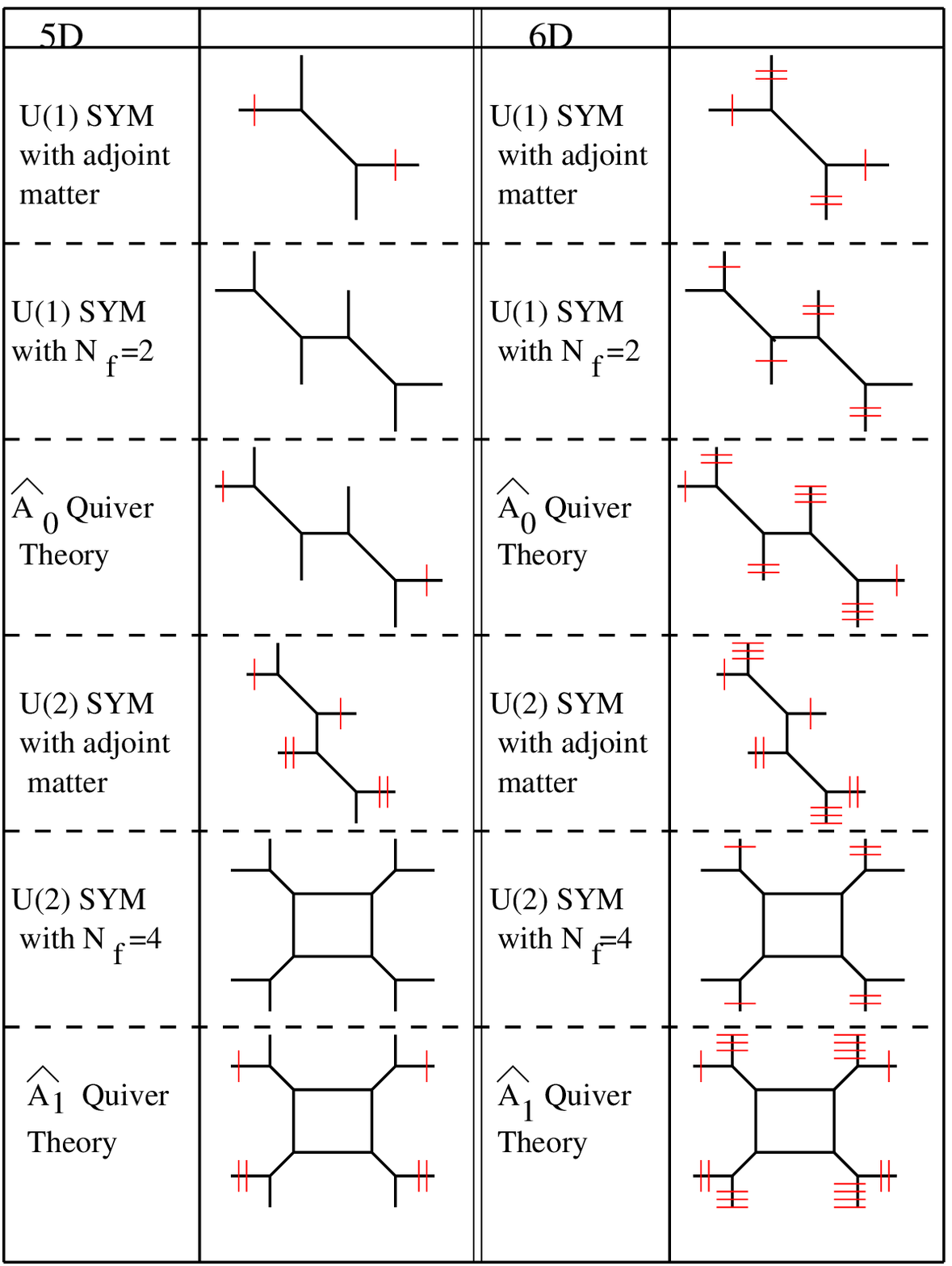}{Field theories in 5 and 6 dimensions with various matter
content and the corresponding web diagrams representing the CY3-fold
geometry.}

\subsection{$\widehat{A}_{N-1}$ theories}
In the last section we saw that 6D theories
with $N_{f}=2N$ can be engineered using
elliptically fibered CY which can be described
by semi-compact web diagrams. In this case one
can ask the question about the relevance, if any, of
completely compact web diagram obtained by identifying both
the NS5-brane direction as well as the D5-brane direction from
the web diagrams of $U(N)$ theory with $N_{f}=2N$.
The case of $U(1)$ with $N_{f}=2N$ is shown in \figref{a0hat}.
\onefigure{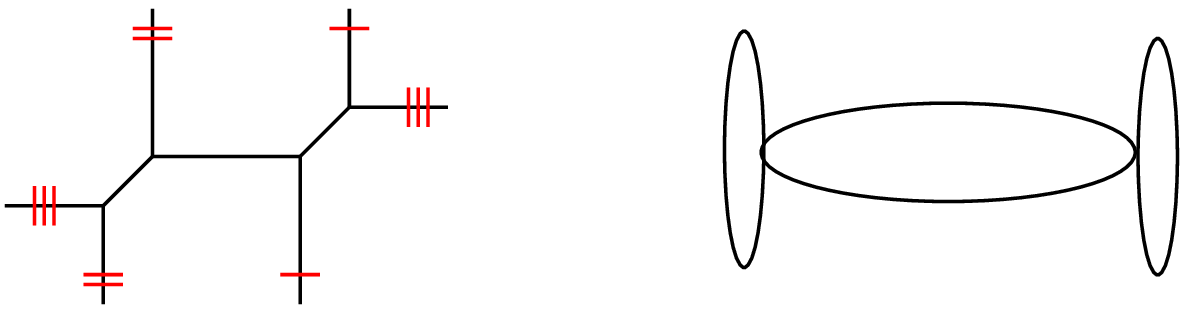}{The web diagram of the 6-dimensional
$\hat A_0$ theory.}
It is easy to see that if the NS5-brane direction is not
compact but the D5-brane direction is compact we get 5D $U(1)\times U(1)$
theory with bifundamental matter. Thus compactifying the NS5-brane
direction gives us 6D $U(1)\times U(1)$ theory compactified
on $\T^{2}$ with bifundamental matter.
In case of N D5-branes we get 6D $U(N)\times U(N)$ theory with
 matter in $(N,\bar{N})+(\bar{N},N)$ compactified on $\T^{2}$
\cite{Brunner:1997gf}.

\section{Topological string amplitudes and BPS degeneracies}

In the previous section we reviewed the geometric engineering of
compactified 4, 5 and 6 dimensional ${\cal N}=2$ theories from CY3-folds
via type IIA, M-theory or F-theory compactifications. An important
ingredient of the geometric engineering recipe is the calculation of
the gauge theory prepotential from the genus zero topological string
amplitude of the corresponding CY3-fold \cite{KKV}. In this section we
will review the interpretation of topological string amplitudes as the
generating functions of the BPS degeneracies of wrapped M2-branes
(or D2-brane and D0-branes) which give rise to particles in the five
dimensional theory.

Consider type IIA strings compactified on a CY3-fold $\X$. The
theory on the transverse four dimensions has ${\cal N}=2$
supersymmetry. This theory in 4 dimensions has certain F-terms
$(g\geq 1)$ \bea \int d^{4}x
F_{g}(t_{i})\,R_{+}^{2}\,F_{+}^{2g-2}\,, \eea which can be
calculated exactly. In the above expression $R_{+}^{2}$ is the
contraction of self-dual part of the Riemann tensor and $F_{+}$ is
the self-dual part of the graviphoton field strength. The function
$F_{g}$ is the A-model topological string amplitude of $\X$  \cite{
Antoniadis:1993ze,BCOV}
and depends on $t_{i}$, the K\"ahler parameters of $\X$. As mentioned
before the prepotential of the theory is given by genus zero
amplitude for which the F-term is given by \bea \int d^{4}x
F_{+}^{i}\wedge F_{+}^{j}\,\partial_{t_{i}}\partial_{t_{j}}F_{0}\ .
\eea Thus the gauge coupling of the 4D theory in terms of the genus
zero amplitude
is given by \bea
\tau_{ij}=\partial_{t_{i}}\partial_{t_{j}}F_{0}(t_{i})\,. \eea

The topological string amplitudes $F_{g}$ arise in the A-twisted
topological theory as integrals over the genus g moduli space of
Riemann surfaces and are related to the generating functions of
the genus $g$ Gromov-Witten invariants. Let us denote by
$\omega\in H^{2}(\X,\CC)$ the complexified K\"ahler class of $\X$.
Then the topological string amplitudes be compactly organized into
the generating function \bea
F(t_{i},\lambda_{s})=\sum_{g=0}^{\infty}\lambda_{s}^{2g-2}F_{g}(t_{i})\,,
\eea where $\lambda_{s}$ is the constant self-dual graviphoton
field strength.

{}From the worldsheet point of view the genus $g$ amplitude,
$F_{g}$, is the generating function of the ``number'' of maps from
a genus $g$ Riemann surface to CY3-fold $\X$. However, the target
space viewpoint provides a more physical interpretation of the
generating function $F(t_{i},\lambda_{s})$, which we now review
\cite{Gopakumar:1998ii,Gopakumar:1998jq}. Recall that in M-theory
compactification on CY3-fold $\X$ we get a 5-dimensional field
theory with eight supercharges. The particles in this theory come
from quantization of the wrapped M2-branes on various 2-cycles of
$\X$. If we consider compactifying one direction then we can
interpret the particles as wrapped D2-branes and the KK modes as
bound D0-branes. In this case integrating out these various
charged particles gives rise to the F-terms in the effective
action. The contribution of a particle of mass $m$ and in
representation ${\cal R}$ of the $SU(2)_{L}\times
SU(2)_{R}$ (the little group of massive particles in 5D) to ${ F}$
is given by \bea
S=\mbox{log}\mbox{det}(\Delta+m^2+2e\,\sigma_{L}{\cal F})\,=
\int_{\epsilon}^{\infty}\frac{ds}{s}
\frac{\mbox{Tr}_{{\cal R}}(-1)^{\sigma_{L}+\sigma_{R}}
e^{-sm^2}e^{-2s e\sigma_{L}{\cal F}}}{(2\sinh (se{\cal F}/2))^{2}}\,, \eea
where $\sigma^{L}$ is the Cartan of $SU(2)_{L}$ and arises because
the graviphoton field strength is self-dual. $e$ is the charge of
the particle and is equal to its mass and we identify the graviphoton
field strength
${\cal F}=\lambda_s$. The mass of the particle is given by the area of the
curve on which the D2-brane is wrapped. An extra subtlety arises
due to D0-branes. In the lift to M-theory we see that a wrapped
M2-brane comes with momentum in the circle direction and therefore
if we denote the mass of the M2-brane wrapping a curve class
$\Sigma\in H_{2}(\X,\Z)$ by $T_{\Sigma}$ then the mass of the
M2-brane with momentum $n$ is given by taking $T_{\Sigma}$ to
$T_{\Sigma}+2\pi in/\lambda$. Let us denote by
$N^{(j_{L},j_{R})}_{\Sigma}$ the number of BPS states coming from
M2-brane wrapped on the holomorphic curve $\Sigma$ and left-right
spin content under $SU(2)_{L}\times SU(2)_{R}$ given by
$(j_{L},j_{R})$. Then the total contribution coming from all
particles is obtained by summing over the momentum, the
holomorphic curves and the left-right spin content, \bea
F&=&\sum_{\Sigma\in H_{2}(\X,\Z)}\sum_{n\in
\Z}\sum_{j_{L},j_{R}}N^{(j_{L},j_{R})}_{\Sigma}
\int_{\epsilon}^{\infty}\frac{ds}{s}\frac{Tr_{(j_{L},j_{R})}
(-1)^{\sigma_{L}+\sigma_{R}}e^{-sT_{\Sigma}-2\pi
in} e^{-2s\sigma_{L}\lambda_s}}{(2\sinh (s\lambda_s/2))^{2}}\,,\\
\nn &=&\sum_{\Sigma\in
H_{2}(\X,\Z)}\sum_{k=1}^{\infty}\sum_{j_{L},j_{R}}N^{(j_{L},j_{R})}_{\Sigma}
e^{-kT_{\Sigma}}\frac{\mbox{Tr}_{(j_{L},j_{R})}(-1)^{\sigma_{L}+\sigma_{R}}
e^{-2k\lambda_s\sigma_{L}}}{k(2\sinh (k\lambda_s/2))^{2}}\,,\\
\nn &=&\sum_{\Sigma\in
H_{2}(\X,\Z)}\sum_{k=1}^{\infty}\sum_{j_{L}}N^{j_{L}}_{\Sigma}
e^{-kT_{\Sigma}}\frac{\mbox{Tr}_{j_{L}}(-1)^{\sigma_{L}}
e^{-2k\lambda_s\sigma_{L}}}{k(2\sinh (k\lambda_s/2))^{2}}\,,\\
\nn \eea where \bea
N^{j_{L}}_{\Sigma}=\sum_{j_{R}}N^{(j_{L},j_{R})}_{\Sigma}(-1)^{2j_{R}}(2j_{R}+1)\,.
\eea
It is useful to define a different basis of $SU(2)_{L}$
representations given by  $I_{g}=(2(0)+(\frac{1}{2}))^{g}$ such
that in terms of this basis \bea
\sum_{j_{L}}N^{j_{L}}_{\Sigma}[j_{L}]=\sum_{g=0}^{\infty}n^{g}_{\Sigma}I_{g}\,.
\eea The coefficients $n^{g}_{\Sigma}$ are integers and given by
\bea
\sum_{g=0}^{\infty}n^{g}_{\Sigma}(-1)^{g}(q^{1/2}-q^{-1/2})^{2g}=
\sum_{j_{L}}N^{j_{L}}_{\Sigma}(q^{-j_{L}}+\cdots +q^{+j_{L}})\,.
\eea
 In terms
of these integers one can write $F$ as \bea F=\sum_{\Sigma\in
H_{2}(\X,\Z)}\sum_{k=1}^{\infty}\sum_{g=0}^{\infty}n^{g}_{\Sigma}
e^{-kT_{\Sigma}}\frac{\mbox{Tr}_{I_{g}}(-1)^{\sigma_{L}}
e^{-2k\lambda_s\sigma_{L}}}{k(2\sinh (k\lambda_s/2))^{2}}\,,\\
\nn \eea It is easy to show that \bea
\mbox{Tr}_{I_{g}}(-1)^{\sigma_{L}}e^{-2k\lambda_s
\sigma_{L}}=\Big(\mbox{Tr}_{I_{1}}(-1)^{\sigma_{L}}
e^{-2k\lambda_s\sigma_{L}}\Big)^{g}=\Big(2\sinh (k\lambda_s/2)\Big)^{2g}\,.
\eea Thus we get \bea F=\sum_{\Sigma\in
H_{2}(\X,\Z)}\sum_{k=1}^{\infty}\sum_{g=0}^{\infty}\frac{n^{g}_{\Sigma}}{k}
(2\sinh (k\lambda_s/2))^{2g-2}e^{-kT_{\Sigma}}\,. \eea The
target space point of view allows the topological string
amplitudes to be written in terms of integers $n^{g}_{\Sigma}$
which give the BPS degeneracies of the states coming from wrapped
D2-branes. The fact that $F$ has this particular form with integer $n^{g}_{\Sigma}$
has been confirmed for many non-compact toric threefolds.

\subsection{Second quantized strings and A-model partition function}

In the previous section we saw that the target space view point
when discussing the topological string amplitudes
is perhaps more interesting than the worldsheet view point since it allows
the amplitudes to be written in terms of invariants which are integers.
However another interesting property, the integrality of $e^{F}$,
becomes clear from this view point as well.

To see this note that we can write $e^{F}$, which we will call
the partition function from now on, as
\bea
Z(\omega,g_{s})=e^{F(\omega,g_{s})}=\prod_{\Sigma\in H_{2}(\X)}\prod_{j_{L}}
\prod_{k=-j_{L}}^{+j_{L}}\prod_{m=0}^{\infty}
(1-q^{2k+m+1}Q^{\Sigma})^{(-1)^{2j+1}(m+1)N^{j_{L}}_{\Sigma}}\,,
\eea
where $Q^{\Sigma}=e^{-T_{\Sigma}}$ and $q=e^{-i\lambda_s}$.
This expression looks very much like `counting'
the states in a Hilbert space (this was also
noted in \cite{Gopakumar:1998ii} for the case of $j_L=0$ BPS states).
In a sense we have already
explained how partition function counts $M2$-branes.  So the
integrality of $Z$ must be directly related to this integrality
in $F$.  We can in fact see $Z$ as the partition function
of a second quantized theory built purely out of fields
creating $M2$-branes.  Let $\Phi_{\Sigma ,m_1,m_2}(z_1,z_2)$ denote
a field creating an $M2$-brane BPS state, where $z_i$ denote
the two complex coordinates of the 4 dimensional space, $\Sigma$
denotes the BPS charge and $m_i$ denote the internal
spins of the BPS particle with respect to $U(1)\times U(1)=
SO(2)\times SO(2)\subset SO(4)$.  Consider only {\it holomorphic}
configurations of the BPS fields.  This is what we usually
do in the context of 2d chiral block of a conformal theory.  If we
do this we have the natural decomposition
$$\Phi_{\Sigma ,m_1,m_2}(z_1,z_2)=z_1^{m_1}z_2^{m_2} \sum_{n_1,n_2\geq 0}
\alpha_{n_1+m_1,n_2+m_2}(\Sigma )z_1^{n_1}z_2^{n_2}
$$
where $\alpha_{n_1+m_1,n_2+m_2}(\Sigma )$ are bosonic
or fermionic modes depending on whether the
field $\Phi$ (which is the lowest
component of a superfield) is bosonic or fermionic respectively.
Note also the prefactor
monomial is the usual mapping of modes from cylinder to the plane
for each $z_i$.  Note that $j^3_L=n_1+n_2+m_1+m_2$  for each
mode $\alpha_{n_1+m_1,n_2+m_2}(\Sigma )$.

Since $N^{j_{L}}_{\Sigma}$ is the
BPS degeneracy of the states with charge $\Sigma$ and
$SU(2)_{L}$ spin $j_{L}$ we can write the above partition function as
\bea
Z:=\mbox{Tr}_{ {\cal H}}(-1)^{2(j_{L}+j_{R})}\,q^{2j^{3}_{L}}\,e^{-T}\,.
\eea
where ${\cal H}$ denotes the subspace of the second quantized Hilbert space
generated by holomorphic modes of the lowest component of the BPS fields and
$T$ denotes the total mass of the BPS states which is the same
as the Hamiltonian of the theory.
It is quite exciting that the partition function of topological
string seems to be counting a second quantized hilbert
space of holomorphic components of BPS states.  It is also
natural to believe there is an interesting algebra related to this
partition function.  In particular we can take the product
of two holomorphic BPS fields as defining an algebra:
$$ \Phi_{\Sigma_1}\Phi_{\Sigma_2}= \sum_{\Sigma_i =\Sigma_1+\Sigma_2}
C_i\Phi_{\Sigma_i}$$
This
would be interesting to study further.  It would also be interesting
to see the connection of this holomorphic OPE of BPS
states to the BPS algebra defined in
\cite{HM}.

\subsection{Non-compact toric threefolds and the topological vertex}
In this section we consider the case of non-compact toric CY3-folds.
These CY3-folds are extremely interesting not only because they are
``simple'' enough so that exact calculation of A-model partition
function can be done but also because they give rise to gauge theories
via geometric engineering \cite{KKV} as we saw in the last section.

{}From the discussion of the last section we see that if the
graviphoton field strength is not self-dual $F:=F_{+}+F_{-}$, then
we can write
the contribution of coming from integrating
out the particle in representation ${\cal R}$ of $SU(2)_{L}\times SU(2)_{R}$
as
\bea
S:=\int_{\epsilon}^{\infty}\frac{ds}{s}
\frac{\mbox{Tr}_{{\cal R}}(-1)^{\sigma_{L}+\sigma_{R}}e^{-sm^2}e^{-2s e(\sigma_{L}F_{+}
+\sigma_{R}F_{-})}}{(2\sinh (seF_{+}/2))(2\sinh (seF_{-}/2))}\,.
\eea
Summing over the contribution from all particles as before we get
\SP{
&F(q_{1},q_{2})=
\\&\sum_{\Sigma\in H_{2}(\X,\Z)}\sum_{n=1}^{\infty}\sum_{j_{L},j_{R}}
\frac{N^{(j_{L},j_{R})}_{\Sigma}\Big((q_{1}q_{2})^{-nj_{L}}+\cdots+
(q_{1}q_{2})^{nj_{L}}\Big)\Big((\frac{q_{1}}{q_{2}})^{-nj_{R}}+\cdots +
(\frac{q_{1}}{q_{2}})^{nj_{R}}\Big)}{n(q_{1}^{n/2}-q_{1}^{-n/2})
(q_{2}^{n/2}-q_{2}^{-n/2})}\,e^{-nT_{\Sigma}}\,,
}
where we have defined $q_{1}=e^{F_{+}},q_{2}=e^{F_{-}}$. The
integers $N^{(j_{L},j_{R})}_{\Sigma}$ give the degeneracy of particles
with spin content $(j_{L},j_{R})$ and charge $\Sigma$ and  are the number
of cohomology classes with spin $(j_{L},j_{R})$ of the moduli space of
D-brane wrapped on a holomorphic curve in the class $\Sigma$
 \cite{Gopakumar:1998ii,Gopakumar:1998jq}.
Because the D-brane has a $U(1)$ gauge field living on its worldvolume
the moduli space of supersymmetric configurations includes not only the
curve moduli but also the moduli of the flat connections on the curve
coming from the gauge field. Since the moduli space of flat connections
on a smooth curve of genus $g$ is $T^{2g}$ therefore the moduli space of
the D-brane is $T^{2g}$ fibration over the moduli space of the curve. The
total space is a K\"ahler manifold and the Lefshetz action by the
K\"ahler class is the diagonal $SU(2)_D\subset SU(2)_{L}\times SU(2)_{R}$ action
on the moduli space.  The $SU(2)_{L}\times SU(2)_{R}$ action on the moduli
space is such that $SU(2)_{L}$ acts on the fiber direction and
the $SU(2)_{R}$ acts in the base direction.

In the previous section when
discussing the generic CY3-folds we summed over the $SU(2)_{R}$ action
by taking the graviphoton field strength to be self-dual. This was
essentially due to the fact that $N^{(j_{L},j_{R})}_{\Sigma}$ can change
as we change the complex structure; the supersymmetry
algebra allows such pairings between neighboring $j_R$'s
to give a non-reduced multiplet. But $N^{j_{L}}_{\Sigma}$, which
sums over all $j_R$'s with alternating signs remains
invariant.  For the case of non-compact toric CY3-folds
there are no complex structure deformations.  Therefore
one would expect no jumps in the  $N^{(j_{L},j_{R})}_{\Sigma}$
degeneracies, and so one would hope to be able
to compute these as well.   We will come back to this
after our discussion of the topological vertex.

\underline{\bf Topological vertex:}
It was shown in \cite{AKMV} (see also
the earlier work \cite{Iqbal,AI2,AMV,DG}) that topological string amplitude
for non-compact 3-folds can be calculated using the corresponding
web diagrams and the topological vertex: A function of $q$,
$C_{R_{1} R_{2} R_{3}}$, depends on three representation,
$R_{1,2,3}$, of $U({\infty})$ associated with each tri-valent
vertex of the web diagram. The topological vertex $C_{R_{1} R_{2} R_{3}}$
is actually an open string amplitude for a certain geometry with
D-branes as we will see later. An expression for the topological vertex in terms of the Hodge integrals was proposed in \cite{DB}. The proposed expression was checked for many nontrivial representations but a general proof remains an open problem. The relation between Hodge integrals and topological vertex with one trivial representation was determined in \cite{zhou0} using localization \footnote{The relation between Hodge integrals and topological vertex was also explored in \cite{zhou00,OP,zhou01,zhou1,zhou2,zhou3}}. For local toric threefolds with one compact 4-cycle (so that one representation of the topological vertex is always trivial) the paritition function can be determined using localization and agrees with the topological vertex calculation \cite{zhou3}. Let us briefly review the idea behind
the topological vertex and the derivation of topological string
amplitudes for a non-compact CY3-fold.

Recall that in the last section we saw that non-compact toric 3-folds
can be represented by web diagrams which captures the non-trivial
aspects of the geometry as a tri-valent graph in two dimensions. The
graph is the locus of degeneration of a $\T^{2}$ fibration over the
plane. Along each edge of the web a 1-cycle of the fiber $\T^{2}$
shrinks and therefore at each point of the edge we have an $\S^{1}$, the
cycle dual to the one shrinking. Given this cycle we can consider a
D-brane with 3 dimensional worldvolume $\S^{1}\times \CC$ which
wraps this cycle and fills two other directions only one of which could
be in the plane of the web diagram. Such a 3-cycle is Lagrangian and
can be used to define the boundary conditions for the open topological
strings \cite{AV,AKV}. As we mentioned before the web diagrams
corresponding to non-singular geometries are tri-valent graphs. All the
vertices of the web diagram are $\sl2z$ transform of each other and
hence the web diagram can be ``built'' using the basic vertex, in which
we have $(1,0),(0,1)$ and $(-1,-1)$ lines coming together, and its
$\sl2z$ transforms joined by edges which are straight lines. Such a web with
only $(1,0), (0,1)$ and $(-1,-1)$ lines corresponds to threefold which
is $\CC^{3}$. In this case the geometry is trivial and the only
contribution  to the topological string amplitude comes from constant maps.
However, as mentioned before we can have D-branes in this geometry which
will provide boundaries for maps from worldsheet with boundaries as
shown in \ref{tv} where $R_{i}$ are the representations in which we take
the holonomy around the circle which the D-branes wrap.
\onefigure{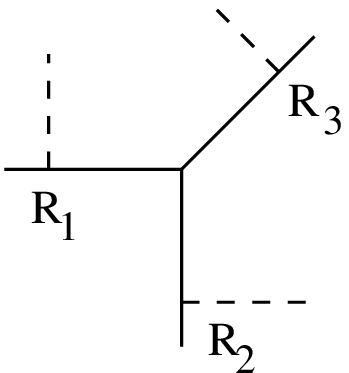}{The topological vertex is defined as the open
string amplitudes in the presence of three stacks of Lagrangian branes.}
The open topological
string amplitude of this geometry  is given by the topological vertex
\bea
C_{R_{1} R_{2} R_{3}}=\sum_{Q_{1},Q_{2}}N^{R_{1}R_{3}^{t}}_{Q_{1} Q_{2}}\,
q^{\kappa_{R_{2}}/2+\kappa_{R_{3}}/2}\,\frac{{\cal W}_{R_{2}^{t} Q_{1}}\,
{\cal W}_{R_{2} Q_{2}}}{{\cal W}_{R_{2}}}\,,
\eea
where
\bea
N^{R_{1} R_{3}^{t}}_{Q_{1} Q_{2}}=\sum_{Q}N^{R_{1}}_{Q
Q_{1}}N^{R_{3}^{t}}_{Q Q_{2}}\,.
\eea
Here $N_{ab}^c$ is the degeneracy of representation $c$ in the tensor
product $a\otimes b$, the ${\cal W}_{R_{1} R_{2}}$ is the
link invariant for Hopf link for $U(\infty)$, and $\kappa_R$
denotes a quadratic casimir for representation $R$.
Another useful representation of the vertex is given using the skew-Schur functions. Let us denote the Young diagram corresponding to representation $R$ as $\mu_{R}$ then \cite{ORV}
\bea
C_{R_{1} R_{2} R_{3}} =
q^{(\kappa_{R_{1}}+\kappa_{R_{3}})/2}
s_{R_{3}^{t}}(q^{\rho})\sum_{R}
s_{R_{1}^{t}/R}(q^{\mu_{R_{3}}+\rho})
s_{R_{2}/R}(q^{\mu_{R_{3}^{t}}+\rho})\,,
\label{tv1}
\eea
where $s_{R_{1}/R}(x)$ is the skew-Schur function defined
as
\bea
s_{R_{1}/R}(x)=\sum_{R_{2}}N^{R_{1}}_{R R_{2}}s_{R_{2}}(x)\,,
\eea
and $q^{\mu+\rho}=\{q^{\mu_{1}-1/2},q^{\mu_{2}-3/2},q^{\mu_{3}-5/2},\ldots\}$.

To calculate the partition function for any non-compact threefold we consider
its web diagram and associate with each leg a representation $R_{i}$ and
with each vertex $C_{R_{i} R_{j} R_{k}}$ where $R_{i,j,k}$ are the
representations on the legs joining the vertex. Then the partition function
is given by multiplying all the vertices together and summing over all
the representations with weights $\prod_{i}e^{-T_{i}\ell_{R_{i}}}$ where
$T_{i}$ is area of the curve associated to the i-th edge and $\ell_{R}$ is the
number of boxes in the Young diagram corresponding to $R$. There are extra
subtleties associated with orientation of the legs which leads to extra
framing factors. For details of this we refer the reader to \cite{AKMV}.

As an example consider the case of the resolved conifold as shown in
\ref{example}.
\onefigure{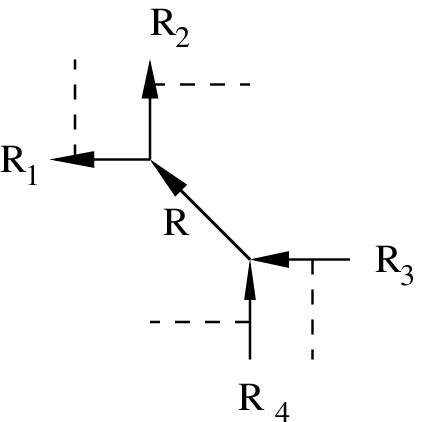}{The web diagram of the resolved conifold with four stacks of
D-branes. The representation $R$, which is summed over, is associated with the
internal line and $R_{1,2,3,4}$ are fixed representations associated with the external lines of the web.}
We will see that using the identities involving Schur and skew-Schur function it is possible to write a simple
expression for the partition function of this geometry. We will use similar identities involving
skew-Schur function to determine the partition functions of gauge theories in the next section.

Denote by $T$ the area of the $\mathbold{P}^{1}$ then partition function is given by
\bea
Z_{R_{1},R_{2},R_{3},R_{4}}&=\sum_{R}e^{-T\ell_{R}}C_{R_{1}R_{2}R^{t}}(q)\,(-1)^{\ell_{R}}C_{R\,R_{3}^{t}\,R_{4}^{t}}(q)\,.
\eea
Using Eq(\ref{tv1}) we can write $Z_{R_{1,2,3,4}}$ as $(Q:=e^{-T}$)
\bea\nn
Z_{R_{1,2,3,4}}&=&q^{(\kappa_{R_{1}}+\kappa_{R_{2}}-\kappa_{R_{3}}-\kappa_{R_{4}})/2}
s_{R_{1}^{t}}(q^{\rho})s_{R_{3}}(q^{\rho})
\sum_{\eta_{1},\eta_{2}}s_{R_{2}^{t}/\eta_{1}}(q^{\mu_{R_{1}}+\rho})s_{R_{4}/\eta_{2}}(q^{\mu_{3}^{t}+\rho})
\\\nn
&&\sum_{R}Q^{\ell_{R}}(-1)^{\ell_{R}}\,s_{R/\eta_{2}}(q^{\mu_{3}+\rho})\,s_{R^{t}/\eta_{1}}(q^{\mu_{R_{1}^{t}}+\rho})\,.
\eea
The sum over $R$ can be carried out exactly using the identity \cite{macdonald}
\bea
\sum_{R}s_{R/\eta_{1}}(x)s_{R^{t}/\eta_{2}}(y)=\prod_{i,j}(1+x_{i}y_{j})\,\sum_{\eta_{3}}s_{\eta_{2}^{t}/\eta_{3}}(x)
s_{\eta_{1}^{t}/\eta_{3}^{t}}(y)\,.
\eea
After summing over $R, \eta_{1,2}$ and denoting the  Schur function corresponding to $R$ in
the variable $(x_{1},x_{2},\cdots,y_{1},y_{2},\cdots)$ by $s_{R}(x,y)$ we get
\bea
Z_{R_{1,2,3,4}}&=&q^{(\kappa_{R_{1}}+\kappa_{R_{2}}-\kappa_{R_{3}}-\kappa_{R_{4}})/2}
s_{R_{1}^{t}}(q^{\rho})s_{R_{3}}(q^{\rho})
\prod_{i,j}(1-Q\,q^{\mu_{R_{3},i}+\rho_{i}+\mu_{R_{1}^{t},j}+\rho_{j}})\,\\\nn
&&\sum_{\eta_{3}}
Q^{\ell_{\eta_{3}}}(-1)^{\ell_{\eta_{3}}}s_{R_{2}/\eta_{3}}(Qq^{-\mu_{R_{1}}-\rho},q^{\mu_{R_{3}}+\rho})
s_{R_{4}/\eta_{3}^{t}}(q^{\mu_{R_{3}^{t}}+\rho},Qq^{-\mu_{R_{1}^{t}}-\rho})\,.
\eea
The sum in the expression above is finite and can be determined easily for any given $R_{2,4}$.
Consider the case $(R_{1},R_{2},R_{3},R_{4})=(\bullet,\bullet,\bullet,R)$ and
$(R_{1},\tableau{1},R_{3},\tableau{1})$ then
\bea
Z_{\bullet \bullet \bullet R}&=&q^{-\kappa_{R}/2}\Big(\prod_{k=1}^{\infty}(1-Qq^{k})^{k}\Big)\,s_{R}(q^{\rho},Qq^{-\rho})\,,\\\nn
Z_{R_{1},\tableau{1},R_{3},\tableau{1}}&=&\Big(\prod_{k=1}^{\infty}(1-Qq^{k})^{k}\Big)
\Big(\prod_{k}(1-Qq^{-k})^{C_{k}(R_{1}^{t},R_{3})}\Big)\,
s_{R_{1}}(q^{\rho})s_{R_{3}^{t}}(q^{\rho})\\\nn
&&\Big(\frac{q}{(1-q)^{2}}(1-Q)^{2}+f_{R_{3},R_{3}^{t}}-Q(f_{R_{1}R_{3}^{t}}+f_{R_{1}^{t}R_{3}})-Q^{2}f_{R_{1}R_{1}^{t}}\Big)\,.
\eea
Where
\bea
f_{R_{1}R_{2}}&=&\sum_{k}C_{k}(R_{1},R_{2})q^{k}=s_{\tableau{1}}(q^{-\mu_{R_{1}}-\rho})s_{\tableau{1}}(q^{-\mu_{R_{2}}-\rho})-s_{\tableau{1}}^{2}(q^{-\rho})\\\nn
&=&q^{-1}(q-1)^{2}f_{R_{1}}f_{R_{2}}+f_{R_{1}}+f_{R_{2}}\,,\,\,\,\mbox{and}\\ \nn
f_{R}&=&\sum_{(i,j)\in R}q^{j-i}\,.
\eea

\underline{\bf Generalized partition function:}
In a previous section we saw that the BPS degeneracies
given by the target space viewpoint of the topological strings are obtained by
summing over the $j_R$ spin content. This is necessary in order to obtain an
index invariant under the complex structure deformations of the CY3-fold.
Hence for a CY3-fold with no complex structure deformation we do not have to
sum over the $SU(2)_{R}$. As mentioned before non-compact CY3-folds are such
spaces for which there are no complex structure deformations and hence a
more general partition function encoding the full $SU(2)_{L}\times SU(2)_{R}$
spin content can be defined.

The generalized Partition function in such cases would be given as a trace
over the second quantized Hilbert space generated by holomorphic
components of the BPS fields:
\SP{
 &Z=\mbox{Tr}_{{\cal H}}(-1)^{2(j_{L}+j_{R})}\,q_{1}^{j^{3}_{L}+j^{3}_{R}}\,
q_{2}^{j^{3}_{L}-j^{3}_{R}}\,e^{-T}=e^{F(q_{1},q_{2})}=\\
&\prod_{\Sigma\in
H_{2}(\X)}\prod_{j_{L},j_{R}}
\prod_{k_{L}=-j_{L}}^{+j_{L}}\prod_{k_{R}=-j_{R}}^{+j_{R}}\prod_{m_{1},m_{2}=0}^{\infty}
(1-q_{1}^{k_{L}+k_{R}+m_{1}+\frac{1}{2}}\,q_{2}^{k_{L}-k_{R}+m_{2}+\frac{1}{2}}
Q^{\Sigma})^{(-1)^{2(j_{L}+j_{R})+1}N^{j_{L},j_{R}}_{\Sigma}}\,.
}
To determine the invariants $n^{(g_{1}, g_{2})}_{\Sigma}$ which correspond
to the basis $(I_{g_{1}},I_{g_{2}})$ we can use the relation
\bea &&\sum_{g_{1},g_{2}\geq
0}n_{\Sigma}^{g_{1},g_{2}}(-1)^{g_{1}+g_{2}}
((q_{1}q_{2})^{1/4}-(q_{1}q_{2})^{-1/4})^{2g_{1}}\,
\big((\tfrac{q_{1}}{q_{2}})^{1/4}-(\tfrac{q_{1}}{q_{2}})^{-1/4}\big)^{2g_{2}}=\\
\nn
&&\sum_{j_{L},j_{R}}N_{\Sigma}^{j_{L},j_{R}}(-1)^{2(j_{L}+j_{R})}\,
((q_{1}q_{2})^{-j_{L}}+\cdots+(q_{1}q_{2})^{+j_{L}})\,\big((\tfrac{q_{1}}{q_{2}})^{-j_{R}}+
\cdots+(\tfrac{q_{1}}{q_{2}})^{+j_{R}}\big)\,.
\eea

Direct computation of such a partition function, say using some
generalized topological vertex, is not known for any non-compact 3-fold.
However, using the geometric engineering relation between the
compactified 5D gauge theories and non-compact 3-folds one can obtain
the generalized partition function for  some cases using the instanton
calculus developed by Nekrasov \cite{Nekrasov:2002qd, NY}.   Note that
in the non-compact case the notion of $SU(2)_R$ is ambiguous, as it
can mix with R-symmetry (there is no normalizable 4d gravity mode).
In fact a particular combination of $SU(2)_R$ and the R-symmetry $SU(2)$
is what is computed in \cite{Nekrasov:2002qd}\ which we find to correspond
to the $SU(2)_R$ defined in the $D2$-brane moduli problem acting
on the base of the moduli space \cite{Gopakumar:1998jq}.

As an example consider the pure five dimensional
$U(2)$ gauge theory which can be obtained via M-theory compactification
on local $\PP^1\times \PP^{1}$. In this case the gauge
theory partition function was calculated in \cite{Nekrasov:2002qd}. This partition
function can also be calculated using the refined vertex formalism developed recently in \cite{refined}.
We can use these results to verify that the generalized partition function gives
the degeneracies which represent the $SU(2)\times SU(2)$ action on
the moduli space of the D-branes.

The partition function for the case of local $\PP^{1}\times \PP^{1}$ is given
by
\bea
Z:=\sum_{R_{1,2}}Q_{b}^{l_{1}+l_{2}}Z_{R_{1},R_{2}}(Q_{f})\ ,
\eea
where $T_{b,f}=-\mbox{log}(Q_{b,f})$ are the K\"ahler parameters
associated with the base and the fiber $\PP^{1}$ and we sum over all
pairs of Young diagrams $R_{1,2}$ such that
\bea\nn
Z_{R_{1},R_{2}}(Q_{f})=\frac{C_{R_{1}}(q_{1},q_{2})C_{R_{2}}(q_{1},q_{2})
C_{R_{1}^{T}}(q_{2},q_{1})C_{R_{2}^{T}}(q_{2},q_{1})\,
q_{1}^{\kappa_{R_{1}}/2+\kappa_{R_{2}}/2}\,(q_{1}/q_{2})^{\sum_{i}
(\mu_{1,i}^{2}-\mu_{2,i}^{2})/2}}{\prod_{k_{1},k_{2}}
(1-q_{1}^{k_{1}}q_{2}^{k_{2}}Q)^{C_{k_{1},k_{2}}(R_{1},R_{2})}}\,.
\eea
In the above expression
\bea
C_{R}(q_{1},q_{2})=(-1)^{l_{R}}(q_{1}q_{2})^{\kappa_{R}/8}\,
q_{1}^{h^{1}_{R}/2}q_{2}^{h^{2}_{R}/2}\,\prod_{(i,j)\in R}
(1-q_{1}^{h^{1}_{R}(i,j)}q_{2}^{h^{2}_{R}(i,j)})^{-1}\,.
\eea
and $\kappa_{R}=\sum_{(i,j)\in R}(j-i),\,h^{1}_{R}(i,j)=\mu^{t}_{j}-i, h^{2}_{R}(i,j)=\mu_{i}-j+1$.
The integers $C_{k_{1},k_{2}}(q_{1},q_{2})$ are given by
\bea\nn
\sum_{k_{1,2}}C_{k_{1},k_{2}}(R_{1},R_{2})q_{1}^{k_{1}}q_{2}^{k_{2}}&=&
\sum_{(i,j)\in R_{1}}q_{1}^{-h^{1}_{R_{2}}(i,j)}q_{2}^{-h^{2}_{R_{1}}(i,j)}+
\sum_{(i,j)\in R^{T}_{2}}
q_{1}^{h^{2}_{R^{T}_{1}}(i,j)}q_{2}^{h^{1}_{R^{T}_{2}}(i,j)}\\ \nn
&&+\sum_{(i,j)\in R_{2}}q_{1}^{h^{1}_{R_{1}}(i,j)}q_{2}^{h^{2}_{R_{2}}(i,j)}+
\sum_{(i,j)\in R_{1}^{T}}q_{1}^{-h^{2}_{R^{T}_{2}}(i,j)}q_{2}^{-h^{1}_{R_{1}^{T}}(i,j)}\,.
\eea

We can use the above partition function to calculate the BPS degeneracies
of various states corresponding to charge $\Sigma \in H_{2}(\X,\Z)$. For
example the term linear in $Q_{b}$, given by
$(R_{1},R_{2})=(\tableau{1},\bullet),(\bullet, \tableau{1})$, determines the
integers $N^{(j_{L},j_{R})}_{\beta}$ for all curves $\beta=B+kF\,k\geq 0$
and gives
\bea
N^{(j_{L},j_{R})}_{B+kF}=\delta_{j_{L},0}\,\,\delta_{j_{R},k+1/2}\,,
\eea
which is consistent with the fact that these are genus zero curves and
therefore action in the fiber direction, which is just a point, is trivial.
The moduli space of these curves is given by $\PP^{2k+1}$ and the
$SU(2)_{R}$ action on this is just the Lefshetz action via K\"ahler class
therefore the cohomology classes decomposes into a spin $[k+1/2]$
representation. If $\omega$ is the K\"ahler class then the $j_{R}^{3}$ on
the cohomology class $\omega^{n}$ is given by $n-k-1/2$. And since there
is one such class for each $n$ we get a single copy of the representation
$[k+1/2]$.

A more interesting example in which both the left and the right spin
content is non-trivial is given by the curve $2B+2F$, the canonical
class of the $F_{0}$. This is a genus one curve and therefore the
corresponding moduli space will admit non-trivial $SU(2)_{L}$ action.
To determine the spin content from the partition function we will have
to expand it to order $Q_{b}^{2}Q_{f}^{2}$, take the log of the
corresponding expression and subtract multicover contribution. In this
case we get
\bea
\sum_{j_{L},j_{R}}N^{(j_{L},j_{R})}_{2B+2F}(j_{L},j_{R})=(\tfrac{1}{2},4)+(0,\tfrac{7}{2})
+(0,\tfrac{5}{2})\,.
\eea
To see that this is the correct result note that the moduli space of
$2B+2F$ together with its jacobian is give by a $\PP^{7}$ bundle over
$\PP^{1}\times \PP^{1}$: pick a point in $\PP^{1}\times \PP^{1}$, the moduli
space of curves passing through that point in the class $2B+2F$ is given by
$\PP^{7}$. Thus the diagonal $SU(2)_{L}\times SU(2)_{R}$ action which just
the Lefshetz action is given by
\bea
(\tfrac{1}{2})\otimes (\tfrac{1}{2})\otimes (\tfrac{7}{2})=
(\tfrac{5}{2})+2(\tfrac{7}{2})+
(\tfrac{9}{2})\,.\eea
Note that since $2B+2F$ is a genus one curve the corresponding jacobian
is also genus one and therefore $j_{L}$ can only be $0,\frac{1}{2}$. From
this restriction on $j_{L}$ and the above diagonal action we see that
the unique left-right spin content is given by
\bea
(\tfrac{1}{2},4)+(0,\tfrac{7}{2})+(0,\tfrac{5}{2})\,,
\eea
exactly as predicted by the partition function calculation.

It would be interesting to generalize the notion of topological
vertex to depend on two parameters $q_1,q_2$ instead of just $q$.
In a sense from \cite{Nekrasov:2002qd}
we already have a prediction for what this should be when
two representations of the vertex are trivial.

\section{Partition functions from the topological vertex}
In this section we determine the A-model partition functions for
the various CY3-fold geometries we discussed in the last section.
The genus zero contribution to the partition function determines
the prepotential of the corresponding gauge theory realized via
geometric engineering.

The partition functions are determined mostly using the topological
vertex
\footnote{In this section we will use slightly different expression for the topological vertex than Eq(\ref{tv1}). Let us denote the expression given
in Eq(\ref{tv1}) by $\widehat{C}_{R_{1}R_{2}R_{3}}(q)$ then in this section
we will use $C_{R_{1}R_{2}R_{3}}(q)=(-1)^{\ell_{R_{1}}+\ell_{R_{2}}+\ell_{R_{3}}}
q^{-(\kappa_{R_{1}}+\kappa_{R_{2}}+\kappa_{R_{3}})/2}
\widehat{C}_{R_{1}R_{2}R_{3}}(q^{-1})$. Since gluing rule involves taking the transpose, which gets rid of
$(-1)^{\ell_{R}}q^{\kappa_{R}}$ factors, and the closed topological string partition function is invariant under $q\mapsto q^{-1}$ therefore both expressions give the same result for geometries with no branes.} \cite{AKMV}. However, in some cases it is easier to use the
Chern-Simons theory\cite{Gopakumar:1998ki, AMV, DG}.

We will discuss in detail the partition function of the CY3-folds
which realize the $U(1)$ and $U(2)$ theory with an adjoint
hypermultiplet as well as the CY3-folds which realize $U(1)$ and
$U(2)$ theory with 2 and 4 fundamental hypermultiplets respectively.
We will also give the expressions for the case of corresponding $U(N)$
theories using the Weyl symmetry present in the geometry.

\subsection{\bf $U(N)$ with massive adjoint}

\subsubsection{$N=1$}We start by discussing the case of 5-dimensional $U(1)$ theory.
The geometry of the corresponding CY3-fold is encoded in the
$(p,q)$ 5-brane web diagram shown in \figref{u1}(a).
\onefigure{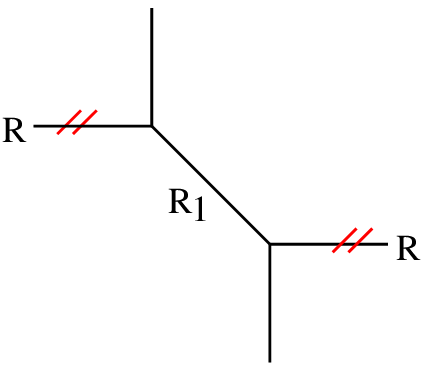}{The web diagram for $U(1)$ theory with an adjoint
field. }
Given the web configuration we can proceed with the partition
function calculation. Using the topological vertex techniques
\cite{AKMV} the partition function in this case is given by \bea
Z(T,T_{m},q):=\sum_{R}e^{-T\,\ell_{R}}\,(-1)^{\ell_{R}}Z_{R}(T_{m},q)\,,
\eea where \bea
Z_{R}(T_{m},q)&=&\sum_{R_{1}}e^{-T_{m}\,\ell_{R_{1}}}\,(-1)^{\ell_{R_{1}}}
C_{\bullet\, R_{1}^{t}\,R}(q)\,C_{\bullet \,R_{1} R^{t}}(q)\,. \eea The two
K\"ahler parameters $T$ and $T_{m}$ are related to the coupling
constant of the gauge theory and the mass of adjoint
hypermultiplet respectively,
\bea
Q_{\tau}:=e^{-T-T_{m}}=e^{2\pi i \tau}\,,\\ \nn
Q_{m}:=e^{-T_{m}}=e^{\beta m}\,.
\eea The partition function
$Z_{R}(T_{m},q)$ can be determined using the expression of the
topological vertex in terms of the Schur  and skew-Schur
polynomials derived in \cite{ORV} and given in the last section.
Let us denote the Young diagram corresponding to $R$ and $R_{1}$
by $\mu_{R}$ and $\mu_{R_{1}}$ respectively. Then $Z_{R}$ is given by, \bea
Z_{R}&=&s_{R}(q^{-\rho})s_{R^{t}}(q^{-\rho})\sum_{R_{1}}\,Q_{m}^{\ell_{R_{1}}}
(-1)^{\ell_{R_{1}}}\,s_{R_{1}^{t}}(q^{-\mu-\rho})\,
s_{R_{1}}(q^{-\mu^{t}-\rho})\,,\\
\nn &=&
s_{R}(q^{-\rho})s_{R^{t}}(q^{-\rho})\sum_{R_{1}}s_{R_{1}^{t}}(-Q_{m}\,q^{-\mu_{R^{t}}-\rho})\,
s_{R_{1}}(q^{-\mu_{R}-\rho})\,,
\eea where in the second equation we used the fact that
$s_{R_{1}}(x)$ is a homogeneous function of degree $l_{R_{1}}$.
Now using the identity \bea
\sum_{R}s_{R^{t}}(x)s_{R}(y)=\prod_{i,j}(1+x_{i}y_{j})\,, \eea we
immediately get \bea Z_{R}&=& s_{R}(q^{-\rho})s_{R^{t}}(q^{-\rho})\prod_{i,j\geq
1}(1-Q_{m} q^{-\mu_{i}-\rho_{i}-\mu^{t}_{j}-\rho_{j}})\,,\\\nn
&=&s_{R}(q^{-\rho})s_{R^{t}}(q^{-\rho})\prod_{i,j\geq 1}(1-Q_{m}
q^{-\mu_{i}+i-\mu^{t}_{j}+j-1})\,,\\\nn
&=&\,s_{R}(q^{-\rho})s_{R^{t}}(q^{-\rho})\,\prod_{k=0}^{\infty}(1-Q_{m}
q^{k+1})^{k+1}\,\,\prod_{(i,j)\in R}(1-Q_{m} q^{h(i,j)})(1-Q_{m}
q^{-h(i,j)})\,, \eea where
\EQ{
h_{R}(i,j)=\mu_{R,i}-j+\mu_{R^{t},j}-i+1
\label{defhook}
}
is
the hook length and we have used the relation \bea \sum_{i,j\geq
1}q^{-h(i,j)}&=&\frac{q}{(1-q)^{2}}+f_{R,R^{t}}(q)=\frac{q}{(q-1)^{2}}+\sum_{(i,j)\in
R} (q^{h(i,j)}+q^{-h(i,j)})\,,\\\nn
f_{R,R^{t}}(q)&=&q^{-1}(q-1)^{2}f_{R}(q)f_{R^{t}}(q)+f_{R}(q)+f_{R^{t}}(q)\,,
\,\,f_{R}(q)=\sum_{(i,j)\in
R}q^{j-i}\,.
 \eea

 Thus the full partition function is given by ($Q=e^{-T}$) \bea
\nn Z&=&\prod_{k\geq 0}^{\infty}(1-Q_{m}
q^{k+1})^{k+1}\,\sum_{R}\,Q^{\ell_{R}}(-1)^{\ell_{R}}\,s_{R}(q^{-\rho})\,s_{R^{t}}(q^{-\rho})\,\prod_{(i,j)\in
R}(1-Q_{m} q^{h(i,j)})(1-Q_{m} q^{-h(i,j)})\,. \eea Using the
definition of $s_{R}(q^{-\rho})=(-1)^{\ell_{R}}C_{R^{t}\,\bullet\,\bullet}$ we get \bea\nn
Z&=&  \prod_{k\geq 0}^{\infty}(1-Q_{m}
q^{k+1})^{k+1}\,\sum_{R}\,Q^{\ell_{R}}(-1)^{\ell_{R}}q^{\sum_{(i,j)\in
R}h(i,j)}\,\prod_{(i,j)\in R}\frac{(1-Q_{m} q^{h(i,j)})(1-Q_{m}
q^{-h(i,j)})}{(1-q^{h(i,j)})^{2}}\,,\\&=&\prod_{k\geq
0}^{\infty}(1-Q_{m}
q^{k+1})^{k+1}\,\sum_{R}\,(QQ_{m})^{l_{R}}\prod_{(i,j)\in
R}\frac{(1-Q_{m} q^{h(i,j)})(1-Q_{m}^{-1}
q^{h(i,j)})}{(1-q^{h(i,j)})^{2}}\,.
\label{ad1}\eea
The first term in the above expression gives the perturbative contribution to the gauge theory prepotential. The
instanton part of the above partition function is given by
\bea
Z_{inst}=\sum_{R}Q^{\ell_{R}}\prod_{(i,j)\in R}\frac{(1-Q_{m}q^{h(i,j)})(1-Q_{m}q^{-h(i,j)})}{(1-q^{h(i,j)})(1-q^{-h(i,j)})}\,.
\label{ins5dadjoint}
\eea
This is the 5-dimensional partition function. To obtain the partition
function of the 4 dimensional gauge theory we take the limit $\beta
\rightarrow 0$
such that $q=e^{-\beta \epsilon}$,
\bea
Z_{inst}=\sum_{R}\,e^{2\pi i \tau \ell_{R}}\prod_{(i,j)\in R}
\frac{(h(i,j)\epsilon+m)(h(i,j)\epsilon-m)}{(\epsilon h(i,j))^{2}}\,.
\eea
which agrees with the results of \cite{Nekrasov:2003rj}.

\underline{\bf From Chern-Simons theory:}
We can calculate the above 5D partition function from the Chern-Simons
theory also. The advantage of this approach is that we get the infinite
product structure of the partition naturally.

Let us briefly review the calculation of the A-model partition function using
geometric transition and the Chern-Simons theory following \cite{AKMV}.
The geometry we are studying admits a geometric transition, {\it
  i.e.\/},
if we take the length
of the internal line in the web diagram , given by $T_{m}$, to zero we can separate the two
lines of the web diagram in the direction transverse to the plane in which the web is embedded.
This is a complex structure deformation of the geometry. Since the lines in the web diagram
are the loci of degeneration of the torus fibered over the plane of the web hence when the
line are separated from each other we get an $\S^3$. Locally this is exactly the conifold transition
the only difference being the globally the geometry is different from that of a conifold. The
closed topological string partition function of the geometry we started from is given by
the topological open string partition function of the geometry obtained after the transition.
In the case of the conifold the open topological string theory partition function is given by
the partition function of the $U(N)$ Chern-Simons theory on $\S^3$. The coupling constant, $k$, of the theory
and rank of the gauge group, $N$, are related to the size of the $\PP^{1}$ and the string coupling as follows
\bea
T_{m}=N\lambda_{s}\,,\,\,\,\lambda_{s}=\frac{2\pi}{N+k}\,.
\eea
But for our geometry there is an extra subtlety because of the compact circle with
boundary on $\S^3$. As discussed in detail in \cite{AKMV,DG} the CS action is modified if there are finite area
holomorphic maps with boundaries on the three cycles. In the case of
 such a holomorphic map the
CS action gets modifies by the operator
\bea
{\cal O}(\tau)&=&\sum_{n=1}^{\infty}\frac{Q_{\tau}^{n}}{n}\,\mbox{Tr}U^{n}\mbox{Tr}V^{n}\,,\\ \nn
&=&\mbox{log}\Big(\sum_{R}Q_{\tau}^{\ell_{R}}\mbox{Tr}_{R}U\,\mbox{Tr}_{R}V\Big)\,.
\eea
where $U,V$ are the holonomies around the two circles of the annulus.
In the case at hand $U=V^{-1}$. Because of the geometry of the annulus
in the target space we also have extra winding numbers.  Thus the operator
that modifies the CS theory action is
given by
\bea
\sum_{k=1}^{\infty}{\cal O}(k\tau)
\eea
 Thus
the partition function is given
by
\bea
Z&=&\int {\cal D}A\,e^{S_{cs}(A)+\sum_{k=1}^{\infty}{\cal O}(k\tau)}=\langle
e^{\sum_{k}{\cal O}(k\tau)}\rangle \,,\\ \nn
&=&\sum_{R_{1,2,\cdots}}Q_{\tau}^{\sum_{k=1}^{\infty}k\ell_{R_{k}}}
\langle \prod_{k=1}^{\infty}\mbox{Tr}_{R_{k}}U\mbox{Tr}_{R_{k}}U^{-1}\rangle\,,\\\nn
&=&\sum_{R_{1,2,\cdots}}Q_{\tau}^{\sum_{k=1}^{\infty}k\ell_{R_{k}}}
\langle \mbox{Tr}_{\otimes_{k}R_{k}}U\mbox{Tr}_{\otimes_{k}R_{k}}U^{-1}\rangle\,,\\\nn
&=&S_{\bullet \bullet}^{-1}(q,q^{N})\sum_{R_{1,2,\cdots}}Q_{\tau}^{\sum_{k=1}^{\infty}k\ell_{R_{k}}}(-1)^{\sum_{k}
\ell_{R_{k}}}\,
W_{\otimes_{k}R_{k}\otimes R_{k},\bullet}(q,q^{N})\,.
\eea
$W_{R,\bullet}(q,q^{N})$ is the quantum dimension of $R$,
\bea
W_{R,\bullet}=\prod_{(i,j)\in R}\frac{q^{(j-i+N)/2}-q^{-(j-i+N)/2}}{q^{h_{R}(i,j)/2}-q^{-h_{R}(i,j)/2}}\,,\,\,\,h_{R}(i,j)
=\mu_{R,i}-i+\mu_{R^{t},j}-j+1\,.
\eea
and $S_{\bullet \bullet}^{-1}$ is the perturbative contribution to the partition function,
\bea
S_{\bullet \bullet}^{-1}=\prod_{k=1}^{\infty}(1-q^{k-N})^{k}\,.
\eea
The partition function can then be written as
\bea
Z:&=&S_{\bullet \bullet}^{-1}\prod_{k=1}^{\infty}\sum_{R}
Q_{\tau}^{k\ell_{R}}(-1)^{\ell_{R}}
W_{R}(q,q^{N})W_{R}(q,q^{N})\,,\\ \nn
&=&\prod_{k=1}^{\infty}\sum_{R}Q_{\tau}^{\ell_{R}}(-1)^{\ell_{R}}W_{R}(q,q^{N})W_{R}(q,q^{N})
\eea
where
\bea
K(Q,q,\lambda)&=&\sum_{R}Q^{\ell_{R}}\,(-1)^{\ell_{R}}W_{R}(q,q^{N})\,W_{R}(q,q^{N})\,\\\nn
&=&\exp\Big(-\sum_{n=1}^{\infty}\frac{Q^{n}}{n}\,W_{\tableau{1}}(q^{n},q^{nN})\,
W_{\tableau{1}}(q^{n},q^{nN})\Big)\,,\\ \nn
&=&\exp\Big(-\sum_{n=1}^{\infty}\frac{Q^{n}}{n}
\Big(\frac{q^{nN/2}-q^{-nN/2}}{q^{n/2}-q^{-n/2}}\Big)^{2}\Big)\,,\\\nn
&=&(1-Q)\,\prod_{r=0}^{\infty}\Big(\frac{(1-q^{r+1+N}Q)(1-q^{r+1-N}Q)}
{(1-q^{r}\,Q)(1-q^{r+2}\,Q)}\Big)^{r+1}\,.
\eea

Since $Q_{m}=e^{-T_{m}}=q^{-N}$ the full partition function is given by
\bea
Z=\Big(\prod_{k=1}^{\infty}(1-Q_{\tau}^{k})\Big)\Big(\prod_{k=1}^{\infty}(1-Q_{m}q^{k})^{k}\Big)
\prod_{k,r=1}^{\infty}\Big(\frac{(1-q^{r}Q_{m}Q_{
\tau}^{k})(1-q^{r}Q_{m}^{-1}Q_{\tau}^{k})}
{(1-q^{r-1}Q_{\tau}^{k})(1-q^{r+1}Q_{\tau}^{k})}\Big)^{r}\,.
\label{pf1}
\eea
The above expression agrees with toplogical vertex computation, Eq(\ref{ad1}),
except for the first term, $\prod_{k=1}^{\infty}(1-Q_{\tau}^{k})$. The reason for this is that in calculating
the partition function we neglected the contribution coming from the annuli which does not end on the three cycle
so that $U,V$ are trivial. The contribution of such annuli is given by
\bea
\sum_{R}Q_{\tau}^{\ell_{R}}=\prod_{k=1}^{\infty}(1-Q_{\tau}^{k})^{-1}\,,
\eea
which cancels the first term in Eq(\ref{pf1}). Thus the Chern-Simons computation agrees with the topological vertex
calculation and moreover it naturally gives the partition function as an infinite product.

\underline{\bf Partition function of the 6-dimensional theory:} Now lets consider the case of geometry giving rise to 6
dimensional gauge theory with massive adjoint. The corresponding web
diagram is shown in \figref{figure1} below.
\onefigure{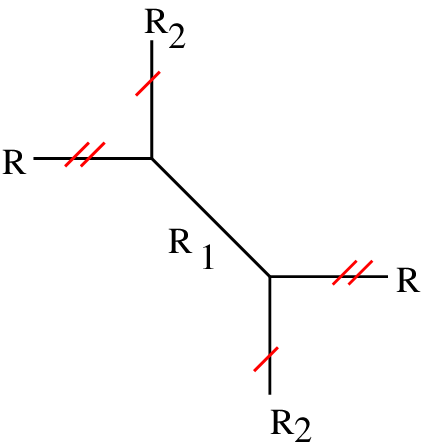}{The web diagram of the 6-dimensional adjoint theory.}
The partition function is given by \bea
Z&:=&\sum_{R,R_{1},R_{2}}Q^{\ell_{R}}Q_{1}^{\ell_{R_{2}}}Q_{m}^{\ell_{R_{1}}}
(-1)^{\ell_{R}+\ell_{R_{1}}+\ell_{R_{2}}}
C_{R\,R_{2}\,R_{1}}\,C_{R^{t}\,R_{2}^{t}\,R_{1}^{t}}\,,\\ \nn
&=&
\sum_{R}Q^{\ell_{R}}(-1)^{\ell_{R}}\,Z_{R}(Q_{1},Q_{m})\ , \eea
where \bea
Z_{R}(Q_{1},Q_{m})=\sum_{R_{1},R_{2}}Q_{1}^{\ell_{R_{2}}}Q_{m}^{\ell_{R_{1}}}
(-1)^{\ell_{R_{1}}+\ell_{R_{2}}}
C_{R\,R_{2}\,R_{1}}\,C_{R^{t}\,R_{2}^{t}\,R_{1}^{t}}
\label{eq1}\eea
Using Eq(\ref{tv1}) $Z_{R}$ is given by
\bea
Z_{R}&=&s_{R}(q^{-\rho})s_{R^{t}}(q^{-\rho})\sum_{R_{1},R_{2}}Q_{1}^{\ell_{R_{1}}}Q_{m}^{\ell_{R_{2}}}\Big(\sum_{R_{3}}s_{R_{2}^{t}/R_{3}}(q^{-\mu_{R}-\rho})
s_{R_{1}/R_{3}}(q^{-\mu_{R^{t}}-\rho})\Big)\\\nn
&&\Big(\sum_{R_{4}}s_{R_{2}/R_{4}}(q^{-\mu_{R^{t}}-\rho})s_{R_{1}^{t}/R_{4}}(q^{-\mu_{R}-\rho})\Big)\,,\\ \nn
&=&s_{R}(q^{-\rho})s_{R^{t}}(q^{-\rho})\sum_{R_{3},R_{4}}s_{R_{2}^{t}/R_{3}}(q^{-\mu_{R}-\rho})s_{R_{2}/R_{4}}(q^{-\mu_{R^{t}}-\rho})
Q_{m}^{\ell_{R_{3}}}(-1)^{\ell_{R_{3}}}\\\nn
&&
\Big(\sum_{R_{1}}s_{R_{1}/R_{3}}(-Q_{m}q^{-\mu_{R^{t}}-\rho})s_{R_{1}^{t}/R_{4}}(q^{-\mu_{R}-\rho})\Big)
\eea

Using the identity (\cite{macdonald} page 93), \bea
\sum_{R}s_{R/R_{3}}(x)s_{R^{t}/R_{4}}(y)=\prod_{i,j}(1+x_{i}y_{j})\,\sum_{\widetilde{R}}s_{R_{4}^{t}/\widetilde{R}}(x)s_{R_{4}/\widetilde{R}}(y)\,.
\eea we get \bea
Z_{R}&=&s_{R}(q^{-\rho})s_{R^{t}}(q^{-\rho})\prod_{i,j}(1-Q_{m}q^{-h_{R}(i,j)})\\\nn
&& \sum_{R_{2},\widetilde{R}}
Q_{1}^{\ell_{R_{2}}}Q_{m}^{\ell_{\widetilde{R}}}(-1)^{\ell_{\widetilde{R}}+\ell_{R_{2}}}
s_{R_{2}^{t}/\widetilde{R}}(q^{-\mu_{R}-\rho},Q_{m}q^{\mu_{R}+\rho})\,s_{R_{2}/\widetilde{R}^{t}}(q^{-\mu_{R^{t}}-\rho},Q_{m}q^{\mu_{R^{t}}+\rho})\,.
\eea Where \bea
s_{R/R_{1}}(x,y)=\sum_{R_{4}}s_{R/R_{4}}(x)s_{R_{4}/R_{1}}(y)\,,
\eea The sum over $R_{2}$ and $\widetilde{R}$ can be determined
exactly using the following identity (\cite{macdonald}, page 93),
\bea\label{identity1}
&&\sum_{A,B}Q_{1}^{\ell_{A}}Q_{2}^{\ell_{B}}(-1)^{\ell_{A}+\ell_{B}}
s_{A/B^{t}}(x,y)s_{A^{t}/B}(z,w)=\\\nn &&\prod_{k=1}^{\infty}
\frac{\prod_{i,j}(1-Q_{1}^{k}Q_{2}^{k-1}x_{i}z_{j})(1-Q_{1}^{k}Q_{2}^{k-1}x_{i}w_{j})(1-Q_{1}^{k}Q_{2}^{k-1}y_{i}z_{j})(1-Q_{1}^{k}Q_{2}^{k-1}y_{i}w_{j})}
{(1-Q_{1}^{k}Q_{2}^{k})}\,. \eea

\bea
Z_{R}&=&\frac{s_{R}(q^{-\rho})s_{R^{t}}(q^{-\rho})}{\prod_{k=1}^{\infty}(1-Q_{1}^{k}Q_{2}^{k})}
\prod_{i,j}(1-Q_{m}q^{-h_{R}(i,j)})\\\nn
&&\prod_{k=1}^{\infty}\prod_{i,j}(1-Q_{1}^{k}Q_{m}^{k-1}q^{-h_{R}(i,j)})(1-Q_{1}^{k}Q_{m}^{k+1}q^{h_{R}(i,j)})(1-Q_{1}^{k}Q_{m}^{k}q^{-\mu_{R,i}-\rho_{i}+\mu_{R^{t},j}+\rho_{j}})
\\\nn
&&(1-Q_{1}^{k}Q_{m}^{k}q^{-\mu_{R^{t},i}-\rho_{i}+\mu_{R,j}+\rho_{j}})\,.
\eea
The above expression can be simplified using
\bea
&&\prod_{i,j}(1-Q_{1}^{k}Q_{m}^{k}q^{-\mu_{R,i}-\rho_{i}+\mu_{R^{t},j}+\rho_{j}})
(1-Q_{1}^{k}Q_{m}^{k}q^{-\mu_{R^{t},i}-\rho_{i}+\mu_{R,j}+\rho_{j}})=\\\nn
&&\prod_{i,j}\frac{1}
{(1-Q_{1}^{k}Q_{m}^{k}q^{-h(i,j)})(1-Q_{1}^{k}Q_{m}^{k}q^{h(i,j)})}\,.
\eea
to obtain
\bea
Z_{R}&=&\frac{s_{R}(q^{-\rho})s_{R^{t}}(q^{-\rho})}{\prod_{k=1}^{\infty}(1-Q_{1}^{k}Q_{2}^{k})}
\prod_{i,j}(1-Q_{m}q^{-h_{R}(i,j)})\\\nn
&&\prod_{k=1}^{\infty}\prod_{i,j}\frac{(1-Q_{1}^{k}Q_{m}^{k-1}\,q^{-h_{R}(i,j)})(1-Q_{1}^{k}Q_{m}^{k+1}q^{h_{R}(i,j)})}
{(1-Q_{1}^{k}Q_{m}^{k}q^{-h_{R}(i,j)})(1-Q_{1}^{k}Q_{m}^{k}q^{h_{R}(i,j)})}\,,\\
&=&Z_{\bullet} (-1)^{\ell_{R}}\prod_{\tableau{1}\in R}\frac{(1-Q_{m}q^{h(\tableau{1})})(1-Q_{m}q^{-h(\tableau{1})})}{(1-q^{h(\tableau{1})})(1-q^{-h(\tableau{1})})}\\\nn
&&
\prod_{k=1}^{\infty}\frac{(1-Q_{\rho}^{k}Q_{m}q^{h(\tableau{1})})
(1-Q_{\rho}^{k}Q_{m}q^{-h(\tableau{1})})(1-Q_{\rho}^{k}Q_{m}^{-1}q^{h(\tableau{1})})
(1-Q_{\rho}^{k}Q_{m}^{-1}q^{-h(\tableau{1})})}{(1-Q_{\rho}^{k}q^{h(\tableau{1})})^{2}(1-Q_{\rho}^{k}q^{-h(\tableau{1})})^{2}}\,,
\eea
Where $Z_{\bullet}$ is the perturbative contribution to the partition function and as discussed
before $Q_{\rho}=Q_{1}Q_{m}$,
\bea
Z_{\bullet}=\prod_{r=0}^{\infty}(1-Q_{m}q^{r+1})^{r+1}\Big(\prod_{k=1}^{\infty}
\frac{(1-Q_{\rho}^{k}Q_{m}^{-1}q^{r+1})(1-Q_{\rho}^{k}Q_{m}q^{-r-1})}{(1-Q_{\rho}^{k}q^{r+1})(1-Q_{\rho}^{k}q^{-r-1})}\Big)^{r+1}\,.
\eea
Thus the instanton part of the partition function (which is zero for $Q=0$)
is given by
\bea
Z_{inst}&=&\sum_{R}(QQ_{m})^{\ell_{R}}\Big(\prod_{\tableau{1}\in
R}\frac{(1-Q_{m}q^{h(\tableau{1})})(1-Q_{m}^{-1}q^{h(\tableau{1})})}
{(1-q^{h(\tableau{1})})^{2}}\Big)\\ \nn
&&\prod_{k=1}^{\infty}\frac{(1-Q_{\rho}^{k}Q_{m}q^{h(\tableau{1})})
(1-Q_{\rho}^{k}Q_{m}q^{-h(\tableau{1})})(1-Q_{\rho}^{k}Q_{m}^{-1}q^{h(\tableau{1})})
(1-Q_{\rho}^{k}Q_{m}^{-1}q^{-h(\tableau{1})})}{(1-Q_{\rho}^{k}q^{h(\tableau{1})})^{2}(1-Q_{\rho}^{k}q^{-h(\tableau{1})})^{2}}\,,
\label{elll}
\eea

Note that for $Q_{m}=1$ i.e., $m=0$ the full partition function is given by
\bea
Z= \prod_{k=0}^{\infty}\frac{(1-q^{k+1})^{k+1}}{(1-Q_{\rho}^{k+1})(1-Q_{\tau}^{k+1})}\,.
\eea
\subsubsection{$N=2$}  In this case the geometry is shown in
\figref{figure3}.
\onefigure{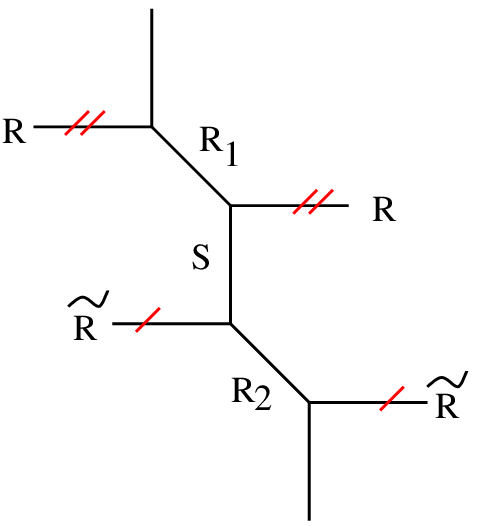}{The web diagram of the 5-dimensional $U(2)$ adjoint theory.}
The partition function is given by
\SP{
Z:&=\sum_{R,\widetilde{R},R_{1,2},S}Q^{\ell_{R}+\ell_{\widetilde{R}}}
Q_{m}^{\ell_{R_{1}}+\ell_{R_{2}}}Q_{f}^{\ell_{S}}(-1)^{\ell_{R}+\ell_{\widetilde{R}}+\ell_{R_{1}}+\ell_{R_{2}}+l_{S}}\,
C_{R^{t}\, \bullet\, R_{1}}\,C_{S^{t}\, R_{1}^{t}\, R}
C_{\widetilde{R}^{t}\, S\, R_{2}} C_{\bullet\, R_{2}^{t}\, \widetilde{R}}\,,\\
&=\sum_{R,\widetilde{R}}Q^{\ell_{R}+\ell_{\widetilde{R}}}(-1)^{\ell_{R}+\ell_{\widetilde{R}}}
K_{R \widetilde{R}}(Q_{m},Q_{f})\, ,
}
where \bea K_{R
\widetilde{R}}(Q_{m},Q_{f})=\sum_{R_{1,2},S}Q_{m}^{\ell_{R_{1}}+\ell_{R_{2}}}Q_{f}^{\ell_{S}}
(-1)^{\ell_{R_{1}}+\ell_{R_{2}}+\ell_{S}} \,C_{R^{t}\, \bullet\,
R_{1}}\,C_{S^{t}\, R_{1}^{t}\, R} C_{\widetilde{R}^{t}\, S\, R_{2}}
C_{\bullet\, R_{2}^{t}\, \widetilde{R}}\,. \label{eq2} \eea

Using the identities involving the skew-Schur functions the above partition function
can be written as a sum over $R,\widetilde{R}$ of a product involving $Q_{m},Q_{f}$. However, we
will use a simpler method which makes use of the fact that the geometry has only a few holomorphic curves
which can contribute.

To determine $K_{R,\widetilde{R}}(Q_{m},Q_{f})$ note that in the
limit $Q_{f}\rightarrow 0$ it is clear from the geometry
(\figref{figure3}) that we get two copies of the partition
function of the $U(1)$ theory therefore we can write
$K_{R,\widetilde{R}}$ as  \bea\nn K_{R
\widetilde{R}}(Q_{m},Q_{f}))&=&\Big(\prod_{k=0}^{\infty}
(1-Q_{m}q^{k+1})^{2(k+1)}\Big)(-Q_{m})^{\ell_{R}+\ell_{\widetilde{R}}}\prod_{(i,j)\in
R,\widetilde{R}}\frac{(1-Q_{m}q^{h(i,j)})(1-Q_{m}^{-1}q^{h(i,j)})}{(1-q^{h(i,j)})^{2}}\,\\
\nn &&\exp\Big(\sum_{n=1}^{\infty}
\frac{Q_{f}^{n}}{n}f_{1}(q^{n})+
\frac{(Q_{f}Q_{m})^{n}}{n}f_{2}(q^{n})+\frac{(Q_{f}Q_{m}^{2})^{n}}{n}f_{3}(q^{n})\Big)\,.
\eea Here, the three terms in the exponential correspond to the
contribution of four holomorphic curves in the geometry. There are three
terms rather than four since we are taking the area of the two
exceptional curves to be equal to $T_{m}$. The contribution of
the two curves which are locally like the conifold is given by the
prefactor in the above equation. The
coefficients $f_{i}(q)$ can be determined from Eq.~\eqref{eq2} by
expanding it to linear order in $Q_{f}$, \bea
f_{1}(q)&=&-\frac{C_{\tableau{1}\,\bullet\, R}}{C_{\bullet\,\bullet\,
R}}\,\frac{C_{\widetilde{R}^{t} \tableau{1}\,
\bullet}}{C_{\widetilde{R}^{t}\, \bullet\, \bullet}}=-\frac{W_{R
\tableau{1}}}{W_{R}}\frac{W_{\widetilde{R}^{t}
\tableau{1}}}{W_{\widetilde{R}^{t}}}=-\frac{q}{(1-q)^{2}}-f_{R,\widetilde{R}^{t}}\,,\\
\nn f_{2}(q)&=& -2f_{1}(q)\,,\\ \nn f_{3}(q)&=&f_{1}(q)\,, \eea
where we used the identity \bea C_{\bullet\, R_{1}\, R_{2}}=W_{R_{1}
R_{2}^{t}}q^{\kappa_{R_{2}}/2}\,.\eea
\bea
K_{R \widetilde{R}}(Q_{m},Q_{f})&=&K_{\bullet \bullet}\,(-Q_{m})^{\ell_{R}+\ell_{\widetilde{R}}}
\prod_{(i,j)\in R,\widetilde{R}}\frac{(1-Q_{m}q^{h(i,j)})(1-Q_{m}^{-1}q^{h(i,j)})}{(1-q^{h(i,j)})^{2}}
\\\nn
&&\prod_{k}\Big(\frac{(1-Q_{f}q^{k})(1-Q_{f}Q_{m}^{2}q^{k})
}{(1-Q_{f}Q_{m}q^{k})^{2}}\Big)^{C_{k}(R,\widetilde{R}^{t})},
\eea
where $\sum_{k}C_{k}(R_{1},R_{2})q^{k}=f_{R_{1}R_{2}}(q)$. $K_{\bullet \bullet}$ contributes to
the perturbative part of the partition function,
\bea
K_{\bullet \bullet}=\prod_{k=0}^{\infty}\Big( (1-Q_{m}q^{k+1})^{2}\,
(1-Q_{f}q^{k+1})(1-Q_{f}Q_{m}^{2})
(1-Q_{f}Q_{m}q^{k+1})^{-2}\Big)^{k+1}\,.
\eea
Define $Q_{F}=Q_{f}Q_{m}=e^{-2a\beta}$, where $a$ is the Coulomb branch moduli,
then the full partition function is given by
\bea
Z&=&K_{\bullet \bullet}\sum_{R, \widetilde{R}}(QQ_{m})^{\ell_{R}+\ell_{\widetilde{R}}}
\prod_{(i,j)\in R,\widetilde{R}}\frac{(1-Q_{m}q^{h(i,j)})(1-Q_{m}^{-1}q^{h(i,j)})}{(1-q^{h(i,j)})^{2}}
\\\nn
&&\prod_{k}\Big(\frac{(1-Q_{F}Q_{m}^{-1}q^{k})(1-Q_{F}Q_{m}q^{k})
}{(1-Q_{F}q^{k})^{2}}\Big)^{C_{k}(R,\widetilde{R}^{t})}.
\eea
Using the following identity \cite{AI1}
\bea\nn
\prod_{k}(1-\,x\,q^{k})^{-C_{k}(R_{1},R_{2}^{t})}=(4x)^{-\frac{(\ell_{R_{1}}+\ell_{R_{2}})}{2}}\,q^{-\frac{\kappa_{R_{1}}-\kappa_{R_{2}}}{4}}
\prod_{i,j=1}^{\infty}\frac{\mbox{Sinh}\frac{\beta}{2}(2a+\hbar(\mu_{1,i}-\mu_{2,j}+j-i))}
{\mbox{Sinh}\frac{\beta}{2}(2a+\hbar(j-i)}
\eea
where $x=e^{-2\beta \,a}$ and $q=e^{-\beta\,\hbar}$ it is easy to show that the above partition function agrees with the results of \cite{Nekrasov:2003rj}.

Generalization to the case of $U(N)$ is simple using the corresponding web diagram discussed
in section 3 and the $A_{N-1}$ Weyl symmetry present in the geometry,
\bea
Z_{inst}&=&\sum_{R_{1},\cdots R_{N}}(QQ_{m})^{\ell_{1}+\cdots +\ell_{N}}\prod_{(i,j)\in R_{1,2,\cdots,N}}
\frac{(1-Q_{m}q^{h(i,j)})(1-Q_{m}^{-1}q^{h(i,j)})}{(1-q^{h(i,j)})^{2}}\\ \nn
&&\prod_{1\leq i<j\leq N}\prod_{k}\Big(\frac{(1-Q_{F_{ij}}Q_{m}^{-1}q^{k})(1-Q_{F_{ij}}Q_{m}q^{k})
}{(1-Q_{F_{ij}}q^{k})^{2}}\Big)^{C_{k}(R_{i},R_{j}^{t})}.
\eea
Where $Q_{F_{ij}}=e^{-\beta(a_{i}-a_{j})}$.

\underline{\bf Partition function of the 6-dimensional theory:} The geometry giving rise
to 6 dimensional theory is shown in \figref{figure4}.
\onefigure{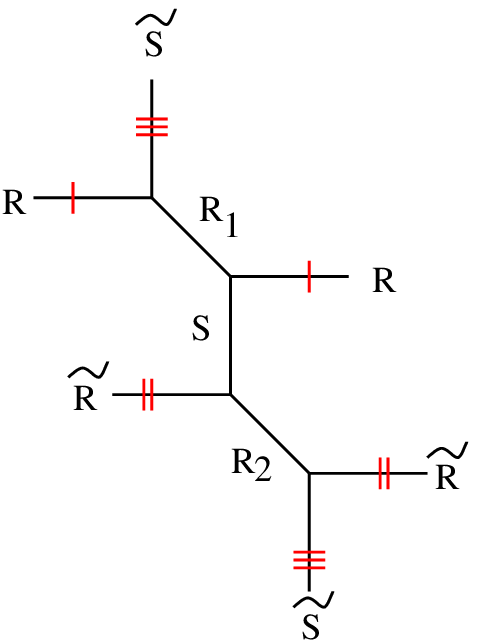}{The web diagram of the 6-dimensional $U(2)$ adjoint theory.}
The partition function is given by
\SP{
Z:&=\sum_{R,\widetilde{R},\widetilde{S},R_{1,2},S}Q^{\ell_{R}+\ell_{\widetilde{R}}}
Q_{1}^{\ell_{\widetilde{S}}}Q_{m}^{\ell_{R_{1}}+\ell_{R_{2}}}
Q_{f}^{\ell_{S}}(-1)^{\ell_{R}+\ell_{\widetilde{R}}+\ell_{R_{1}}+\ell_{R_{2}}+\ell_{S}+\ell_{\widetilde{S}}}\,
C_{R^{t}\, \widetilde{S}\, R_{1}}\,C_{S^{t}\, R_{1}^{t}\, R}
C_{\widetilde{R}^{t}\, S \,R_{2}} C_{\widetilde{S}^{t}\, R_{2}^{t}\,
  \widetilde{R}}
\,,\\
&=\sum_{R,\widetilde{R}}Q^{\ell_{R}+\ell_{\widetilde{R}}}(-1)^{\ell_{R}+\ell_{\widetilde{R}}}
{\cal K}_{R \widetilde{R}}(Q_{m},Q_{f},Q_{1})\, ,} where \SP{ &{\cal
K}_{R \widetilde{R}}(Q_{m},Q_{f},Q_{1})\\
&=\sum_{R_{1,2},S,\widetilde{S}}
Q_{m}^{\ell_{R_{1}}+\ell_{R_{2}}}Q_{f}^{\ell_{S}}Q_{1}^{\ell_{\widetilde{S}}}
(-1)^{\ell_{R_{1}}+\ell_{R_{2}}+\ell_{S}+\ell_{\widetilde{S}}}
\,C_{R^{t} \widetilde{S} R_{1}}\,C_{S^{t}\, R_{1}^{t}\, R}
C_{\widetilde{R}^{t}\, S\, R_{2}} C_{\widetilde{S}^{t}\, R_{2}^{t}\,
\widetilde{R}}\,. \label{eq3}} We can write ${\cal
K}_{R\widetilde{R}}$ as \bea \frac{{\cal K}_{R
\widetilde{R}}(Q_{m},Q_{f},Q_{1})}{{\cal K}_{R
\widetilde{R}}(Q_{m},Q_{f},Q_{1}=0)}=
\exp\Big(\sum_{n,k=1}^{\infty}\frac{(Q_{1}Q_{m}^{2}Q_{f})^{nk}}{n}f_{n,k}(Q_{m},Q_{f},q)\Big)\,.
\label{6de1} \eea The coefficients $f_{n,k}$ can be determined
easily by comparing the above expression with Eq.~\eqref{eq3}. It
turns out that $f_{n,k}$ is independent of $k$ and has the form \bea
f_{n,k}(Q_{m},Q_{f},q)=f(Q_{m}^{n},Q_{f}^{n},q^{n})\,. \eea The
function $f(Q_{m},Q_{f},q)$ is given by \bea\label{6de3}
&&f(Q_{m},Q_{f},q)=1-Q_{f}f_{R\,\widetilde{R}^{t}}-Q_{f}^{-1}f_{R^{t}\,\widetilde{R}}+2Q_{f}Q_{m}f_{R
\widetilde{R}^{t}}\\\nn && +2(Q{f}Q_{m})^{-1}f_{R^{t}
\widetilde{R}}-Q_{f}Q_{m}^{2}f_{R
\widetilde{R}^{t}}-(Q_{f}Q_{m}^{2})^{-1}f_{R^{t} \widetilde{R}}-
(2-Q_{m}-Q_{m}^{-1})(f_{R,R^{t}}(q)+
f_{\widetilde{R},\widetilde{R}^{t}}(q))\\\nn
&&-\frac{q}{(1-q)^{2}}\Big(Q_{f}+Q_{f}^{-1}+2Q_{m}+\frac{2}{Q_{m}}
-2Q_{f}Q_{m}-\frac{2}{Q_{f}Q_{m}}+Q_{f}Q_{m}^{2}+\frac{1}{Q_{f}Q_{m}^{2}}-4\Big)\,.
\eea It is easy to see from Eq(\ref{6de1}) and Eq(\ref{6de3}) that
the partition function of the 6-dimensionaltheory is given by the
following substitution in the corresponding 5-dimensional partition
function, \bea (1-z\,q^{k})\mapsto
(1-z\,q^{k})\prod_{r=1}^{\infty}\frac{(1-Q_{\rho}zq^{k})(1-Q_{\rho}z^{-1}q^{-k})}{(1-Q_{\rho}^{k})^{k}}\,.
\eea

\subsection{$U(N)$ with $N_{f}=2N$}

\subsubsection{ $N=1$}
We start by discussing the case of $U(1)$ theory with two
hypermultiplets in the fundamental representation. The CY geometry
which gives rise to this theory via geometric engineering
\cite{KKV} is well known  and is blowup of $T^{*}\PP^{1}\times
\CC$ at two points. The toric geometry of this CY space is
encoded in the toric web shown below (for more details about toric
web see \cite{Leung:1998tw,AMV}).
\onefigure{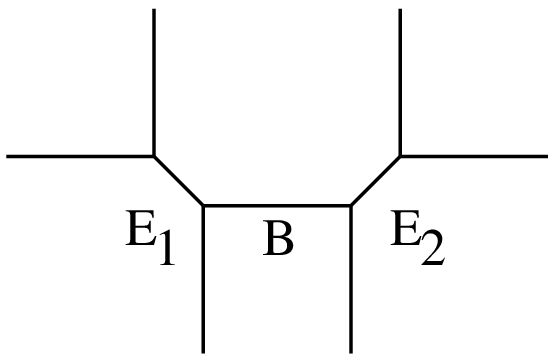}{The web diagram of the 5-dimensional $U(1)$ theory with two hypermultiplets.}
Since this geometry is so simple it is possible to write down the partition function
almost without any calculation using expression for the free energy in terms of
integer invariants \cite{Gopakumar:1998ii,Gopakumar:1998jq}. The only
holomorphic curves in the geometry are
$B,E_{1},E_{2},B-E_{1},B-E_{2},B-E_{1}-E_{2}$ with integer invariants
\cite{AMV}
\bea
N^{g}_{B}&=&N^{g}_{B-E_{1}-E_{2}}=-\delta_{g,0}\,,\\ \nn
N^{g}_{E_{i}}&=&N^{g}_{B-E_{i}}=\delta_{g,0}\,.
\eea
It is easy to derive the above expression from the definition of the integer invariants.
To see this note that the curves $E_{i}$ and $B-E_{i}$ are rigid and therefore the
corresponding moduli space is just a point. On the other hand the moduli space
of $B$ and $B-E_{1}-E_{2}$ is $\CC$. Hence, since the genus zero integer invariant
of a curve $C$ with moduli space ${\cal M}$ is $(-1)^{\mbox{dim}{\cal M}}\chi({\cal M})$
therefore $N^{0}_{B}=N^{0}_{B-E_{1}-E_{2}}=-1$ and $N^{0}_{E_{i}}=N^{0}_{B-E_{i}}=1$.
Thus the instanton part of the free energy is given by
\bea
F=\sum_{n=1}^{\infty}\frac{Q_{b}^{n}-\lambda_{1}^{-n}-\lambda_{2}^{-n}-Q_{b}^{n}\lambda_{1}^{-n}-
Q_{b}^{n}\lambda_{2}^{-n}+Q_{b}^{n}\lambda_{1}^{-n}\lambda_{2}^{-n}}{n(q^{n/2}-q^{-n/2})^{2}}\,,
\eea
where $Q_{b}=e^{-T_{b}},\lambda_{i}=e^{T_{E_{i}}}$ and $T_{b},T_{E_{i}}$ are the area of curves $B$ and
$E_{i}$ respectively.

The partition function $Z=e^{F}$ can be written easily using the
multicovering structure of the free energy and is given by \bea
Z&=&
Z_{pert}\prod_{k=0}^{\infty}\frac{(1-\lambda_{1}^{-1}Q_{b}q^{k+1})^{k+1}
(1-\lambda_{2}^{-1}Q_{b}q^{k+1})^{k+1}} {(1-Q_{b}q^{k+1})^{k+1}
(1-Q_{b}\lambda_{1}^{-1}\lambda_{2}^{-1}q^{k+1})^{k+1}}\,,
\label{pfu1direct} \eea where \bea
Z_{pert}=\prod_{k=0}^{\infty}(1-\lambda_{1}^{-1}q^{k+1})^{k+1}
(1-\lambda_{2}^{-1}q^{k+1})^{k+1}\,,
\eea and gives the perturbative contribution to the prepotential
in the 4D field theory limit because in this limit $\beta\rightarrow
0$ such that \bea Q_{b}=(\frac{\beta
\Lambda}{2})^{2}\,,\,\,\lambda_{i}=e^{\beta
m_{i}}\,,\,\,\,q^{-\beta \hbar}\,. \eea The partition function of
the pure $5D$ $U(1)$ theory is recovered in the limit
$\lambda_{i}\rightarrow \infty$.

For the $U(1)$ theory we are discussing we saw that the
corresponding geometry is simple enough to allow us to write down
the partition function directly. However, for $U(N)$ with $N>1$
the corresponding geometries are such that the partition functions
can not be derived so easily. For this reason we now derive the
partition function using the open-closed duality via geometric
transition \cite{AMV} since this method can be extended to the case of
geometries giving rise to $U(N)$ theories with $N_{f}=2N$. The
geometry shown in \figref{u1hyper2} has two exceptional curves
$E_{1,2}$ with normal bundle ${\cal O}(-1)\oplus {\cal O}(-1)$ and
therefore the geometry in the neighborhood of these curves in that
of resolved conifold. Thus these curves can be shrunk and deformed
into two three cycles which are topologically $\S^3$. The A-model
partition function is then given by the partition function of
$U(N_{1})\times U(N_{2})$ Chern-Simons theory on the two three
cycles modified by the holomorphic maps between the two three
cycles as shown in \figref{u1hyper2} below.
\onefigure{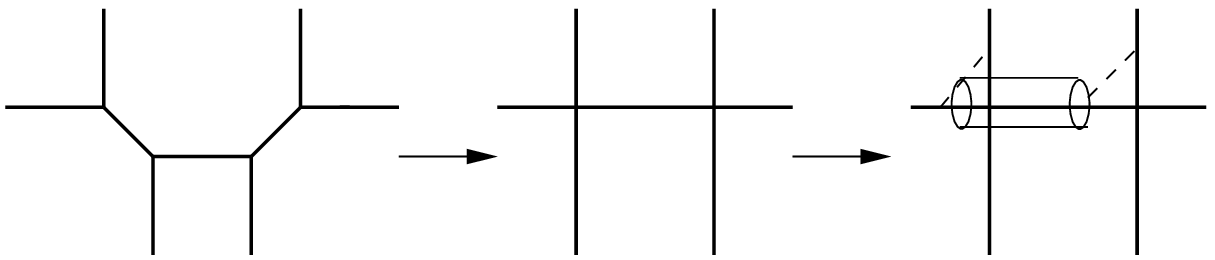}{Transition from closed string geometry to open
string geometry by a geometric transition.}
The partition function in this case is given by
\bea
Z=\int e^{S_{cs}(A_{1})+S_{cs}(A_{2})+{\cal O}(r)}\,.
\eea
Here, $A_{1,2}$ are the $U(N_{1,2})$ gauge fields on the two three cycles respectively and
${\cal O}(r)$ is the contribution from the annuli shown in \figref{u1hyper2} of length $r$,
\bea
e^{{\cal O}(r)}=\sum_{R}e^{-r\,\ell_{R}}\,\mbox{Tr}_{R}U_{1}\,\mbox{Tr}_{R}U_{2}\,,
\eea
where $U_{1,2}$ is the holonomy of $A_{1,2}$ around the two boundary components of the annuli.
Thus the partition function is given by
\bea
Z&=&\sum_{R}e^{-r\,\ell_{R}}\langle \mbox{Tr}_{R}U_{1}\rangle \langle \mbox{Tr}_{R}U_{2}\rangle\,,\\ \nn
&=& (S^{-1})_{00}(q,\lambda_{1})\,(S^{-1})_{00}(q,\lambda_{2})\,\sum_{R}e^{-r\,\ell_{R}}\,
W_{R}(q,\lambda_{1})\,W_{R}(q,\lambda_{2})\,,
\eea
where $\langle \mbox{Tr}_{R}U_{1,2}\rangle =(S^{-1})_{0R}(q,\lambda_{1,2})=
(S^{-1})_{00}\,W_{R}(q,\lambda_{1,2})$
and
\bea
W_{R}(q,\lambda_{i})=\prod_{(i,j)\in R}\frac{[j-i]_{\lambda_{i}}}{[h(i,j)]}\,,\,\,
[x]_{\lambda}=q^{x/2}\lambda^{1/2}-q^{-x/2}\lambda^{-1/2}\,,
\eea
is the quantum dimension of the representation $R$ with $\lambda_{i}=q^{N_{i}}$. Thus
the partition function is
given by
\bea
Z&=&Z_{pert}\sum_{R}Q^{\ell_{R}}W_{R}(q,\lambda_{1})\,W_{R}(q,\lambda_{2})\,,\\ \nn
&=&Z_{pert}\,Z_{inst}\eea
where $Q=e^{-r}$ and $Z_{pert}=(S^{-1})_{00}(q,\lambda_{1})(S^{-1})_{00}(q,\lambda_{2})$
gives the perturbative contribution to the prepotential in the field theory
limit such that $\lambda_{i}=e^{\beta m_{i}}$.
In terms of the 4D gauge theory $\beta \rightarrow 0$ and
\bea Q\sqrt{\lambda_{1}\lambda_{2}}&=&(\frac{\beta
\Lambda}{2})^{2}\,,\\ \nn q&=&e^{-\beta \hbar}\,,\\ \nn
\lambda_{i}&=&e^{\beta m_{i}}\,. \eea The sum giving the partition
function can be evaluated to get a product formula
\bea
Z_{pert}=\prod_{k=0}^{\infty}(1-\lambda^{-1}_{1}q^{k+1})^{k+1}
(1-\lambda^{-1}_{2}q^{k+1})^{k+1}\,,
\eea
and
\bea
Z_{inst}&=&\sum_{R}Q^{\ell_{R}}W_{R}(q,\lambda_{1})\,W_{R}(q,\lambda_{2})\,,\\ \nn
&=&\exp\Big(\sum_{n\geq 1}\frac{Q^{n}}{n}\,W_{\tableau{1}}(q^n,\lambda_{1}^n)
W_{\tableau{1}}(q^n,\lambda_{2}^n)\Big)\,,\\\nn
&=&\prod_{k=0}^{\infty}\frac{(1-Q\sqrt{\frac{\lambda_{2}}{\lambda_{1}}}q^{k+1})^{k+1}
(1-Q\sqrt{\frac{\lambda_{1}}{\lambda_{2}}}q^{k+1})^{k+1}}
{(1-Q\sqrt{\lambda_{2}\lambda_{1}}q^{k+1})^{k+1}
(1-\frac{Q}{\sqrt{\lambda_{2}\lambda_{1}}}q^{k+1})^{k+1}}\,,\\ \nn
&=& \prod_{k=0}^{\infty}\frac{(1-\lambda_{1}^{-1}Q_{b}q^{k+1})^{k+1}
(1-\lambda_{2}^{-1}Q_{b}q^{k+1})^{k+1}}
{(1-Q_{b}q^{k+1})^{k+1}
(1-Q_{b}\lambda_{1}^{-1}\lambda_{2}^{-1}q^{k+1})^{k+1}}\,,\eea
where $Q_{b}=Q\sqrt{\lambda_{1}\lambda_{2}}$ and
the partition function of the pure 5D $U(1)$ theory
is recovered in the limit \bea
\lambda_{i}\rightarrow\infty\,,\,\,\,\,Q_{b}=\text{fixed}\,.
\eea
It is easy to see that the above partition function agrees with the one
given by Nekrasov and with the one given in Eq.~\eqref{pfu1direct}.

\underline{\bf Partition function of the 6-dimensional theory:} To discuss the six dimensional case and
the corresponding geometries we will have to make
toric web diagrams periodic around one extra direction, as discussed
in section 3.
 Consider the case of ${\cal O}(-1)\oplus {\cal O}(-1)
\rightarrow \PP^{1}$ blown up at two points, {\it i.e.\/}~the geometry we
considered in the previous section. In this case we can glue the
external lines to obtain the geometry shown in \figref{hyperu16d}
below.
\onefigure{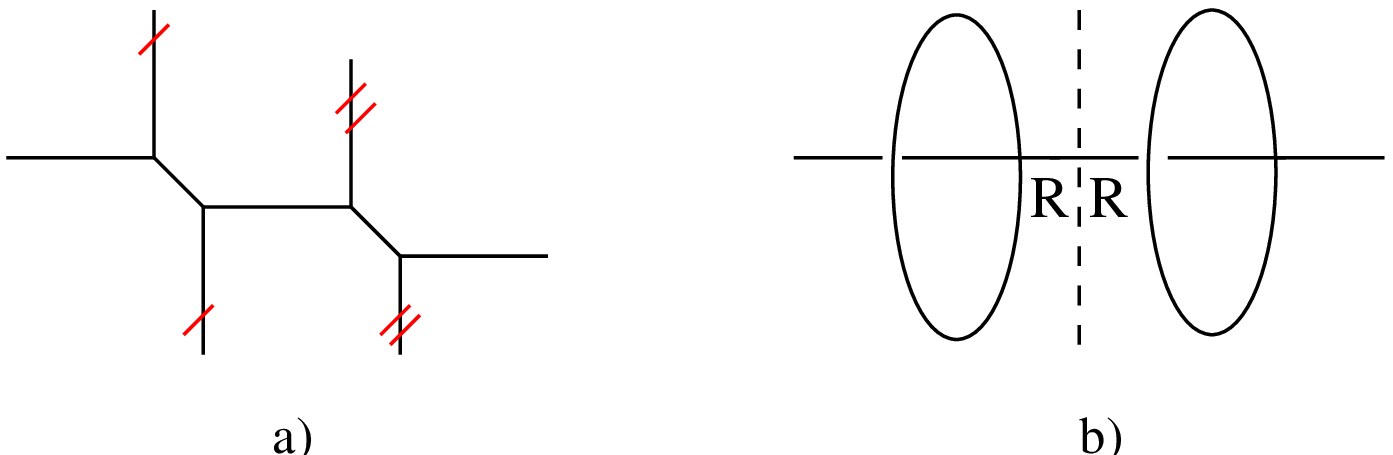}{The web diagram of the 6-dimensional $U(1)$ theory with
two fundamental hypermultiplets.}

To obtain the partition function we divide the geometry in two parts as shown in
\figref{hyperu16d}(b). Then the partition function is given by
\bea
Z=\sum_{R}Q_{b}^{\ell_{R}}(-1)^{\ell_{R}}{\cal K}_{R}(Q_{m_{1}},Q_{1})\,
{\cal K}_{R^{t}}(Q_{m_{2}}, Q_{1})
\,,
\eea
where
\bea
{\cal K}_{R}(Q_{m},Q_{1})&=&\sum_{R_{1},S}Q_{1}^{\ell_{S}}Q_{m}^{\ell_{R_{1}}}(-1)^{\ell_{S}+
\ell_{R_{1}}}C_{\bullet\, S^{t}\, R_{1}^{t}}\,C_{R_{1}\, R^{t}\, S}\,,\\ \nn
&=&(-1)^{\ell_{R}}s_{R}(q^{-\rho})\sum_{S,R_{1}}Q_{1}^{\ell_{S}}Q_{m}^{\ell_{R_{1}}}(-1)^{\ell_{R_{1}}+\ell_{S}}
\\\nn
&&\sum_{R_{3},R_{4}}s_{S/R_{3}}(q^{-\rho})s_{S^{t}/R_{4}}(q^{-\mu_{R^{t}}-\rho})s_{R_{1}^{t}/R_{3}}(q^{-\rho})s_{R_{1}/R_{4}}(q^{-\mu_{R}-\rho})\,.
\label{eq5}
\eea
Summing over $R_{1}$ we get
\bea
{\cal K}_{R}&=&(-1)^{\ell_{R}}s_{R}(q^{-\rho})\prod_{i,j}(1-Q_{m}q^{-\mu_{R,i}-\rho_{i}-\rho_{j}})
\\\nn
&&\sum_{S,R_{3},R_{4},R_{5}}Q_{1}^{\ell_{S}}Q_{m}^{\ell_{R_{4}}}(-1)^{\ell_{R_{4}}+\ell_{S}}s_{S/R_{3}}(q^{-\rho})s_{S^{t}/R_{4}}(q^{-\mu_{R^{t}}-\rho})
s_{R_{3}^{t}/R_{5}}(-Q_{m}q^{-\mu_{R}-\rho})s_{R_{4}^{t}/R_{5}^{t}}(q^{-\rho})\,.
\eea Simplifying the above expression by summing over $R_{3,4}$
gives \bea {\cal
K}_{R}&=&(-1)^{\ell_{R}}s_{R}(q^{-\rho})\prod_{i,j}(1-Q_{m}q^{-\mu_{R,i}-\rho_{i}-\rho_{j}})\,\\\nn
&&\sum_{S,R_{5}}Q_{1}^{\ell_{S}}Q_{m}^{\ell_{R_{5}}}(-1)^{\ell_{S}+\ell_{R_{5}}}
s_{S/R_{5}^{t}}(q^{-\rho},Q_{m}q^{\mu_{R}+\rho})
s_{S^{t}/R_{5}}(q^{-\mu_{R^{t}}-\rho},Q_{m}q^{-\rho})\,. \eea Using
the identity Eq(\ref{identity1}) we get \bea &&{\cal
K}_{R}=(-1)^{\ell_{R}}s_{R}(q^{-\rho})\prod_{i,j}(1-Q_{m}q^{-\mu_{R,i}-\rho_{i}-\rho_{j}})\\\nn
&&
\prod_{k=1}^{\infty}\frac{\prod_{i,j}(1-Q_{\rho}^{k}Q_{m}^{-1}q^{-\mu_{R^{t},i}-\rho_{i}-\rho_{j}})(1-Q_{\rho}^{k}q^{-\rho_{i}+\rho_{j}})
(1-Q_{\rho}^{k}q^{\mu_{R,i}+\rho_{i}-\mu_{R^{t},j}-\rho_{j}})(1-Q_{\rho}^{k}Q_{m}q^{\mu_{R,i}+\rho_{i}+\rho_{j}})}
{(1-Q_{\rho}^{k})}\,. \eea The above expression can be further
simplified to\footnote{This simplification requires the identity:
$\sum_{i,j}q^{\mu_{R,i}+\rho_{i}+\rho_{j}}=\frac{q}{(1-q)^{2}}+\sum_{(i,j)\in
R}q^{j-i}\,.$}  \bea\nn {\cal K}_{R}= {\cal
K}_{\bullet}(-1)^{\ell_{R}}s_{R}(q^{-\rho})\prod_{(i,j)\in
R}(1-Q_{m}q^{i-j})
\Big(\prod_{k=1}^{\infty}\frac{(1-Q_{\rho}^{k}Q_{m}^{-1}q^{j-i})(1-Q_{\rho}^{k}Q_{m}q^{i-j})}
{(1-Q_{\rho}^{k}q^{h_{R}(i,j)})(1-Q_{\rho}^{k}q^{-h_{R}(i,j)})}\Big)
\,.\eea where  ${\cal K}_{\bullet}$ is the perturbative contribution
\bea {\cal
K}_{\bullet}(Q_{m})=\prod_{r=0}^{\infty}(1-Q_{m}q^{r+1})^{r+1}\Big(\prod_{k=1}^{\infty}
\frac{(1-Q_{\rho}^{k}Q_{m}^{-1}q^{r+1})^{r+1}(1-Q_{\rho}^{k}Q_{m}q^{-r-1})^{r+1}}{(1-Q_{\rho}^{k})(1-Q_{\rho}^{k}q^{r+1})^{2r+2}}\Big)\,\,
\eea Thus the full partition function is given by \bea Z&=&{\cal
K}_{\bullet}(Q_{m_{1}}){\cal
K}_{\bullet}(Q_{m_{2}})\sum_{R}Q_{b}^{\ell_{R}}\prod_{(i,j)\in
R}\frac{(1-Q_{m_{1}}q^{j-i})(1-Q_{m_{2}}q^{i-j})}{(1-q^{h_{R}(i,j)})(1-q^{-h_{R}(i,j)})}\\\nn
&&\prod_{k=1}^{\infty}\frac{(1-Q_{\rho}^{k}Q_{m_{1}}^{-1}q^{j-i})(1-Q_{\rho}^{k}Q_{m_{1}}q^{i-j})(1-Q_{\rho}^{k}Q_{m_{2}}^{-1}q^{i-j})(1-Q_{\rho}^{k}Q_{m_{2}}q^{j-i})}{(1-Q_{\rho}^{k}q^{h_{R}(i,j)})^{2}
(1-Q_{\rho}^{k}q^{-h_{R}(i,j)})^{2}}\,. \eea The 5D limit is given
by $Q\to 0$ and the 4D limit is given by $\beta\to 0$.

\subsubsection{$N=2$}
The Calabi-Yau
geometry giving rise to this theory is shown in \figref{hypersu2} below.
\onefigure{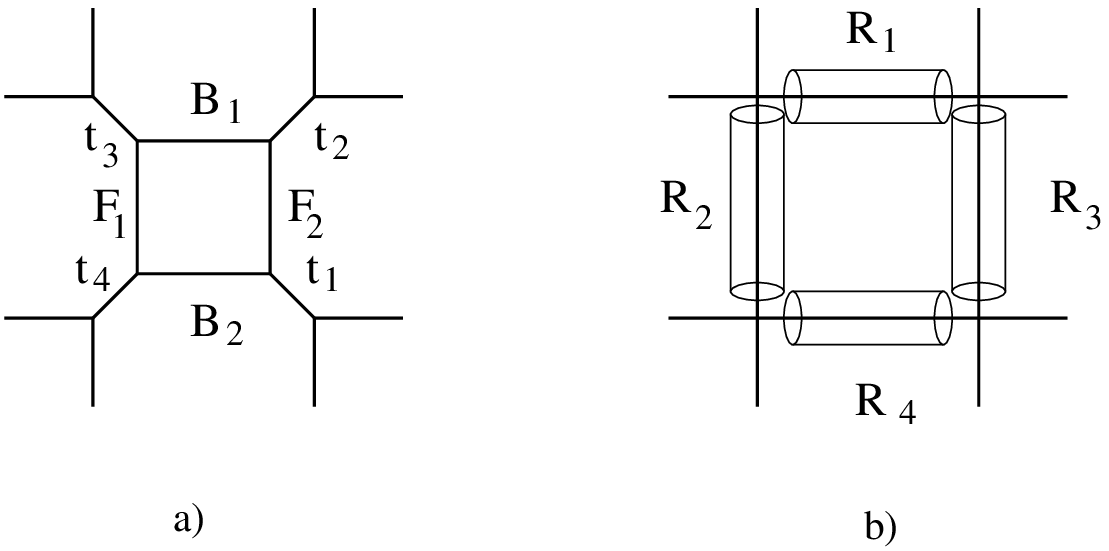}{a) The web diagram of the 5-dimensional $U(2)$ theory
with four fundamental hypermultiplets, b) the corresponding
open string geometry obtained by geometric transition.}
The partition function for this case was calculated in  \cite{AMV} and is given by
\noindent \bea \nn Z
&=&
\sum_{R_{1,2,3,4}}Q_{B_{1}}^{\ell_{1}}Q_{B_{2}}^{\ell_{3}}Q_{F_{1}}^{\ell_{2}}Q_{F_{2}}^{\ell_{4}}
W_{R_{1}R_{4}}(\lambda_{4},q)W_{R_{4}R_{3}}(\lambda_{3},q)W_{R_{3}R_{2}}
(\lambda_{2},q)W_{R_{2}R_{1}}(\lambda_{1},q)\,,
\eea
where
\bea
Q_{B_{1,2}}=e^{-T_{B_{1,2}}}\,,\,\,\,Q_{F_{1,2}}=e^{-T_{F_{1,2}}}\,,\,\,\,
\lambda_{1,2,3,4}=e^{t_{1,2,3,4}}\,.
\eea
$T_{B_{i}}$ and $T_{F_{i}}$ are the lengths of the annuli shown in \figref{hypersu2}(b)
and $t_{a}$ are the area of the four exceptional curves and
\bea
W_{R_{1}R_{2}}=\sum_{R}N^{R}_{R_{1}R_{2}}q^{\frac{1}{2}(\kappa_{R}-\kappa_{R_{1}}-\kappa_{R_{2}})}W_{R}\,.
\eea The partition function can be written
as $Z=Z_{pert}Z_{inst}$ where $Z_{pert}$ is the perturbative contribution to the
partition function and $Z_{inst}$ is the instanton contribution. From the discussion of
section 3 it follows that the instanton contribution arises from the terms involving
$Q_{B_{1,2}}$. In the following we will focus our attention on the instanton contribution
only.

To determine the partition function note that it can be written as
\bea
Z=\sum_{R_{1},R_{3}}Q_{B_{1}}^{\ell_{R_{1}}}Q_{B_{2}}^{\ell_{R_{3}}}G_{R_{1}R_{3}}(Q_{F_{1}},\lambda_{1},\lambda_{2})G_{R_{1}R_{2}}(Q_{F_{2}},\lambda_{4},\lambda_{3})\,,
\eea where \bea
G_{R_{1}R_{2}}(Q,\lambda_{1},\lambda_{2})&=&\sum_{R}W_{R_{1}R}(q,\lambda_{1})W_{RR_{2}}
(q,\lambda_{2})Q^{\ell_{R}}\,. \label{eqx} \eea As discussed in
\cite{AMV} before the Hopf link invariants
$W_{R_{1}R_{2}}(\lambda,q)$ are given by Schur functions,
$s_{R}(x)$. Using the identity \bea
\sum_{R}s_{R}(x)s_{R}(y)&=&\prod_{i,j}(1-Qx_{i}y_{j})^{-1}\\\nn
&=&\mbox{Exp}\Big(\sum_{n=1}^{\infty}\frac{Q^{n}}{n}f(x^n,y^n)\Big)\,,\,\,f(x,y)=\sum_{i,j}x_{i}y_{j}\,.
\eea we get \bea
G_{R_{1}R_{2}}(Q,\lambda_{1},\lambda_{2})&=&W_{R_{1}}(q,\lambda_{1})W_{R_{2}}(q,\lambda_{2})
\mbox{Exp}\Big(\sum_{n=1}^{\infty}
\frac{Q^{n}}{n}F^{R_{1}R_{2}}(q^{n},\lambda_{1}^{n},\lambda_{2}^{n})\Big)\,,
\label{conj}\eea The function $F^{R_{1}R_{2}}$ can be determined
easily by expanding Eq(\ref{eqx}) to first order in $Q$, \bea
F^{R_{1}R_{2}}-F^{\bullet,\bullet}&=&\sqrt{\lambda_{1}\lambda_{2}}f_{R_{1}R_{2}}-\sqrt{\frac{\lambda_{2}}
{\lambda_{1}}}f_{R_{2}}-\sqrt{\frac{\lambda_{1}}{\lambda_{2}}}f_{R_{1}}
\eea Where the function $f_{R}$ and $f_{R_{1}R_{2}}$ are given by
\bea f_{R}(q)&=&\sum_{(i,j)\in R}q^{j-i},\\\nn
f_{R_{1}R_{2}}(q)&:=&\sum_{k}C_{k}(R_{1},R_{2})q^{k}=
(q+q^{-1}-2)f_{R_{1}}f_{R_{2}}+f_{R_{1}}+f_{R_{2}}\,.\eea Using the
above definitions in Eq(\ref{eqx}) we get \bea\nn
G_{R_{1}R_{2}}(Q,\lambda_{1},\lambda_{2})=G_{00}(Q,\lambda_{1},\lambda_{2})
\frac{\prod_{k}(1-q^{k}\sqrt{\frac{\lambda_{1}}{\lambda_{2}}}Q)^{C_{k}(R_{1},\bullet)}
\prod_{k}(1-q^{k}\sqrt{\frac{\lambda_{2}}{\lambda_{1}}}Q)^{C_{k}(R_{2},\bullet)}}
{\prod_{k}(1-q^{k}\sqrt{\lambda_{1}\lambda_{2}}Q)^{C_{k}(R_{1},R_{2})}}\,.
\eea The full partition function is given by \SP{
&Z=Z_{pert}\sum_{R_{1,2}}Q_{B_{1}}^{\ell_{1}}Q_{B_{2}}^{\ell_{2}}W_{R_{1}}
(q,\lambda_{1})W_{R_{2}}(q,\lambda_{2})W_{R_{2}}(q,\lambda_{3})W_{R_{1}}(q,\lambda_{4})\,
\\
&\frac{\prod_{k}(1-q^{k}\sqrt{\frac{\lambda_{1}}{\lambda_{2}}}
Q_{F_{1}})^{C_{k}(R_{1},\bullet})(1-q^{k}\sqrt{\frac{\lambda_{2}}{\lambda_{1}}}Q_{F_{1}})^{C_{k}(R_{2},\bullet)}
(1-q^{k}\sqrt{\frac{\lambda_{3}}{\lambda_{4}}}Q_{F_{2}})^{C_{k}(R_{2},\bullet)}
(1-q^{k}\sqrt{\frac{\lambda_{4}}{\lambda_{3}}}Q_{F_{2}})^{C_{k}(R_{1},\bullet)}}
{\prod_{k}(1-q^{k}\sqrt{\lambda_{1}\lambda_{2}}Q_{F_{1}})^{C_{k}(R_{1},R_{2})}
(1-q^{k}\sqrt{\lambda_{3}\lambda_{4}}Q_{F_{2}})^{C_{k}(R_{1},R_{2})}}\
,
}

Define the renormalized K\"ahler parameters $T_{b,f}$
of the base and the fiber $\PP^{1}$,
\bea T_{B_{1}}&=&T_{b}-\frac{1}{2}(t_{1}+t_{4})\,,\\ \nn
T_{B_{2}}&=&T_{b}-\frac{1}{2}(t_{3}+t_{4})\,,\\ \nn
T_{F_{1}}&=&T_{f}-\frac{1}{2}(t_{1}+t_{2})\,,\\ \nn
T_{F_{2}}&=&T_{f}-\frac{1}{2}(t_{3}+t_{4})\,. \eea
Then in terms of the renormalized parameters we get ($C_{k}(R)=C_{k}(R,\bullet)$)
\SP{
&Z_{inst}=\\
&\sum_{R_{1,2}}Q_{b}^{\ell_{1}+\ell_{2}}Z^{(0)}_{R_{1},R_{2}}\prod_{k}(1-q^{k}
\lambda_{1}^{-1})^{C_{k}(R_{1})}(1-q^{k}\lambda_{1}^{-1}Q_{f})^{C_{k}(R_{2})}
(1-q^{-k}\lambda_{2}^{-1})^{C_{k}(R_{2})}
(1-q^{k}\lambda_{2}^{-1}Q_{f})^{C_{k}(R_{1})}\,\\
&(1-q^{-k}\lambda_{3}^{-1})^{C_{k}(R_{2})}
(1-q^{k}\lambda_{3}^{-1}Q_{f})^{C_{k}(R_{1})}
(1-q^{k}\lambda_{4}^{-1})^{C_{k}(R_{1})}(1-q^{k}\lambda_{4}^{-1}Q_{f})^{C_{k}(R_{2})}\,.
}
Where
\bea
Z^{(0)}_{R_{1}R_{2}}=\frac{C_{R_{1}}(q)^{2}C_{R_{2}}(q)^{2}}
{\prod_{k}(1-q^{k}Q)^{C_{k}(R_{1},R_{2})}}\,.
\label{eqy}
\eea
The renormalized parameters are define such that in the limit
$\lambda_{i}\rightarrow \infty$ we get the
partition function of pure 5D gauge theory, {\it i.e.\/}~the
A-model partition function of local $\PP^{1}\times \PP^{1}$.

To obtain the partition function of the 4-dimensional gauge theory we have to
take the limit
\bea
Q_{f}=e^{-2a\beta}\,,\,\,\lambda_{a}=e^{-\beta m_{a}}\,,\,\,q=e^{-\beta \epsilon}\,,\,\,\beta \mapsto 0\,.
\eea
In this limit it is easy to show that Eq(\ref{eqy}) agrees with the results
of \cite{Nekrasov:2002qd}. The case of $N_{f}=0,1$ was discussed recently
in \cite{EK}.

\underline{\bf $U(N)$ with $N_{f}=2N$:} The Calabi-Yau geometry in this case is
shown in \figref{hyperun} below.
\onefigure{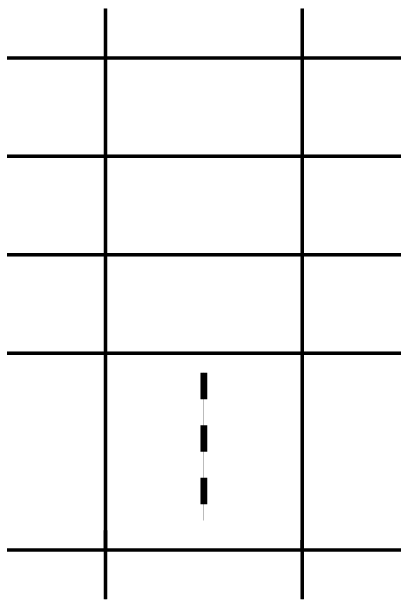}{The web diagram of the 5-dimensional $U(N)$ theory with
$2N$ fundamental hypermultiplets.}
The partition function can be calculated using either the topological vertex or
the Chern-Simons theory. We just state the result which can also be obtained from
the Weyl symmetry present in the geometry and the result of the $U(2)$ partition
function calculated before.
In this case the partition
function is given by \bea
Z:=\sum_{R_{1,2,\ldots,N}}Q_{B_{1}}^{\ell_{1}}Q_{B_{2}}^{\ell_{2}}\cdots
Q_{B_{N}}^{\ell_{N}}\,K_{R_{1}\cdots R_{N}}(\lambda_{1,\ldots,
N},Q_{F_{1},\ldots,F_{N}})\,K_{R_{1}\cdots
R_{N}}(\lambda_{N+1,\ldots,2N},Q_{F_{1},\ldots,F_{N}})\,. \eea where \bea
K_{R_{1}\cdots R_{N}}(\lambda_{1,\ldots,N},Q_{F_{1},\ldots,F_{N}})=
\prod_{1\leq i<j\leq
  N}G_{R_{i},R_{j}}(Q_{ij},\lambda_{i},\lambda_{j})\ ,\eea
where $Q_{ij}=\prod_{k=i}^{j-1}Q_{f_{k}}$.
Define
\bea
Q_{ij}=Q_{F_{ij}}/\sqrt{\lambda_{i}\lambda_{j}}\,,\\ \nn
Q_{B_{i}}=Q_{b}/\sqrt{\lambda_{i}\lambda_{N+i}}\,.
\eea
Then we get
\bea
Z=\sum_{R_{1,\ldots,N}}Q_{b}^{\ell_{1}+\cdots+\ell_{N}}Z^{(0)}_{R_{1},\ldots,R_{N}}
\prod_{i=1}^{N}\prod_{k}(1-q^{k}\lambda_{i}^{-1})^{C_{k}(R_{i})}
(1-q^{k}\lambda^{-1}_{i+N})^{C_{k}(R_{i})}\\ \nn
\times\prod_{1\leq i<j\leq N}
\prod_{k}(1-q^{k}\lambda_{i}^{-1}Q_{F_{ij}})^{C_{k}(R_{j})}
(1-q^{k}\lambda_{j}^{-1}Q_{F_{ij}})^{C_{k}(R_{i})}\\ \nn
\times(1-q^{k}\lambda_{N+i}^{-1}Q_{F_{ij}})^{C_{k}(R_{j})}
(1-q^{k}\lambda_{N+j}^{-1}Q_{F_{ij}})^{C_{k}(R_{i})}\ .
\eea
In the limit $\lambda_{i}\rightarrow \infty$ we get the partition function of the
pure 5D $SU(N)$ gauge theory with zero Chern-Simons term.

\underline{\bf 6D case:} In this case the partition function can
be calculated from the geometry shown in \figref{unhyper6d}(a)
below.
\onefigure{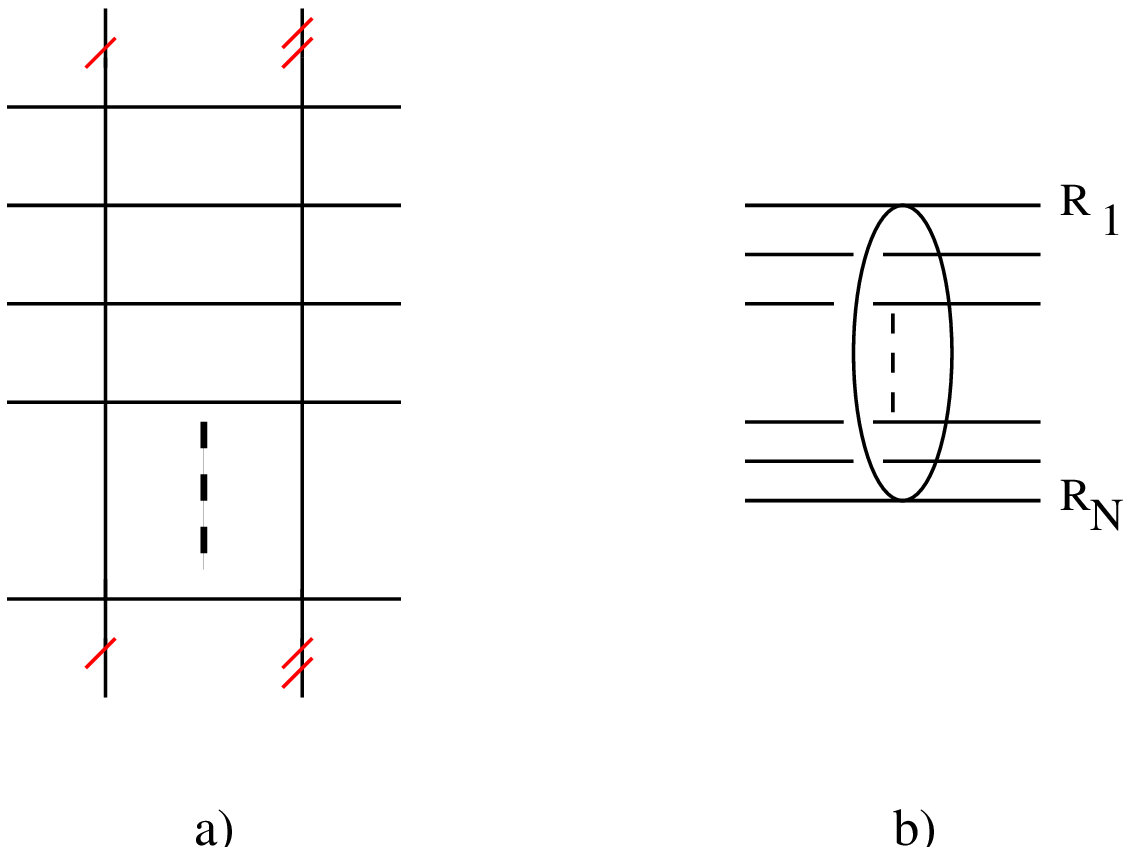}{a) The web diagram of the 6D $U(N)$ theory
with $2N$ hypermultiplets, b) the half of the web diagram used to calculate
the partition function.}
To calculate the partition function
we will slice the geometry in two parts (shown in
\figref{unhyper6d}(b)) and after calculating the partition
function of each part we will glue them together. If we denote the
partition function of the geometry in \figref{unhyper6d}(b) by
$K_{R_{1}\cdots R_{N}}$ then the full partition function is given
by \bea Z=\sum_{R_{1,\ldots,N}}Q_{B_{1}}^{l_{1}}\cdots
Q_{B_{N}}^{l_{N}}\,K_{R_{1}\cdots R_{N}}(Q_{F_{1,2,\ldots,
N}},\lambda_{1,\ldots,N},q)\,K_{R_{1}\cdots
R_{N}}(Q_{F_{1,2,\ldots,N}},\lambda_{i+N},q)\,.
\label{eqw}\eea In
calculating $K_{R_{1}\cdots R_{N}}$ we will have to take into
account the contributions from annuli which start and end on the
same three cycle and wind around the circle arbitrary number of
times. Also contribution from annuli which start and end on
different three cycles after winding around the circle arbitrary
number of times have to be considered. It is easy to see that \SP{
K_{R_{1}\cdots
R_{N}}&=\sum_{i=1}^{N}\sum_{R^{(i)}_{1,2,\ldots}}Q_{\rho}^{\sum_{k\geq
1}kl_{R^{(i)}_{k}}}W_{\prod_{k}R^{(i)}_{k}\otimes
R^{(i)}_{k},R_{i}}\\
&\times\sum_{i<j,R^{(ij)}_{0,1,2,\ldots}}Q_{\tau}^{\sum_{k\geq
0}kl_{R^{(ij)}_{k}}}Q_{ij}^{\sum_{k\geq
0}l_{R^{(ij)}_{k}}}\,W_{\prod_{k}R^{(ij)}_{k},R_{i}}\,W_{\prod_{k}R^{(ij)}_{k},R_{j}}\,.
}
Using the above expression in Eq(\ref{eqw}) we can evaluate the partition function
as a series in $Q_{\rho}$.

\section{Instanton moduli spaces and partition functions}

The most direct connection between geometric engineering
and the gauge theory perspective appears in 5 dimensional
theories with geometry $\R^4\times\S^1$, where we view $\R^4$ as
space and $\S^1$ as the Euclidean time with radius $\beta$.  In particular
consider M-theory on $\X\times\R^4\times\S^1$ where $\S^1$ has radius
$\beta$ in M-theory units.  Let $T^M_i$ denote the K\"ahler moduli
of CY in M-theory units.  Assume $\X$ is such that it engineers an ${\cal N}=1$
supersymmetric $U(N)$ gauge theory in 5d.  Moreover consider breaking
$U(N)\rightarrow U(1)^N$ by going to a generic point on Coulomb
branch given by K\"ahler moduli $a_i^M$.  The Yang-Mills coupling constant is
\EQ{
{1\over g_{YM}^2}=T_B^M\ ,
}
where $T^M_B$ is the K\"ahler moduli of the base measured in M-theory
units.
Instantons are BPS particles of this theory and they can
carry $U(1)$ charges.  The BPS mass
of such an instantons is given by
\EQ{
m=k T^M_B+ n_i a_i^M\ ,
}
where $k$ denotes the instanton number and $n_i$ denotes the
$U(1)$ charges.  Compactification on a circle of radius
$\beta$ lead to computations of the form
\EQ{
{\rm Tr}\,\exp(-\beta H)\ ,
}
where for BPS states
\EQ{
\beta H=\beta m =\beta (k T^M_B+n_i a_i^M)\ .}
From the perspective of 4d type IIA string on CY, this can be viewed
as computing the partition function of topological string because the
metric, or K\"ahler form, as measured in type IIA strings and M-theory differ by\EQ{
\beta k_M= k_{II}\ ,}
thus $\beta m=\beta (k T^M_B+ n_i a_i^M)=kT^{II}_B+n_ia^{II}_i$.  This
explains the fact that from type II string perspective the topological
string partition function was related to ${\rm Tr}\exp(-T)$ where $T$ measures
the size of the cycles in type IIA units.  Thus topological string
partition function for type IIA strings can be viewed as an M-theory
partition function on a circle, or a 5d gauge theory
BPS partition function on
a circle.  Thus the second quantized partition function
of BPS states we have computed in the context of topological
string should be related to some suitable partition function
of second quantized BPS states involving instantons of the gauge theory.
This then makes contact with the work of Nekrasov \cite{Nekrasov:2002qd}
where he developed an instanton calculus precisely for such cases.
The link between the topological string computations and the 5d gauge
theory computation of Nekrasov has been proven in \cite{AI1,AI2}.
Our main aim in reviewing aspects of it here is twofold: First we want
to generalize these to gauge theories in 5d involving adjoint fields.
secondly, we wish to generalize these to 6d gauge theories compactified
on $\T^2$.

Before describing the calculations in detail we first make some
general comments about the meaning of the ``instanton partition
function''. In the conventional setting charge $k$ instanton effects are
calculated in terms of a zero-dimensional (matrix) supersymmetric
sigma model with the $k$-instanton moduli space of $U(N)$ $\ms_{k,N}$ as
target (see \cite{review}).
The exact details will be somewhat different for the theory with
an adjoint as opposed to fundamental hypermultiplets.
The sigma models are coupled to various isometries of the target
space. Firstly, to the abelian subgroup $U(1)^N\subset U(N)$ of the
global gauge group which acts
on $\ms_{k,N}$. This gives coupling which depend on $N$ parameters which
are identified with the VEVs $a_i$ of the parent theory. These coupling
imply that the integrals over $\ms_{k,N}$ localize over fixed points of
global gauge transformations. In particular, the fixed-point set
consists of the moduli space of point-like instantons
\cite{Hollowood:2002zv,Hollowood:2002ds}. This is still a complicated
space to integrate over. The additional insight of
\cite{Nekrasov:2003rj} was that if one, in addition, coupled to
isometries corresponding to the
abelian parts of the Lorentz group, involving two parameters
$\epsilon_1$ and $\epsilon_2$,\footnote{In
the following we shall for the most part make the simplifying choice
$\epsilon_1=-\epsilon_2=\epsilon$.} then the fixed-point set becomes
discrete. The instanton partition function can then be expressed as a
sum over these discrete points and each contribution is a ratio of
the usual fermionic and bosonic fluctuation determinants.

If we now lift the theory to five-dimensions compactified on a
circle, instantons in four-dimensions are now solitons in
five-dimensions whose world-lines can wrap around the circle. The
instanton partition function now involves quantum mechanics on the
instanton moduli space and due to the couplings to the isometries
defines an equivariant generalization of an index on $\ms_{k,N}$. The
localization techniques are still valid the only difference being
that the fermionic and bosonic determinants now include a product
over all the Kaluza-Klein modes of fluctuations.

We will also be interested in F-theory compactifications
on elliptic threefolds times a $\T^2$. From the viewpoint
of 6d theory, we now have a 1+1 dimensional sigma model
from $\T^2$ to the instanton moduli space.  It is now
clear that whatever index one is computing will be
replaced by the corresponding elliptic index, where the complex
structure of $\T^2$ will enter the elliptic index.  We will discuss
this in more detail below in the context of our main example which
is the mass deformed $(1,1)$ supersymmetric
theory in 6D compactified on $\T^2$ (giving $\N=2^*$ in 4D). In this case the
localization procedure naturally involves replacing the weights
of the circle actions
$x_i$ by its elliptic generalization
$\theta_1(\tfrac{\beta}{2i}x_i|\rho)$.\footnote{The fact
that
  this depends on only $\rho$ rather than its conjugate is a
  reflection of the fact that only the anti-holomorphic modes
  contribute: the ratio of determinants of the fermionic and bosonic holomorphic modes cancel.}

\subsection{Calculation of 5D partition functions}

In this section we use the instanton calculus to compute
the mass deformed ${\cal N}=2^*$ BPS partition function in 5D.
As has been shown in \cite{Nekrasov:2002qd} the relevant index
computation involves the $\chi_y$ genus of the instanton moduli space.

For a closed complex manifold $\ms$, its $\chi_{y}$ genus is defined
as \bea \chi_{y}(\ms)&=&\sum_{p,q\geq 0} y^{p}(-1)^{p} \text{dim}\,
H^{q}(\ms,\Lambda^{p}T^{*}\ms)=\sum_{p\geq
0}y^{p}\chi(\ms,\Lambda^{p}T^{*}\ms)\,\\ \nn
&=&\int_{\ms} \text{ch}\,\Lambda_{-y}(T^{*}\ms)\text{Td}(\ms)=
\int_{\ms} \prod_{j=1}^{d}(1-ye^{-x_{j}})\frac{x_{j}}{1-e^{-x_{j}}}\ ,
\eea
where $\{x_{1},\ldots,x_{d}\}$ denote the Chern roots of
$T\ms$, the tangent bundle.
 If $\ms$ has a torus action with isolated fixed points
 $\{p_{1},\ldots,p_{n}\}$ and weights $\{w_{i,1},\ldots,w_{i,d}\}$
 at $p_{i}$ then it follows from
 localization theorem
 \bea
 \chi_{y}(\ms)&=&\sum_{i=1}^{n}\frac{\prod_{j=1}^{d}(1-ye^{-w_{i,j}})\frac{
 w_{i,j}}{1-e^{-w_{i,j}}}}{\prod_{j=1}^{d}w_{i,j}}\,\\ \nn
 &=&
 \sum_{i=1}^{n}\prod_{j=1}^{d}\frac{1-ye^{-w_{i,j}}}{1-e^{-w_{i,j}}}\,.
 \label{chiy}
 \eea
For the case we are interested in, $\ms=\ms_{k,N}$, $y=e^{-\beta m}$, and the fixed points and weights at each fixed point
of $\ms_{k,N}$ under the $U(1)^{N}\times U(1)\times U(1)$ action were
calculated in \cite{Nekrasov:2002qd,NY}.
The group $U(1)^{N}\times U(1)\times U(1)$ mentioned above is the Cartan of the gauge group and the spacetime
rotation group. The fixed points of $\ms_{k,N}$ are in one to one
correspondence with the partitions of $k$ into $N$ colors {\it i.e.\/},
the fixed points are labelled by $N$ representations $R_{\alpha}$ of $U(\infty)$ such that
\EQ{
k=\ell_{R_{1}}+\ell_{R_{2}}+\cdots +\ell_{R_{N}}\,.
\label{colpar}
}
Let us denote the corresponding Young diagrams by $\mu^{i}$ (and the transpose diagram by $\mu^{t,i}$) then given a fixed points of $\ms_{k,N}$ labelled by
$(\mu^{1},\ldots, \mu^{N})$ the corresponding weights are given by \cite{Nekrasov:2002qd,fucito1, FP, NY}
\bea
\sum_{i,j}e^{w_{i,j}}= \sum_{\alpha,\gamma=1}^{N}e^{\beta(a_{\alpha}-a_{\gamma})}\Big(\sum_{(i,j)\in R_{\alpha}} \,q^{\mu^{\alpha}_{i}+\mu^{t,\gamma}_{j}-i-j+1}+\sum_{(i,j)\in R_{\gamma}}\,q^{-\mu^{\gamma}_{i}+\mu^{t,\alpha}-i-j+1}\Big)\,.
\label{weights}
\eea
For the case of $N=1$ we see that the above expression simplifies to
\bea
\sum_{(i,j)\in R}(q^{h(i,j)}+q^{-h(i,j)})\,,\,\,h(i,j)=\mu_{i}+\mu^{t}_{j}-i-j+1
\label{weights1}
\eea
Using the above weights for the $N=1$ case in Eq.~(\ref{chiy}) we get
\bea
\sum_{k}Q^{k}\chi(\ms_{k,1})&=&\sum_{k}Q^{k}\sum_{R,\ell_{R}=k}\prod_{(i,j)\in R}\frac{(1-yq^{h(i,j)})(1-yq^{-h(i,j)})}{(1-q^{h(i,j)})(1-q^{-h(i,j)})}\,,\\ \nn
&=&\sum_{R}Q^{\ell_{R}}\prod_{(i,j)\in R}\frac{(1-yq^{h(i,j)})(1-yq^{-h(i,j)})}{(1-q^{h(i,j)})(1-q^{-h(i,j)})}\,.
\eea
This agrees exactly with the Eq.~(\ref{ins5dadjoint}) which was calculated using the
topological vertex if we identify $y=Q_{m}$.

For $N>1$ using the weights given above in Eq.~(\ref{weights}) in Eq.~(\ref{chiy}) we get
\SP{
&\sum_{k}Q^{k}\chi(\ms_{k,N})=\\
&\sum_{R_{1,\ldots,N}}
Q^{\sum_{i=1}^{N}\ell_{R_{i}}}\prod_{\alpha,\gamma=1}^{N}
\prod_{(i,j)\in R_{\alpha}}\frac{(1-y\,e^{\beta(a_{\alpha}-a_{\gamma})}\,q^{\mu^{\alpha}_{i}+\mu^{t,\gamma}_{j}-i-j+1})}
{(1-e^{\beta(a_{\alpha}-a_{\gamma})}\,q^{\mu^{\alpha}_{i}+\mu^{t,\gamma}_{j}-i-j+1})}
\prod_{(i,j)\in R_{\gamma}}\frac{(1-y\,e^{\beta(a_{\alpha}-a_{\gamma})}\, q^{-\mu^{\gamma}_{i}+\mu^{t,\alpha}-i-j+1})}{(1-e^{\beta(a_{\alpha}-a_{\gamma})}\,q^{-\mu^{\gamma}_{i}+\mu^{t,\alpha}-i-j+1})}\,,
}
which also agrees with the $U(2)$ case discussed in the last section for $N=2$.

\subsection{Calculation of 6D partition function}
In this case, the instanton partition of the six-dimensional theory
can be interpreted as the generating functional for an elliptic genus
of the instanton moduli space,
\EQ{
Z=\sum_kQ^k\chi(\ms_{k,N})\ .
}
The elliptic genus $\chi(\ms)$ is defined as the partition
function in the Ramond-Ramond sector of the $\N=2$ two-dimensional
sigma-model with $\ms$ as target on the torus $\T^2$ \cite{SW,LSW}:
\EQ{
\chi(\ms)={\rm Tr}\,\Big((-1)^F\,y^{F_L}
Q_{\rho}^{L_0}\bar{Q_{\rho}}^{\bar L_0}
e^{\sum_{i=1}^Na_iJ_i+\epsilon_1K_1+\epsilon_2K_2}\Big)\ ,
}
where
\EQ{
 Q_\rho=e^{2\pi i\rho}\ ,\qquad y:=e^{-\beta m}\ ,
}
$F=F_L+F_R$, the sum of the left and
right fermion numbers.
 The remaining
terms correspond to coupling to abelian isometries of $\ms$. The charges
$J_i$ corresponding to the $U(1)^N\subset U(N)$ of the gauge
group,\footnote{Only $SU(N)$ acts non-trivially on $\ms$ so we can fix
$\sum_{i=1}^Na_i=0$.} while $K_{1,2}$ are the charges corresponding to
the abelian subgroup $U(1)^2\subset SU(2)_L\times SU(2)_R$ of the
Lorentz group of $\R^4$.  Note that naively $Z$ vanishes
if we include the right-moving fermionic zero mode on $\R^4$ which
is always a factor for $\ms_{k,N}$, and so this trace is meant with
the zero mode deleted. This reflects the same condition for
BPS partition function, namely we only include the lowest component
for each BPS multiplet.

The simplest example is given by taking $N=1$. The $U(1)$ theory does not have
smooth instanton solutions. In fact
one way to think about instantons in an abelian theory is to turn on
spacetime non-commutativity. In that case, there are instanton
solutions whose only moduli correspond to the positions of the
individual instantons. A single instanton has a moduli space ${\bf
  R}^4$ which represents its position in Euclidean spacetime. For
charge $k$, the moduli space is a smoothed version of the symmetric
product
\EQ{
\ms_{k,1}\thicksim\text{Sym}^k\big(\R^4\big)\ .
}
For the six-dimensional theory, we can work directly in terms of the
symmetric product.

The elliptic genus for one instanton, for which $\ms_{1,1}\simeq{\bf
R}^4$, can be written down straightforwardly. Choosing
$\epsilon_1=-\epsilon_2=\epsilon$,
\SP{
\chi(\ms_{1,1},Q_{\rho},y,q)&=\prod_{n=1}^\infty\frac{
(1-Q_{\rho}^{n}\alpha q)
(1-Q_{\rho}^{n-1}y^{-1}q^{-1})(1-Q_{\rho}^{n}
y^{-1}q)(1-Q_{\rho}^{n-1}y q^{-1})}
{(1-Q_{\rho}^{n-1}q)^2
(1-Q_{\rho}^{n}q^{-1})^2}\\
&=\frac{\theta_1\big(\tfrac{\beta}{2i}(\epsilon+m)\big|\rho\big)
\theta_1\big(\tfrac{\beta}{2i}(\epsilon-m)\big|\rho\big)}
{\theta^2_1\big(\tfrac{\beta}{2i}\epsilon\big|\rho\big)}\ ,
}
where
\EQ{
y=e^{-\beta m}\ ,\qquad q=e^{-\beta\epsilon}=e^{-\lambda_s}\ .
}
This expression as a function $\phi(\rho,z_1,z_2)$,
$y=e^{2\pi iz_1}$, $q=e^{2\pi iz_2}$
is a weak Jacobi form of weight 0 and indices 2 and 0, for $z_1$ and
$z_2$, respectively.

For higher instanton number we can apply the formula of
\cite{Dijkgraaf:1996xw} for
the elliptic genus of a symmetric product. If
\EQ{
\chi(\ms)=\sum_{n\geq0,p_1,p_2,\ldots}c(n,p_1,p_2,\ldots)
Q_{\rho}^ny_1^{p_1}y_2^{p_2}\cdots\ ,
\label{defch}
}
then
\EQ{
\sum_{k=1}^\infty
Q^k\chi(\text{Sym}^k(\ms))=\prod_{k=1,n=0}^\infty\prod_{p_1,p_2,\ldots}
\frac1
{(1-Q^kQ_{\rho}^ny_1^{p_1}y_2^{p_2}\cdots)^{c(nk,p_1,p_2,\ldots)}}\ ,
\label{genf}
}
where we have allowed for coupling to an arbitrary
number of conserved quantities
though $y_1,y_2,\ldots$. In the present case, we have coupling to two
charges through $\alpha$ and $q$.
The instanton partition function is then equal to the
generating function \eqref{genf} with the $c(n,p_1,p_2)$ extracted from
\eqref{defch}.  The fact that the relevant moduli space
is that of a genus 2 curve, as we have found before, was already
noted in \cite{Dijkgraaf:1996xw} in the context of elliptic genus
of symmetric products.  The problem of interest there was related
to computation of entropy of 5D black holes \cite{Strominger:1996sh}
where the relevant space is the moduli space of instantons on $K3$ or $\T^4$,
as opposed to the case of interest here which is $\R^4$.

In the last section we saw that localization allows us to write the
$\chi_{y}$ genus of $\ms$ as a sum of contributions from the fixed points.
The elliptic genus of $\ms$ can similarly be written as the sum over
the fixed points with weights $w_{i,j}$,
\bea
\chi(\ms)=\sum_{i=1}^{n}\prod_{j=1}^{d} \Big(\frac{1-ye^{-w_{i,j}}}{1-e^{-w_{i,j}}}
\Big(\prod_{k=1}^{\infty}\frac{(1-Q_{\rho}^{k}\,y\,e^{-w_{i,j}})(1-Q_{\rho}^{k}\,y^{-1}\,e^{w_{i,j}})}
{(1-Q_{\rho}^{k}e^{-w_{i,j}})(1-Q_{\rho}^{k}\,e^{w_{i,j}})}\Big)\Big)\,.
\eea

Let's first consider the case of $N=1$. In this the weights are give by Eq.~(\ref{weights1}). Substituting then in
in the above equation we get
\bea\nn
\sum_{k}Q^{k}\chi(\ms_{k,1})=\sum_{R}Q^{\ell_{R}}\prod_{(i,j)\in R}
\frac{(1-yq^{h(i,j)})(1-yq^{-h(i,j)})}{(1-q^{h(i,j)})(1-q^{-h(i,j)})}
\,\\ \nn\prod_{n=1}^{\infty}\frac{(1-Q_{\rho}^{n}yq^{h(i,j)})(1-Q_{\rho}^{n}y^{-1}q^{-h(i,j)})(1-Q_{\rho}^{n}yq^{-h(i,j)})(1-Q_{rho}^{n}y^{-1}q^{h(i,j)})}
{(1-Q_{\rho}^{n}q^{h(i,j)})^{2}(1-Q_{\rho}^{n}q^{-h(i,j)})^{2}}\,.
\label{el2}
\eea
This agrees with the topological vertex computation of the last section, Eq.~(\ref{elll}), once we use the identification $y=Q_{m}$.
The above expression can also be written as
\EQ{
Z=\sum_{R}Q^{\ell_R}\prod_{(i,j)\in R}
\frac{\theta_1\big(\tfrac{\beta}{2i}(h(i,j)\epsilon+m)\big|\rho\big)
\theta_1\big(\tfrac{\beta}{2i}(h(i,j)\epsilon-m)\big|\rho\big)}
{\theta_1\big(\tfrac{\beta}{2i}h(i,j)\epsilon\big|\rho\big)^2}\ .
}
One can easily check by hand
that the two expressions Eq.~(\ref{genf}) and Eq.~(\ref{el2}) agree for
the first few terms.

\subsection{Extracting the Curves From Instantons}

In this section, we show how the curves of our two six-dimensional
theories compactified on a torus can be extracted by using
instantons.  In a sense this is already done
in the previous section, when we showed that A-model
topological string amplitudes agree with the instanton calculus
computations.  Since one knows how to extract the mirror
curve in the A-model setup, this is a proof of how
one can extract the curve from the instanton calculus.
However, one can also do this directly as was done in \cite{Nekrasov:2003rj}.

The central quantity is the instanton partition function which is a
given by a sum over instanton numbers of the form
\EQ{
Z=\sum_kQ^kZ_k\ ,\qquad Q=e^{2\pi i\tau}\ ,
}
where $\tau$ is the complexified coupling of the theory.
The associated free energy has an expansion which includes
the prepotential ${\cal F}_0$ as well as a whole series of
gravitational couplings:
\EQ{
Z=\exp\Big(\sum{\cal F}_g\lambda_s^{2g-2}\Big)\ .
}
The curve appears in the limit of $\lambda_s\rightarrow 0$.

\subsubsection{The theory with an adjoint}

From our discussion of the last section where we computed the instanton
partition functions from the topological vertex and the localization
calculation (Eq(236)) we see that the instanton partition function for the $U(N)$ theory with an adjoint 
can be expressed as
\cite{Nekrasov:2003rj}
\SP{
&Z(a_i)=\exp\Big(\sum_{ij}\big(\gamma_\epsilon(a_i-a_j)-
\gamma_\epsilon(a_i-a_j+m)\big)\Big)\\
&\times\sum_{R_1,\ldots,R_N}Q^{\ell_{R_1}+\cdots+\ell_{R_N}}
\prod_{(i,p)\neq(j,q)}\frac{
\sigma(a_i-a_j+\epsilon(\mu^i_p-\mu^j_q+q-p))\sigma(a_i-a_j+\epsilon(q-p)+m)}
{\sigma(a_i-a_j+\epsilon(q-p))\sigma(a_i-a_j+\epsilon(\mu^i_p-\mu^j_q+q-p)+m)}\
. \label{pfn} } The sum is over colored partitions
of $k$, the instanton charge, as in
\eqref{colpar}.
This is a partition of $k$ into $N$ Young Tableau $\{R_1,\ldots,R_N\}$
with a total of $k$ boxes described by the data
\EQ{ \mu^i_1\geq \mu^i_2\geq\cdots\geq
\mu^i_{n_i}\qquad\text{with}\qquad\sum_{p=1}^{n_i}
\mu^i_p=\ell_{R_i},\quad \sum_{i=1}^N\ell_{R_i}=N\ , }
for $i=1,\ldots,N$.
The instanton partition function given above although looks different but is
exactly the equal to the one calculated in the last section (for the case of $U(1)$ and $U(2)$) if
we use the following two identities \cite{AI1},
\bea
\prod_{i,j=1}^{\infty}\frac{\sigma(\mu_{i}-\mu_{j}+j-i)}
{\sigma(j-i)}&=&\prod_{(i,j)\in R}\frac{1}{\sigma^{2}(h(i,j))}\,,\\ \nn
\prod_{\alpha\neq \gamma}\prod_{i,j=1}^{\infty}\frac{\sigma(a_{\alpha \gamma}+\epsilon(\mu^{\alpha}_{i}-\mu^{\gamma}_{j}+j-i))}
{\sigma(a_{\alpha \gamma}+\epsilon(j-i))}&=&
\prod_{k}\frac{1}{(\sigma(a_{12}+\epsilon k))^{2C_{k}(R_{1},R_{2}^{t})}}\,,\,\,\alpha,\gamma=1,2.
\eea
The integers $C_{k}(R_{1},R_{2})$ were defined in Section 4,
$\sum_{k}C_{k}(R_{1},R_{2})q^{k}=f_{R_{1},R_{2}^{t}}$.

For the four, five and six-dimensional theory,
\bea
\sigma_{4D}(x)=x\ ,\qquad
\sigma_{5D}(x)=\sinh(\tfrac{\beta x}{2})\ ,\qquad
\sigma_{6D}(x)=
\theta_1\big(\tfrac{\beta x}{2i}\big|\rho\big)\,
\eea
respectively. In addition, the kernel
$\gamma_\hbar(x)$ is defined by the finite difference equation
\EQ{
\gamma_\epsilon(x+\epsilon)+\gamma_\epsilon(x-\epsilon)-2\gamma_\epsilon(x)=
\log\sigma(x)\ .
\label{find}
}

In the four-dimensional case, the partition function can be simply
written in terms of a sum over certain ``paths'' $f(x)$ which are
associated to the colored partitions. In concrete terms, it is
simpler to work on terms of the second derivative of $f''(x)$:
\EQ{ f''(x)=2\sum_{i=1}^N\Big(\sum_{p=1}^{n_i}\big(
\delta(x-a_i-\epsilon
(\mu^i_p-p+1))-\delta(x-a_i-\epsilon(\mu^i_p-p))\big)+\delta(x-\epsilon
n_i)\Big)\ . \label{denf} } In the limit $\epsilon\to0$, $f''(x)$
becomes a positive density with $N$ intervals of support along $N$
open contours ${\cal C}_i$ with end-points \EQ{ {\cal
C}_i=[r_i,s_i]\qquad i=1,\ldots,N \label{defC} } located in the
vicinity of $a_i$. This picture naturally extends by continuity to
the six-dimensional theory where now $f''(x)$ is defined as a
density in $\tilde \T^2$ with support along the $N$ contours
\eqref{defC}.

The following identities arise from
\eqref{denf}. First of all, one has the normalization condition
\EQ{
\int_{{\cal C}_i}dx\,f''(x)=2\ .
\label{ppa}
}
Secondly, the VEVs are
recovered via
\EQ{
a_i=\tfrac12\int_{{\cal C}_i}dx\,x\,f''(x)\ ,
\label{ppb}
}
while the instanton charge is
\EQ{
k=-\frac1{2\epsilon^2}\sum_{i=1}^Na_i^2+\frac1{4\epsilon^2}\int_{{\cal
    C}}dx\,x^2\,f''(x)\ ,
\label{ppc}
}
where the union of all the contours is
\EQ{
{\cal C}=\bigcup_{i=1}^N{\cal C}_i\ .
}

In the $\epsilon\to0$ limit, $Z$ is dominated by a saddle-point determined
by minimizing the functional ${\cal E}[f'']$
\SP{
{\cal E}[f'']&=-\tfrac14\int_{{\cal C}} dx\,dy\,f''(x)f''(y)
\big(\gamma_0(x-y)-\tfrac12\gamma_0(x-y+m)-\tfrac12
\gamma_0(x-y-m)\big)\\
&\qquad-\frac{i\pi\tau}2\int_{{\cal C}} dx\,x^2\,f''(x)
+\sum_{i=1}^N\lambda_i
\big(a_i-\tfrac12\int_{{\cal C}_i}dx\,x\,f''(x)\big)\ .
\label{fex}
}
in which case
\EQ{
Z\thicksim\exp\Big(-\epsilon^{-2}{\cal E}[f'']\Big)\ .
}
In the above we, have included Lagrange multipliers $\lambda_i$ to enforce the
fact that the $a_i$ are fixed. The kernel $\gamma_0(x)$ is the first
term in the small $\epsilon$ expansion
\EQ{
\gamma_\epsilon(x)=\sum_{g=0}^\infty\gamma_g(x)\epsilon^{2g-2}\ .
}
It follows from \eqref{find} that
\EQ{
\gamma''_0(x)=\log\sigma(x)\ .
\label{rsg}
}

We can now calculate $Z$ in this limit by finding the saddle-point. In
fact rather than calculate $Z$ we shall see that the Seiberg-Witten
curve arises in the description of the critical density $f''(x)$. To
start with, the saddle-point equation is, for $x\in{\cal C}_i$,
\EQ{
\int_{{\cal C}}dy\, f''(y)\big(\gamma_0(x-y)-\tfrac12\gamma_0(x-y+m)-
\tfrac12\gamma_0(x-y-m)\big)+i\pi\tau x^2+\lambda_ix=0\ .
\label{SpE}
}
In order to solve this equation, it is convenient to introduce a
resolvent
\EQ{
\tilde\omega(x)=\int_{{\cal C}} dy\,f''(y)\,\partial_x\log\theta_1
\big(\tfrac{\beta}{2i}(x-y)\big|\rho\big)\ .
\label{defresi}
}
This is a multi-valued analytic function on $\tilde \T^2$, since it
picks up an additive piece under continuation around the $B$-cycle,
\EQ{
\tilde\omega(x+2\pi i/\beta)=\tilde\omega(x)\ ,\qquad
\tilde\omega(x+2\pi i\rho/\beta)=\tilde\omega(x)-2\beta^{-1}\ ,
}
with $N$ branch cuts ${\cal C}_i$. As usual with a resolvent,
the discontinuity across a cut is proportional to the density:
\EQ{
\tilde\omega(x+\epsilon)-\tilde\omega(x-\epsilon)=2\pi if''(x)\
,\qquad x\in{\cal C}\ ,
\label{disc}
}
where $\epsilon$ is a suitable infinitesimal chosen so that
$x\pm\epsilon$ lie infinitesimally above and below the cut at $x$.
The third derivative of \eqref{SpE} with respect to $x$ can then be written
\EQ{
\tilde\omega(x+\epsilon)+\tilde\omega(x-\epsilon)-\tilde\omega(x+m)-
\tilde\omega(x-m)=0\ ,\qquad x\in{\cal C}\ .
\label{iuut}
}
This equation is identical the
equation for the resolvent in the matrix model
\eqref{spe}, except that there is no potential on the right-hand
side. However, the similarity suggests that we define the function
\EQ{
\tilde G(x)=\tilde\omega(x+\tfrac{m}2)-\tilde\omega(x-\tfrac{m}2)\ ,
}
to match \eqref{defg}. This function is now an
analytic function on $\tilde \T^2$ with $N$ pairs of branch cuts
\EQ{
{\cal C}_i^\pm={\cal C}_i\pm\tfrac{m}2\ .
}
The equation \eqref{iuut} then becomes a gluing condition
\EQ{
\tilde G(x+ \tfrac{m}2\pm\epsilon)=\tilde G(x-\tfrac{m}2\mp
\epsilon)\qquad x\in{\cal C}\ .
\label{gluei}
}
Pictorially, the top/bottom of ${\cal C}_i^+$ is glued to the bottom/top of
${\cal C}_i^-$. So $\tilde G$ is single-valued on a Riemann surface of genus
$N+1$ just as in the matrix model.

We now prove that this curve is the Seiberg-Witten curve $\Sigma$.
In order to do this, we prove that the period matrix has the form
\eqref{psw}. This itself follows from the existence of the
multi-valued function $z$ with the monodromies \eqref{multiz}. In
the present context, we will identify \EQ{ z(P)=\frac1{4\pi
i}\int_{P_0}^P\tilde G(x)dx\ . } It follows from \eqref{disc} that
\EQ{ \oint_{A_j}dz=\frac1{4\pi i}\int_{{\cal C}_j}dx\,f''(x)=1\ ,
\label{zai} } where $A_j$ is a cycle encircling the top cut ${\cal
C}_j^+$ as in Fig.~\ref{mm1}. Now we consider the integral of $dz$ over
the conjugate cycle $B_j$ which goes from a point on the lower cut
$x-\tfrac{m}2\in{\cal C}_j^-$ to the point $x+\tfrac{m}2$ on the
upper cut ${\cal C}_j^+$. For $x\in{\cal C}_j$, \SP{
&\oint_{B_j}dz=\int_{x-\tfrac{im}2}^{x+\tfrac{im}2}\tilde G(x')dx'\\
&\qquad =\frac1{2\pi i}\int_{{\cal C}}dy\,f''(y)\big(\gamma''_0(x-y)-
\tfrac12\gamma''_0(x-y+m)-\tfrac12\gamma''_0(x-y-m)\big)=\tau
}
independent of $x$,
where the last equality follows from taking the second derivative of
the saddle-point equation \eqref{SpE} for $x\in{\cal C}_j$ and using
the relation \eqref{rsg}.

{}From \eqref{zai}, it follows that $dz=\sum_{i=1}^N\omega_i$
and therefore that the first $N$ rows and columns of
the period matrix satisfy\footnote{In principal, one could have
$dz=\sum_{i=1}^N\omega_i+\lambda\omega_{N+1}$, for arbitrary
$\lambda$. The only effect of this is to re-define the coupling
$\tau$ and so we choose $\lambda=0$.}
\EQ{
\sum_{j=1}^N\Pi_{ij}=\tau\qquad\forall i=1,\ldots,N\ .
}
These are precisely the conditions on the period matrix \eqref{psw}.
We can view these $N$ equations as $N$ conditions on the moduli
$\{r_i,s_i\}$ (the ends of the contours ${\cal C}_i$).
However, there are $N$ additional conditions that arise
from the constraints
\EQ{
a_j=\tfrac12\int_{{\cal C}_j}dx\,xf''(x)=\oint_{A_j}x\,dz\ .
}
We can therefore think of the $a_j$ as the moduli of the curve
$\Sigma$ and notice that $x\,dz$ is the Seiberg-Witten
differential.

So we have shown that the Seiberg-Witten geometry that we engineered
out of the matrix model in Section 2
also describes the $\epsilon\to0$ limit of the instanton partition function.

\subsubsection{The theory with fundamentals}

In this section, we follow the same procedure using instantons to extract the
Seiberg-Witten curve for the six-dimensional $\N=(1,0)$ theory with
fundamental hypermultiplets compactified on the torus $\T^2$.

The instanton partition function with $N_f$ fundamental
hypermultiplets is \cite{Nekrasov:2002qd},
\SP{
&Z(a_i)=\exp\Big(\sum_{ij}\gamma_\epsilon(a_i-a_j)+\sum_{if}\gamma_\epsilon(a_i-m_f)
\Big)\\
&\times\sum_{R_1,\ldots,R_N}Q^{\ell_{R_1}+\cdots+\ell_{R_N}}
\prod_{(i,p)\neq(j,q)}\frac{\sigma(a_i-a_j+\epsilon(\mu^i_p-\mu^j_q+q-p))}
{\sigma(a_i-a_j+\epsilon(q-p))}\prod_{ipqf}
\frac{\sigma(a_i-m_f+\epsilon(\mu^i_p+q-p))}{\sigma(a_i-m_f+\epsilon(q-p))}\ .
\label{pfnf}
}
In the last section we calculated this partition function (Eq(222) and Eq(223)) 
using the topological vertex formalism and the Chern-Simons theory.
In order to take the $\epsilon\to0$, we follow exactly the same steps as
for the adjoint theory. We assume that in the limit, $f''(x)$ is a
density with $N$ intervals of support along the contours \eqref{defC} in
$\tilde \T^2$. The conditions \eqref{ppa}-\eqref{ppc} continue to hold.
The functional to be extremized, replacing \eqref{fex}, is
\SP{
{\cal E}[f'']&=-\tfrac14\int_{{\cal C}} dx\,dy\,f''(x)f''(y)
\gamma_0(x-y)+\tfrac12\sum_{f=1}^{N_f}\int_{{\cal C}}dx\,f''(x)
\gamma_0(x-m_f)\\
&\qquad-\frac{i\pi\tau}2\int_{{\cal C}} dx\,x^2\,f''(x)
+\sum_{i=1}^N\lambda_i
\big(a_i-\tfrac12\int_{{\cal C}_i}dx\,x\,f''(x)\big)\ .
}
This yields the saddle-point equation
\EQ{
\int_{{\cal C}}dy\, f''(y)\gamma_0(x-y)
-\sum_{f=1}^{N_f}\gamma_0(x-m_f)+i\pi\tau x^2+\lambda_ix=0\qquad
x\in{\cal C}_i\ .
\label{SpEf}
}
As in the $\N=2^*$ case, it is convenient to introduce a resolvent
defined by
\EQ{
\tilde\omega(x)=\int_{{\cal C}} dy\,f''(y)\,\partial_x\log\theta_1
\big(\tfrac{\beta}{2i}(x-y)\big|\rho\big)-
\sum_{f=1}^{N_f}\partial_x\log\theta_1
\big(\tfrac{\beta}{2i}(x-m_f)\big|\rho\big)\ ,
\label{defresf}
}
in which case the third derivative of the saddle-point equation has the form
\EQ{
\tilde\omega(x+\epsilon)+\tilde\omega(x-\epsilon)=0\
,\qquad x\in{\cal C}\ .
\label{iuutf}
}
Notice that the resolvent \eqref{defresf} is only well-defined on the
torus $\tilde \T^2$ if $N_f=2N$, otherwise it picks up an additive
ambiguity around the $B$-cycle of the $\tilde \T^2$ torus. This is
presumably related to the anomaly of the six-dimensional theory unless
$N_f=2N$.

\begin{figure}
\begin{center}
\includegraphics[scale=0.7]{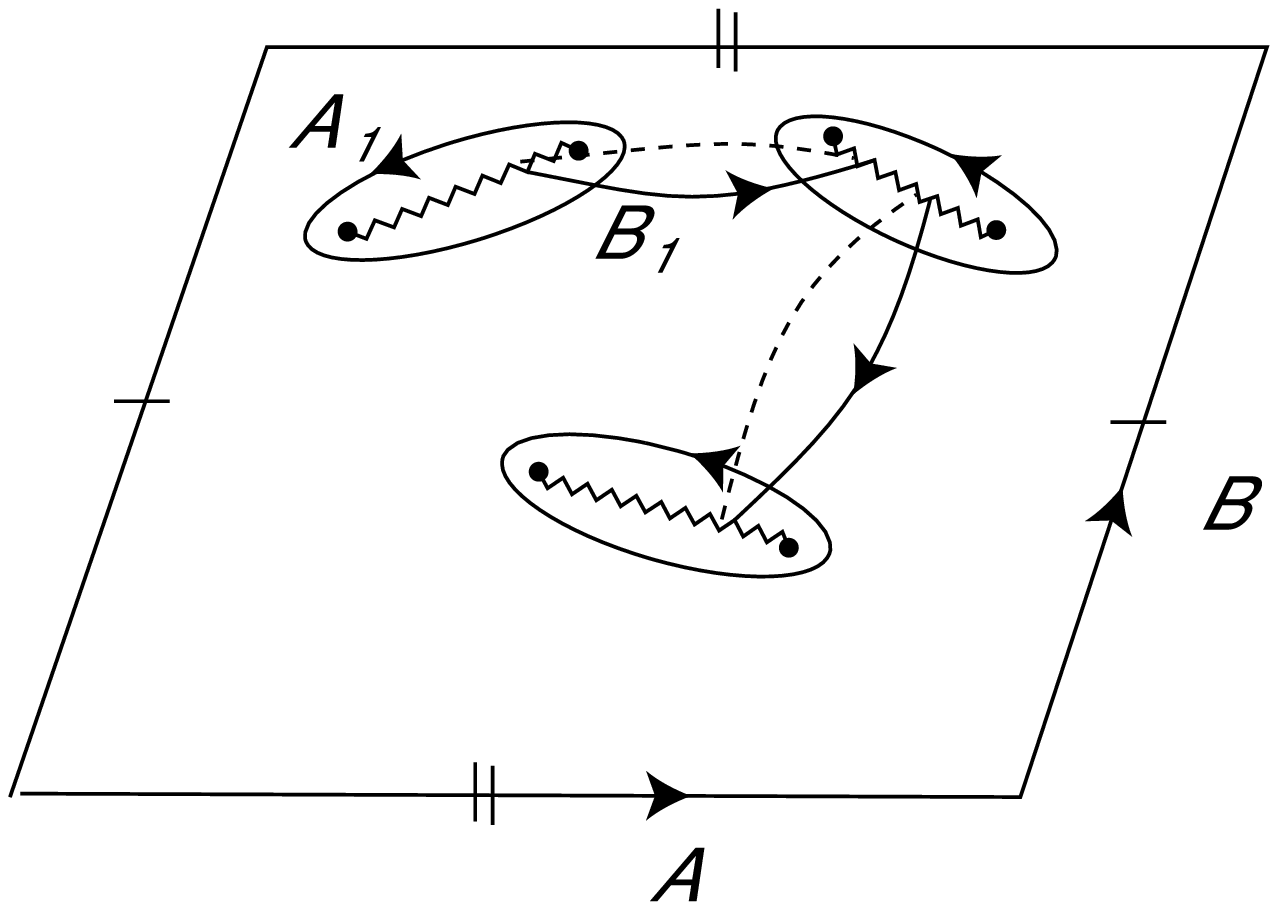}
\end{center}
\caption{\small The cut $\tilde \T^2$ on which the resolvent
$\tilde\omega(x)$ is defined. The contours $A_j$, $j=1,\ldots,N$
encircle the
$j^{\rm th}$ cut, while the contours $B_j$, $j=1,\ldots,N-1$
join
the $j^{\rm th}$ and $j+1^{\rm th}$ cuts and return on the lower
sheet.}
\label{mm5}\end{figure}

The normalization condition \eqref{ppa} requires
\EQ{
\oint_{A_j}\tilde\omega(x)dx=-2\pi i\int_{{\cal C}_j}f''(x)dx=-4\pi i\
,
\label{aint}
}
where $A_j$ is a cycle that encircles the $j^{\rm th}$ cut, as
illustrated in Fig.~\ref{mm5}.
In addition, for $x_j\in{\cal C}_j$ consider the integral
\EQ{
\int_{x_j}^{x_{j+1}}\tilde\omega(x)dx=\int_{{\cal C}}dy\,f''(y)\big(
\log\theta_1\big(\tfrac{\beta}{2i}(x_{j+1}-y)\big|\rho\big)-
\log\theta_1\big(\tfrac{\beta}{2i}(x_j-y)\big|\rho\big)\big)=0\
,
\label{bint}
}
by the second derivative of \eqref{SpEf}.

\begin{figure}
\begin{center}
\includegraphics[scale=0.5]{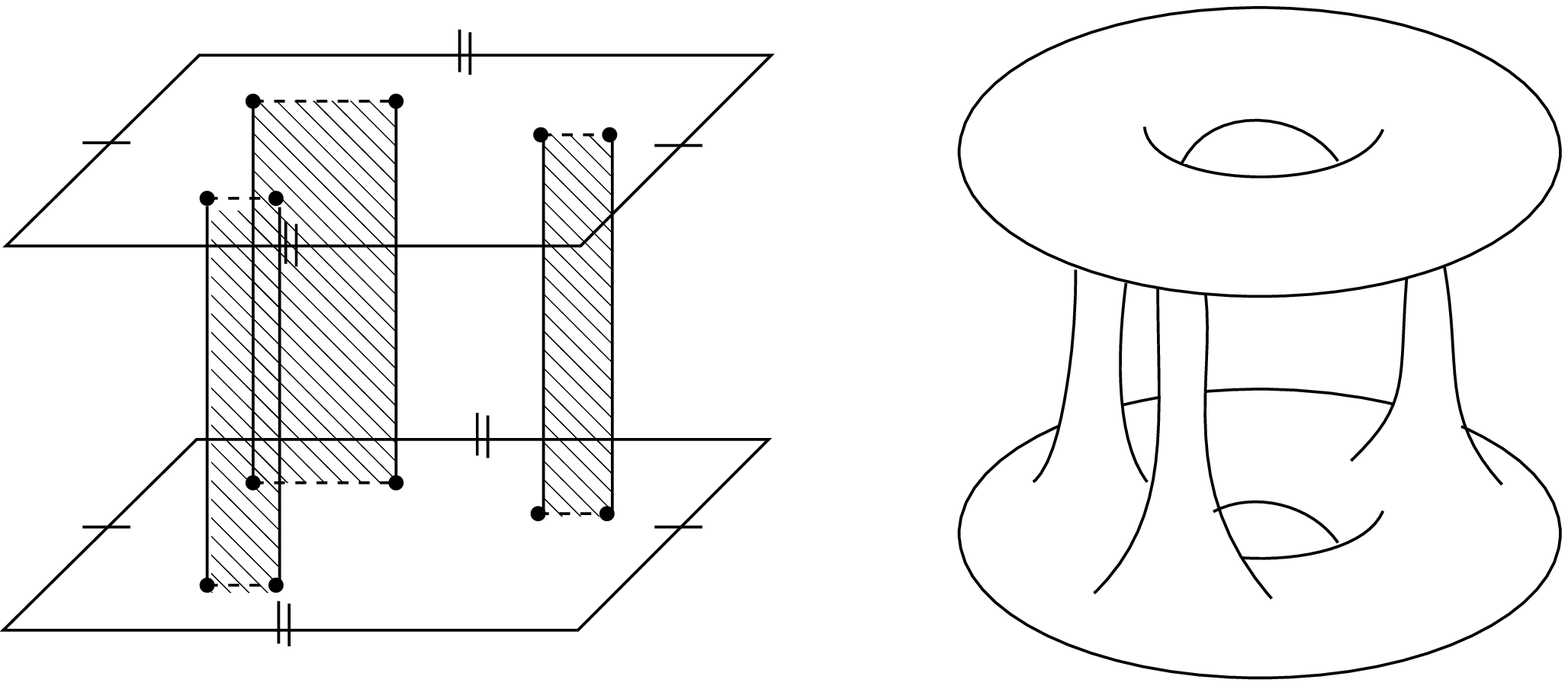}
\end{center}
\caption{\small The solution involves a double cover of the $\tilde
\T^2$ torus joined by $N$ branch cuts to create a surface of
genus $N+1$.}
\label{mm6}\end{figure}

The solution of these conditions naturally leads to a curve
which is the
double cover of the torus $\tilde \T^2$ for which the ${\cal C}_i$ are
$N$ square-root branch cuts joining the 2 sheets. This geometry is
illustrated in Fig.~\ref{mm6}.
There is a natural
involution which exchanges the two sheets. In particular, we
can trivially solve \eqref{iuutf} if $\tilde\omega(x)$ is a meromorphic
function which is odd under the involution and which has, in view of
\eqref{defresf}, simple poles at $x=m_f$ of the form
\EQ{
\tilde\omega(x)=\mp\frac1{x-m_f}+{\cal O}(1)
\label{resmf}
}
on the top and bottom sheets, respectively.
So $\tilde\omega(x)dx$ is
a 1-form on $\Sigma$ whose only singularities are simple poles at
$m_f$, with residues $\mp2\pi i$, on the bottom and top sheet,
respectively, and whose integrals around the
cycles $A_j$, $j=1,\ldots,N$ and $B_j$, $j=1,\ldots,N-1$ are
\EQ{
\oint_{A_j}\tilde\omega(x)dx=-4\pi i\ ,\qquad
\oint_{B_j}\tilde\omega(x)dx=0\ ,
}
where $B_j$ is the cycle illustrated in Fig.~\ref{mm5} and the latter
integral follows from \eqref{bint}.

Since the contour $\cup_{j=1}^NA_j$ can
be pulled off the back of the top sheet, at the expense of picking up
residues at the simple poles $x=m_f$, we have
\EQ{
-4\pi iN=
\sum_{j=1}^N\oint_{A_j}\tilde\omega(x)dx=-\sum_{f=1}^{N_f}\oint_{m_f}
\tilde\omega(x)dx=-2\pi iN_f\ ,
}
by \eqref{resmf}. Hence, for consistency we find
\EQ{
N_f=2N\ ,
}
as we noted previously.

We claim that the unique solution to these conditions is
\EQ{
\tilde\omega(x)dx=-2d\log t\ ,
}
where $t$ is the function
\EQ{
t=\frac{P(x)}{\sqrt{Q(x)}}+\sqrt{\frac{P(x)^2}{Q(x)}-c}
}
with
\EQ{
P(x)=\prod_{i=1}^N\theta_1\big(\tfrac{\beta}{2i}
(x-\zeta_i)\big|\rho\big)\ ,\qquad
Q(x)=\prod_{f=1}^{2N}\theta_1\big(\tfrac{\beta}{2i}(x-m_f)
\big|\rho\big)\ .
}
Hence,
\EQ{
\tilde\omega(x)=-\frac2y\Big(P'(x)-\frac{P(x)Q'(x)}{2Q(x)}\Big)\ ,
}
where
\EQ{
y^2=P(x)^2-cQ(x)\ .
\label{geom6a}
}
Notice that \eqref{geom6a} is precisely the curve we found from the
web diagram in \eqref{geom6}. Notice that we
must also choose our $x$-origin so that
\EQ{
\sum_{i=1}^N\zeta_i=\sum_{f=1}^{2N}m_f
}
in order that $\tilde\omega(x)$ is valued on $\tilde \T^2$.
{}From this solution, the density $f''(x)$ is determined by \eqref{disc} once the
cuts ${\cal C}_j$ are identified and
where $c$ a constant which is fixed in terms of the coupling
constant $\tau$ by substituting $f''(x)$ in the second derivative of
the saddle-point equation \eqref{SpEf}.
The cuts are identified as follows. In the limit of weak coupling
$c\to0$ and the roots of $y=0$ come in pairs on $\tilde \T^2$
located in the
vicinity of each $x=\zeta_j$. The roots near $\zeta_j$ are the
ends of the cut ${\cal C}_j$.

The geometry \eqref{geom6}, \eqref{geom6a}, is the Seiberg-Witten curve of
the $U(N)$ six-dimensional $\N=(1,0)$ theory with $N_f=2N$ hypermultiplets.
The $\zeta_j$ are moduli which are determined in terms of the $a_j$'s by
the conditions \eqref{ppb}
\EQ{
a_j=\tfrac12\int_{{\cal C}_j}xf''(x)\,dx=
-\tfrac1{4\pi i}
\oint_{A_j}x\tilde\omega(x)dx=\tfrac1{2\pi i}
\oint_{A_j}x\frac{dt}t\ ,
}
from which we deduce that $\lambda=xdt/(2\pi it)$ is the Seiberg-Witten
differential.

Notice that the resulting curve $\Sigma$
is embedded holomorphically in
$\T^3\times\R$ defined by the coordinates $(x,z=\tfrac1{2\pi
  i}\log t)$. In the M-theory picture, $x$ and $z$ are identified with
the spacetime coordinates as in \eqref{coords} and the M5-brane is
wrapped on the curve.

\section{6D SYM and the 5-brane}

We have seen that 6d $(1,1)$ supersymmetric gauge theory
compactified on $\T^2$ and mass deformed by the mass parameter
$m$ has an interesting moduli space.  The moduli space is
three dimensional, given by the complex structure of $\T^2$, $\rho$,
the K\"ahler class of $\T^2$, $\tau$, and the mass parameter $m$.
The two natural $SL(2,{\bf Z})$ symmetries of $\tau, \rho$ are
combined to an $Sp(4,{\bf Z})$ symmetry when $m\not =0$.  Note that
this is a mass deformed NS 5 brane of type IIB compactified on $\T^2$.  By
a T-duality on one of the circles
of $\T^2$ this can be viewed as NS5-brane of type IIA compactified on
$\T^2$ with complex structure $\tau$ and K\"ahler structure $\rho$. Or,
lifted up to M-theory, this can be viewed as a mass deformed
M5 brane wrapped on a $\T^2$.  The dual description we have
found can also be given an M5 brane description:  namely,
we have given the dual description as an M5 brane wrapped on a
genus 2 curve embedded in $\T^4$, where the $\tau $ and $\rho$
are both complex moduli of this genus 2 curve.  This is an amusing
duality involving M5 brane where K\"ahler and complex structure on one
side are mapped to complex parameters on the other side.

It is also noteworthy that we have found  a triality symmetry
between $(\hat \tau, \hat \rho ,\tfrac{\beta m}{2\pi i} )=
(\tau -\tfrac{\beta m}{2\pi i},\rho -\tfrac{\beta m}{2\pi
  i},\tfrac{\beta m}{2\pi i})$.  The interpretation of this triality symmetry
for the M5 brane theory wrapped on a
$\T^2$ would be interesting to understand directly.

\section*{Acknowledgments}

We would like to thank N. Nekrasov for valuable discussions.
We would also like to thank the Simons Workshop on Mathematics and Physics
which resulted in this work. TH would also like to thank H. Braden, K. Ohta,
N. Dorey and P. Kumar for valuable discussions.

The research of AI and CV is supported in part by NSF grant DMS-0074329.
CV is additionally supported by NSF grant PHY-9802709.

\bibliography{physics}

\begin{thebibliography}{99}

\bibitem{KKV}
S.~Katz, A.~Klemm and C.~Vafa, ``Geometric engineering of quantum
field theories,'' {\em Nucl. Phys.} {\bf B497}, (1997) 173--195,
{{\tt hep-th/9609239}}.

\bibitem{Katz:1997eq}
S.~Katz, P.~Mayr and C.~Vafa, ``Mirror symmetry and exact solution
of 4D N = 2 gauge theories. I,'' Adv.\ Theor.\ Math.\ Phys.\  {\bf
1}, 53 (1998) {\tt hep-th/9706110}.

\bibitem{DV1}
R.~Dijkgraaf and C.~Vafa, ``Matrix models, topological strings,
and supersymmetric gauge theories,'' Nucl.\ Phys.\ B {\bf 644}
(2002) 3, {\tt hep-th/0206255}.

\bibitem{Cachazo:2001jy}
F.~Cachazo, K.~A.~Intriligator and C.~Vafa, ``A large N duality
via a geometric transition,'' Nucl.\ Phys.\ B {\bf 603}, 3 (2001)
{\tt hep-th/0103067}.

\bibitem{Nekrasov:2002qd}
N.~A.~Nekrasov, ``Seiberg-Witten prepotential from instanton
counting,'' {\tt hep-th/0206161}.

\bibitem{Vafa:1996xn}
C.~Vafa, ``Evidence for F-Theory,'' {\em Nucl. Phys.} {\bf B469},
(1996) 403--418, {{\tt hep-th/9602022}}.

\bibitem{Morrison:1996na}
D.~R.~Morrison and C.~Vafa, ``Compactifications of F-Theory on
Calabi--Yau Threefolds -- I,'' Nucl.\ Phys.\ B {\bf 473}, 74
(1996) {\tt hep-th/9602114}.

\bibitem{Morrison:1996pp}
D.~R.~Morrison and C.~Vafa, ``Compactifications of F-Theory on
Calabi--Yau Threefolds -- II,'' Nucl.\ Phys.\ B {\bf 476}, 437
(1996) {\tt hep-th/9603161}.

\bibitem{DV4}
R.~Dijkgraaf and C.~Vafa, ``N = 1 supersymmetry, deconstruction,
and bosonic gauge theories,'' {\tt hep-th/0302011}.

\bibitem{Gopakumar:1998ii}
R.~Gopakumar and C.~Vafa, ``M-theory and topological strings. I,''
{\tt hep-th/9809187}.

\bibitem{Gopakumar:1998jq}
R.~Gopakumar and C.~Vafa, ``M-theory and topological strings.
II,'' {\tt hep-th/9812127}.

\bibitem{AKMV}
M. Aganagic, A. Klemm, M. Marino, C. Vafa, ``The Topological Vertex,''
{\tt hep-th/0305132}.

\bibitem{Dijkgraaf:1996xw}
R.~Dijkgraaf, G.~W.~Moore, E.~Verlinde and H.~Verlinde, ``Elliptic
genera of symmetric products and second quantized strings,''
Commun.\ Math.\ Phys.\  {\bf 185} (1997) 197 {\tt hep-th/9608096}.

\bibitem{Strominger:1996sh}
A.~Strominger and C.~Vafa, ``Microscopic Origin of the
Bekenstein-Hawking Entropy,'' Phys.\ Lett.\ B {\bf 379}, 99 (1996)
{\tt hep-th/9601029}.

\bibitem{Katz:1999xq}
S.~Katz, A.~Klemm and C.~Vafa,
``M-theory, topological strings and spinning black holes,''
Adv.\ Theor.\ Math.\ Phys.\  {\bf 3}, 1445 (1999)
[arXiv:hep-th/9910181].


\bibitem{DV3}
R.~Dijkgraaf and C.~Vafa,
``A perturbative window into non-perturbative physics,''
{\tt arXiv:hep-th/0208048}.

\bibitem{Cachazo:2002pr}
F.~Cachazo and C.~Vafa, ``N = 1 and N = 2 geometry from fluxes,''
{\tt arXiv:hep-th/0206017}.

\bibitem{Hollowood:2003gr}
T.~J.~Hollowood, ``Five-dimensional gauge theories and quantum
mechanical matrix models,'' JHEP {\bf 0303} (2003) 039
{\tt[arXiv:hep-th/0302165]}.

\bibitem{mm1}
N.~Dorey, T.~J.~Hollowood, S.~P.~Kumar and A.~Sinkovics, ``Exact
superpotentials from matrix models,'' JHEP {\bf 0211} (2002) 039
{\tt hep-th/020908}.

\bibitem{mm2}
N.~Dorey, T.~J.~Hollowood, S.~P.~Kumar and A.~Sinkovics, ``Massive
vacua of N = 1* theory and S-duality from matrix models,'' JHEP
{\bf 0211} (2002) 040 {\tt hep-th/0209099}.


\bibitem{DV2} R.~Dijkgraaf and
C.~Vafa, ``On geometry and matrix models,'' Nucl.\ Phys.\ B {\bf
644} (2002) 21, {\tt hep-th/0207106]}.

\bibitem{GH}
P.~Griffiths and J.~Harris, ``Principles of Algebraic Geometry,'' Wiley 1978.

\bibitem{LB}
H.~Lange and Ch.~Birkenhake, ``Complex Abelian Varieties,''
Springer-Verlag 1992.

\bibitem{bh}
H.~W.~Braden and T.~J.~Hollowood, ``The Curve of Compactified 6D
Gauge Theories and Integrable Systems," JHEP {\bf 0312} (2003) 023,
{\tt hep-th/0311024}.

\bibitem{Ganor:2000un}
O.~J.~Ganor, A.~Y.~Mikhailov and N.~Saulina, ``Constructions of
non-commutative instantons on T(4) and K(3),'' Nucl.\ Phys.\ B
{\bf 591} (2000) 547 {\tt hep-th/0007236}.

\bibitem{Cheung:1998wj}
Y.~K.~Cheung, O.~J.~Ganor, M.~Krogh and A.~Y.~Mikhailov,
``Instantons on a non-commutative T(4) from twisted (2,0) and  little-string theories,''
Nucl.\ Phys.\ B {\bf 564} (2000) 259
{\tt hep-th/9812172}.

\bibitem{Dorey:2001qj}
N.~Dorey, T.~J.~Hollowood and S.~Prem Kumar, ``An exact elliptic
superpotential for N = 1* deformations of finite  N = 2 gauge
theories,'' Nucl.\ Phys.\ B {\bf 624} (2002) 95 {\tt
hep-th/0108221}.


\bibitem{Hollowood:2003ds}
T.~J.~Hollowood, ``Critical points of glueball superpotentials and
equilibria of  integrable systems,'' {\tt hep-th/0305023}.




\bibitem{Fock:1999ae}
V.~Fock, A.~Gorsky, N.~Nekrasov and V.~Rubtsov, ``Duality in
integrable systems and gauge theories,'' JHEP {\bf 0007} (2000)
028 {\tt hep-th/9906235}.


\bibitem{Braden:2001yc}
H.~W.~Braden, A.~Gorsky, A.~Odessky and V.~Rubtsov,
``Double-elliptic dynamical systems from generalized
Mukai-Sklyanin  algebras,'' Nucl.\ Phys.\ B {\bf 633} (2002) 414
{\tt hep-th/0111066}.

\bibitem{Donagi:1995cf}
R.~Donagi and E.~Witten, ``Supersymmetric Yang-Mills Theory And
Integrable Systems,'' Nucl.\ Phys.\ B {\bf 460} (1996) 299 {\tt
hep-th/9510101}.

\bibitem{Martinec:1995qn}
E.~J.~Martinec, ``Integrable Structures in Supersymmetric Gauge
and String Theory,'' Phys.\ Lett.\ B {\bf 367} (1996) 91 {\tt
hep-th/9510204}.


\bibitem{D'Hoker:1997ha}
E.~D'Hoker and D.~H.~Phong, ``Calogero-Moser systems in SU(N)
Seiberg-Witten theory,'' Nucl.\ Phys.\ B {\bf 513} (1998) 405 {\tt
hep-th/9709053}.


\bibitem{nekrasov4}
N. Nekrasov, ``Five dimensional gauge theories and relativistic integrable
systems,'' Nucl. Phys. {\bf B531} (1998) 323-344, {\tt hep-th/9609219}.

\bibitem{Witten:1997sc}
E.~Witten, ``Solutions of four-dimensional field theories via
M-theory'', {\em Nucl. Phys.} {\bf B500}, (1997) 3--42, {{\tt
hep-th/9703166}}.

\bibitem{Douglas:1996xp}
M.~R.~Douglas, S.~Katz and C.~Vafa, ``Small instantons, del Pezzo
surfaces and type I' theory,'' Nucl.\ Phys.\ B {\bf 497}, 155
(1997) {\tt hep-th/9609071}.

\bibitem{Morrison:1996xf}
D.~R.~Morrison and N.~Seiberg, ``Extremal transitions and
five-dimensional supersymmetric field  theories,'' Nucl.\ Phys.\ B
{\bf 483}, 229 (1997) {\tt hep-th/9609070]}.


\bibitem{Intriligator:1997pq}
K.~A.~Intriligator, D.~R.~Morrison and N.~Seiberg,
``Five-dimensional supersymmetric gauge theories and degenerations
of  Calabi-Yau spaces,'' Nucl.\ Phys.\ B {\bf 497}, 56 (1997) {\tt
hep-th/9702198}.



\bibitem{Leung:1998tw}
N.~C. Leung and C.~Vafa, ``Branes and toric geometry,'' {\em Adv.
Theor. Math. Phys.} {\bf 2} (1998) 91--118, {{\tt
hep-th/9711013}}.


\bibitem{Kol:1997fv}
B.~Kol, ``5d field theories and M theory,'' {\em JHEP} {\bf 9911},
026 (1999),{{\tt hep-th/9705031}},\\
O.~Aharony, A.~Hanany and B.~Kol, ``Webs of (p,q) 5-branes, five
dimensional field theories and grid diagrams,'' {\em JHEP} {\bf
9801}, 002 (1998),{{\tt hep-th/9710116}},\\
B.~Kol and J.~Rahmfeld, ``BPS spectrum of 5 dimensional field
theories, (p,q) webs and curve counting,'' {\em JHEP} {\bf 9808},
006 (1998), {{\tt hep-th/9801067}}.

\bibitem{Ooguri:1995wj}
H.~Ooguri and C.~Vafa, ``Two-Dimensional Black Hole and
Singularities of CY Manifolds,'' Nucl.\ Phys.\ B {\bf 463}, 55
(1996) {\tt hep-th/9511164}.

\bibitem{Klemm:1996bj}
A.~Klemm, W.~Lerche, P.~Mayr, C.~Vafa and N.~P.~Warner,
``Self-Dual Strings and N=2 Supersymmetric Field Theory,''
Nucl.\ Phys.\ B {\bf 477}, 746 (1996) {\tt hep-th/9604034}

\bibitem{Witten:1993yc}
E.~Witten, ``Phases of {${\cal N} = 2$} theories in two
dimensions,'' {\em Nucl. Phys.} {\bf B403} (1993) 159--222, {{\tt
hep-th/9301042}}.

\bibitem{Hori:2000kt}
K.~Hori, C.~Vafa, ``Mirror Symmetry'', {{\tt hep-th/0002222}}.

\bibitem{Hori:2000jk}
K. Hori, A Iqbal, C. Vafa, "D-branes and Mirror Symmetry", {{\tt
hep-th/0005247}}.

\bibitem{Bershadsky:1997sb}
M.~Bershadsky and C.~Vafa, ``Global anomalies and geometric
engineering of critical theories in six  dimensions,'' {\tt
hep-th/9703167}.

\bibitem{Brunner:1997gf}
I.~Brunner and A.~Karch,
``Branes at orbifolds versus Hanany Witten in six dimensions,''
JHEP {\bf 9803}, 003 (1998)
{\tt hep-th/9712143}.


\bibitem{Antoniadis:1993ze}
I.~Antoniadis, E.~Gava, K.~S.~Narain and T.~R.~Taylor,
``Topological amplitudes in string theory,''
Nucl.\ Phys.\ B {\bf 413}, 162 (1994)
{\tt hep-th/9307158}.

\bibitem{BCOV}
M.~Bershadsky, S.~Cecotti, H.~Ooguri and C.~Vafa,
``Kodaira-Spencer theory of gravity and exact results for quantum string amplitudes,''
Commun.\ Math.\ Phys.\  {\bf 165}, 311 (1994)
{{\tt hep-th/9309140}}.

\bibitem{HM}
J.~A.~Harvey and G.~W.~Moore,
``On the algebras of BPS states,''
Commun.\ Math.\ Phys.\  {\bf 197}, 489 (1998)
{{\tt hep-th/9609017}}.

\bibitem{AV}
M.~Aganagic and C.~Vafa,
``Mirror symmetry, D-branes and counting holomorphic discs,''
{{\tt hep-th/0012041}}.

\bibitem{AKV}
M.~Aganagic, A.~Klemm and C.~Vafa,
``Disk instantons, mirror symmetry and the duality web,''
Z.\ Naturforsch.\ A {\bf 57}, 1 (2002)
{{\tt hep-th/0105045}}.
\bibitem{Gopakumar:1998ki}
R.~Gopakumar and C.~Vafa, ``On the gauge theory/geometry
correspondence,'' Adv.\ Theor.\ Math.\ Phys.\  {\bf 3}, 1415
(1999) {\tt hep-th/9811131}.

\bibitem{AMV}
M. Aganagic, M. Marino, C. Vafa, ``All Loop Topological String Amplitudes
from Chern-Simons Theory,'' {\tt hep-th/0206164}.

\bibitem{DG}
D.~E.~Diaconescu, B.~Florea and A.~Grassi,
``Geometric transitions, del Pezzo surfaces and open string instantons,''
Adv.\ Theor.\ Math.\ Phys.\  {\bf 6}, 643 (2003)
{{\tt hep-th/0206163}}.

\bibitem{ORV}
A.~Okounkov, N.~Reshetikhin and C.~Vafa,
``Quantum Calabi-Yau and classical crystals,''
{\tt hep-th/0309208}.

\bibitem{Nekrasov:2003rj}
A.~Okounkov, N.~Nekrasov,
``Seiberg-Witten theory and random partitions,''
{\tt hep-th/0306238}.

\bibitem{review}
N.~Dorey, T.~J.~Hollowood, V.~V.~Khoze and M.~P.~Mattis, ``The
calculus of many instantons,'' Phys.\ Rept.\  {\bf 371} (2002) 231
{\tt hep-th/0206063}.

\bibitem{Hollowood:2002zv}
T.~J.~Hollowood, ``Testing Seiberg-Witten theory to all orders in
the instanton expansion,'' Nucl.\ Phys.\ B {\bf 639} (2002) 66
{\tt hep-th/0202197}.

\bibitem{Hollowood:2002ds}
T.~J.~Hollowood,
``Calculating the prepotential by localization on the moduli space of  instantons,''
JHEP {\bf 0203} (2002) 038 {\tt hep-th/0201075}.


\bibitem{Gorsky:2000px}
A.~Gorsky and A.~Mironov, ``Integrable many-body systems and gauge
theories,'' {\tt hep-th/0011197}.

\bibitem{Iqbal}
A.~Iqbal,
``All genus topological string amplitudes and 5-brane webs as Feynman  diagrams,''
{{\tt hep-th/0207114}}.

\bibitem{DB}
D. -E. Diaconescu, B. Florea, "Localization and Gluing of Topological
Amplitudes," {\tt hep-th/0309143}.


\bibitem{zhou0}
C. -C. M. Liu, K. Liu, J. Zhou, ``A Formula of Two-Partition Hodge Integrals,"
{\tt math.AG/0310272}.


\bibitem{zhou00}
C. -C. M. Liu, K. Liu, J. Zhou, ``A Proof of a Conjecture of Marino-Vafa on Hodge Integrals," {\tt math.AG/0306434}.

\bibitem{OP}
A. Okounkov, R. Pandharipande, ``Hodge integrals and invariants of the unknot",
{\tt math.AG/0307209}.

\bibitem{zhou01}
C. -C. M. Liu, K. Liu, J. Zhou, ``Marino-Vafa Formula and Hodge Integral Identities",
{\tt math.AG/0308015}.

\bibitem{zhou1}
J. Zhou, ``Hodge integrals, Horwitz numbers, and symmetric groups,''
{\tt math.AG/0308024}.

\bibitem{zhou2}
J. Zhou, ``A Conjectue on Hodge integrals,'' {\tt math.AG/0310282}.

\bibitem{zhou3}
J. Zhou, ``Localization on moduli spaces and free field realization of Feynman rules,''
{\tt math.AG/0310283}.

\bibitem{AI2}
A. Iqbal, A. -Kian. Kashani-Poor, ``$SU(N)$ geometries and topological string
amplitudes,'' {\tt hep-th/0306032}.

\bibitem{AI1}
A. Iqbal, A. -Kian. Kashani-Poor, ``Instanton counting and Chern-Simons theory,''
{\tt hep-th/0212279}.

\bibitem{fucito1}
U. Bruzzo, F. Fucito, J. F. Morales, A. Tanzini, ``Multi-instanton calculus
and equivariant cohomology,'' JHEP {\bf 0305} (2003) 054,
{\tt hep-th/0211108}.

\bibitem{FP}
R. Flume, R. Poghossian, ``An algorithm for the microscopic evaluation
of the coefficients of the Seiberg-Witten prepotential,'' Int. J. Mod. Phys. {\bf A18} (2003)
2541, {\tt hep-th/0208176}.

\bibitem{NY}
H. Nakajima, K. Yoshioka, ``Instanton counting on blowup, I,"
{{\tt math.AG/0306198}}.

\bibitem{refined}
A. Iqbal, C. Koz\c{c}az, C. Vafa, ``The refined topological vertex," {\tt arXiv:hep-th/0701156v1}.

\bibitem{me1}
N.~Dorey, T.~J.~Hollowood, S.~P.~Kumar and A.~Sinkovics,
``Exact superpotentials from matrix models,''
JHEP {\bf 0211} (2002) 039
{\tt hep-th/0209089}.


\bibitem{me2}
N.~Dorey, T.~J.~Hollowood, S.~P.~Kumar and A.~Sinkovics,
``Massive vacua of N = 1* theory and S-duality from matrix models,''
JHEP {\bf 0211} (2002) 040
{\tt hep-th/0209099}.

\bibitem{EK}
T. Eguchi, H. Kanno, ``Topological Strings and Nekrasov's formulas,''
{\tt hep-th/0310235}.

\bibitem{SW},
A. N. Schellekens, N. P. Warner, ``Anomalies, Characters and Strings,''
Nucl. Phys. {\bf B287}: 317, 1987.

\bibitem{LSW}
E.~Witten,
``The Index Of The Dirac Operator In Loop Space,''
\,PUPT-1050\,\,\,,
{\it Proc. of Conf. on Elliptic Curves and Modular Forms in
Algebraic Topology, Princeton, N.J., Sep 1986.}


\bibitem{macdonald}
I. G. Macdonald, ``Symmetric functions and Hall polynomials,''
(second edition, 1995), Oxford Mathematical Monographs,
 Oxford Science Publications.

\end{thebibliography}

\end{document}